\documentclass[%
 reprint,
 superscriptaddress,
 amsmath,amssymb,
 aps, pre,nolongbibliography
]{revtex4-2}

\usepackage[usenames,dvipsnames]{xcolor}
\usepackage{amsthm}
\usepackage{amssymb}

\usepackage{graphicx}% Include figure files
\usepackage{dcolumn}% Align table columns on decimal point
\usepackage{bm}% bold math
\usepackage[utf8]{inputenc}
\usepackage[english]{babel} 
\usepackage{comment}

\usepackage{bigstrut}
\usepackage{float}

\usepackage[unicode, colorlinks, urlcolor=blue, linkcolor=blue, pagecolor=blue, citecolor=blue]{hyperref} 

\let\oldcfrac\cfrac
\renewcommand{\cfrac}[2]{\oldcfrac{#1}{#2}\,}
\usepackage{cancel}

\usepackage{diagbox}

\addto\captionsenglish{}
\begin{document}

\title{Stability of a Nondegenerate Two--Component Weakly Coupled Plasma}

\newcommand\JIHT{Joint Institute for High Temperatures, Izhorskaya 13 Bldg 2, Moscow 125412, Russia}
\newcommand\MIPT{Moscow Institute of Physics and Technology, Institutskiy Pereulok 9, Dolgoprudny, Moscow Region, 141701, Russia}

\author{G. S. Demyanov}
\email[Corresponding author: ]{demyanov.gs@phystech.edu}
\affiliation{\JIHT}
\affiliation{\MIPT}
\author{P. R. Levashov}
\email{pasha@jiht.ru}
\affiliation{\JIHT}
\affiliation{\MIPT}

\date{\today}

\begin{abstract}
The paper discusses the problem of stability of a two-component plasma and proposes a consistent consideration of quantum and long-range effects 
to calculate the thermodynamic properties of such a plasma.
We restrict ourselves by the case of a non-degenerate plasma to avoid the fermionic sign problem and consider the weakly coupled regime. Long-range interaction effects are taken into account using the angular-averaged Ewald potential (AAEP). 
To calculate thermodynamic properties, we apply both classical Monte Carlo (CMC) and path integral Monte Carlo (PIMC) methods.
%We apply both classical Monte Carlo and path integral Monte Carlo (PIMC) methods to calculate thermodynamic properties. 
A special method is developed to correctly calculate potential energy in PIMC simulations with long-range interaction effects. Our theoretical estimations show that the probability of a bound state formation is very low at a coupling parameter $\Gamma~\le~0.01$, so both classical and quantum simulations give the same energy at $\Gamma~\leq~0.01$, and the thermodynamic limit coincides with the Debye--H\"uckel theory. 
%Convergence to the limit improves if long-range interaction effects are included into the potential. 
At higher $\Gamma$, the $N$-convergence is lost due to the formation of bound states.

\end{abstract}

\maketitle

\section{Introduction}

Stability of matter has been a long--standing problem in physics since the discovery of electrons and atomic nuclei. The long--range nature of the Coulomb potential as well as its singularity at zero distance between two unlike charges make it really difficult to prove that Coulomb systems do not collapse or explode. Indeed, for a classical system of $N$ particles interacting with a two-body potential $v(\mathbf{r})$, the following two conditions should be satisfied \cite{Baus:PR:1980}: \eqref{eq:stab} the condition of stability:

\begin{equation}\tag{i}
   \label{eq:stab}
   V_N =\frac{1}{2}\sum\limits_{i, j = 1}^N v(\mathbf{r}_i - \mathbf{r}_j) \ge -AN;\ A\ge 0,
\end{equation}
where the potential energy per particle $V_N/N$ should be bounded from below by a constant $A$, and \eqref{eq:tempering} the condition of weak tempering:

\begin{equation}\tag{ii}
   \label{eq:tempering}
   v(\mathbf{r}) \le \frac{B}{|\mathbf{r}|^{3 + \varepsilon}}\ \text{for}\ |\mathbf{r}|\ge R,\ \varepsilon > 0,\ B > 0,
\end{equation}
in which the potential should tend to zero sufficiently fast at large interparticle distances.

The conditions \eqref{eq:stab} and \eqref{eq:tempering} are satisfied for short-range potentials such as Lennard--Jones ensuring the thermodynamic stability, the existence of the thermodynamic limit, and the equivalence of different statistical ensembles.

However, for the Coulomb potential, both conditions \eqref{eq:stab} and \eqref{eq:tempering} are not valid. In particular, this means that a classical system of charged particles is unstable (a well-known result of Earnshaw's theorem \cite{earnshaw1848nature}). For a quantum system of particles, the condition \eqref{eq:stab} should be replaced by the so-called condition of $H$-stability: the ground state of the system is bounded from below by a constant times the first power of the particle number.

The $H$-stability of a Coulomb system of particles was proved for the first time by Dyson and Lenard \cite{Dyson_Lenard:JMP:1967,Lenard_Dyson:JMP:1968}. A simpler proof was proposed by Lieb and Thirring \cite{Lieb:PRL:1975} in the framework of nonrelativistic quantum mechanics. One more remarkable fact was revealed by Dyson \cite{Dyson:JMP:1967}, who stated that at least one of the charged species should be fermions  to prevent collapse. Thus, both the uncertainty and Pauli exclusion principles provide the stability of matter. Notably, for the one-component plasma (OCP), the $H$-stability condition is fulfilled classically \cite{Lieb1975}.

The existence of the thermodynamic limit for Coulomb systems, despite the violation of condition \eqref{eq:tempering}, was proved by Lieb and Lebowitz \cite{Lieb:AM:1972} for a stable system (i.e., for a system with the energy bounded from below). This consideration is purely classical and is based on the electrical neutrality of a system and screening effects. Since an OCP satisfies these conditions ($H$-stability and neutrality), it has a thermodynamic limit.
%The OCP being a classical system has a thermodynamic limit. 
%Although OCP is a classical system, it has a thermodynamic limit.
The classical two-component plasma (TCP), however, is unstable; so no thermodynamic limit exists for this system.

It follows from the above that a proper simulation of a TCP should take into account both quantum effects and the Fermi statistics. This is possible with modern supercomputers but requires plenty of computing resources. For this reason, it is a common practice to replace a classical TCP by a system for which the condition \eqref{eq:stab} or both conditions \eqref{eq:stab}, \eqref{eq:tempering} are satisfied. 

For example, the attractive and repulsive Coulomb potentials in a TCP can be replaced by the repulsive Debye potential; the attractive Coulomb potential can be replaced by some restricted from below quantum pseudopotential (or simply by a truncated Coulomb potential). It is even possible to simulate a TCP with the Coulomb potential, while prohibiting somehow the collisions of unlike charges. Each of the listed techniques leads to a stable system with a thermodynamic limit but obviously any such system differs from a two-component Coulomb system of charges. Below we overview some particular simplifications that are often used in computer simulations.

Two major methods for simulating a TCP are based on Monte Carlo (MC) and molecular dynamics (MD) approaches. In the MC method,  thermodynamic properties are calculated by averaging them over sampled configurations in the phase space, whereas in MD, the equations of particles motion are solved. The main advantage of MD is that it allows studying both equilibrium and non-equilibrium properties of the system.

To prevent the collapse of particles in Coulomb systems, one needs to consider the uncertainty principle at small distances between the particles \cite{Hansen1987}. For this purpose, the Coulomb potential is usually replaced by a ``pseudopotential'' to satisfy the condition \eqref{eq:stab} \cite{PhysRevLett.41.1379, PhysRevA.23.2041}. Some examples of pseudopotentials include the Deutsch pseudopotential \cite{DEUTSCH1977317}, which is mostly the same as the one with the repulsive core \cite{Tiwari:PRE:2017} (see Eq.~(14) in Ref.~\cite{PhysRevA.23.2041}), the shelf Coulomb potential \cite{Zelener1977equation, Zelener_2016} (see Eq. (9) in Ref. \cite{Norman_1979}), correction at the origin by adding a very small number to the distance between the particles (see Eq. (1) in Ref. \cite{Kuzmin:PhysPlasma:2002}) or by smooth decay to a finite value  \cite{Maiorov1991, Maiorov_1995}, and others \cite{gabdullin2016thermodynamic}.

One of the most justified ways of constructing a pseudopotential taking into account the uncertainty principle is to calculate the Slater sum \cite{Butlitsky2008, Bonitz2004}.
Evaluating the sum in the first order of perturbation theory leads to the well-known Kelbg pseudopotential \cite{Kelbg:AnnDerPhys:1963, Bonitz:ContrPlasPhys:2023}, which has been used in numerous calculations \cite{Ebeling:CPP:1999, Lavrinenko:2018, Filinov:PRE:2020, Filinov:CPP:2001}.

To account for exchange effects between electrons in MD simulations, an effective repulsion is added to the interaction potential \cite{DEUTSCH1978381} (see Eq. (16) in Ref. \cite{PhysRevA.23.2041}, Eq. (5) in Ref. \cite{Lavrinenko:2018}, and Eq. (7) in Ref. \cite{Benedict:PRE:2012}). Such calculations yield both thermodynamic \cite{Tiwari:PRE:2017} and non-equilibrium \cite{Maiorov1991, Kurilenkov:1984, Khomkin2020, Zelener_2018, Morozov:PRE:2005} properties, as well as study relaxation processes \cite{Dumin:PlasPhys:2022}.
Classical Monte Carlo (CMC) modeling of a TCP with various pseudopotentials is also applied \cite{Barker1965, Barker:PR:1968, ZelNorFil72, ValNorFil74, Hansen1985}, although less frequently than MD.

Quantum molecular dynamics (QMD) in the framework of density functional theory is used to accurately account for quantum effects \cite{Redmer:CPP:2010, Holst:PRB:2008}. This approach allows calculating both thermodynamic \cite{Holst:PRB:2008, Norman:ContrPlasPhys:2019} and optical properties of plasma using the Kubo--Greenwood formula \cite{Knyazev:PhysPlas:2016}.

An alternative method to avoid pseudopotentials is to represent particles as wave packets (wave-packet MD) \cite{Zwicknagel:CPP:1993}. This method allows calculating the thermodynamic properties of a non-ideal plasma \cite{Lavrinenko:CPP:2019} and has a computational complexity of $N^2$ (vs. $N^3$ for QMD). So, compared to QMD, it takes less time (by several orders of magnitude) to compute one simulation step (see Table~I in Ref.~\cite{Lavrinenko:PRE:2021}).

A more common practice is to use the Path Integral Monte Carlo (PIMC) method \cite{Filinov_2001, Filinov:CPP:2001, Militzer:PRE:2001, Bezkrovniy:PRE:2004}. In theory, it allows obtaining exact results in any coupling and degeneracy regimes. However, the main challenge of PIMC is the fermion sign problem \cite{Alexandru:RevModPhys:2022}, which can be approximately solved by reducing the density matrix to a determinant form \cite{Filinov:PhysPlas:2022, Larkin:PhysPlas:2021} or a product of determinants \cite{DORNHEIM20181}. Another method is the fixed node approximation developed by Ceperley \cite{Ceperley1991}.

The properties of a hydrogen plasma were also studied using the hypernetted chain approximation \cite{Ichimaru:PhysRevA:1985, Ramazanov:PhysPlasmas:2014}, which takes into account the degeneracy of electrons. Based on these calculations, an equation of state was constructed \cite{Tanaka:PRA:1985, Ichimaru:PhysRep:1987}, which can be used to develop a broad-band plasma model \cite{ChabrierPotekhin:PRE:1998, Potekhin:PRE:2000, Potekhin:PRE:2009, PotekhinAddendum:PRE:2009, BaikoChugunov:2021}.

In MD simulations, it is crucial to monitor the formation and decay of classical bound states in a TCP. Bound states can be determined by analyzing either the particle energy \cite{Tiwari:PRE:2017} or trajectory \cite{MorozovIV:2006, Lankin2008}. Reference~\cite{Tiwari:PRE:2017} shows that the number of bound states can vary significantly depending on the depth of a  pseudopotential at small distances. Furthermore, the formation of bound states is critical to study recombination  processes \cite{Kuzmin:PhysPlasma:2002}. In CMC, the fraction of bound states can be affected by a sampling process \cite{Barker:PR:1968}. Also, in theoretical considerations, bound states pose several challenges due to the divergence of the atomic partition function \cite{Ebeling:book:1976,Lankin_2009, Starostin_2006, Starostin_2009}.

Another challenge to properly simulate a TCP is long-range effects. To take them into account, the periodic boundary conditions (PBC) are assumed \cite{Benedict:PRE:2012, Tiwari:PRE:2017} and Ewald-based techniques are applied (particle-mesh-based methods \cite{Eastwood:JCP:1974}, fast multipole \cite{Greengard:JCP:1987}, and a smooth particle mesh Ewald method \cite{Essmann:1995}). However, long-range interactions are often disregarded \cite{Zelener_2016, Lavrinenko:2018}; some authors discuss ambiguities that arise when using the Ewald procedure and PBC \cite{Maiorov1991, Morozov:PRE:2005, Lavrinenko:2016}. Interestingly,  long-range effects are almost always taken into account in OCP simulations \cite{Brush:JChemPhys:1966, Hansen:PRA:1973, Lucco:PRE:2022}.

In PIMC simulations, long-range effects are usually not considered for a TCP \cite{Filinov_2001, Militzer:PRE:2001, Storer:JMP:1968, Filinov:PhysPlas:2022, Benedict:PRE:2012}; although some recent papers address these issues \cite{Filinov:PRE:2020, MILITZER201688}.
The neglect of long-range interaction leads to slow $N$-convergence to the thermodynamic limit for structures with a short- and long-range order.
Contrary to a TCP, the Ewald potential is widely used in simulations of uniform electron gas systems \cite{Fraser:PRB:1996, Larkin:PhysPlas:2021, DORNHEIM20181, math10132270, Bohme:PRE:2023}.

In this paper, we consider an almost classical weakly coupled TCP.
Typically, these conditions arise at high temperatures or low densities.
In the latter case, such a state is called an \emph{ultracold plasma}; it is studied both experimentally \cite{Killian:PRL:1999, Kulin:PRL:2000, KILLIAN200777} and by simulations \cite{Guo:PRE:2010, Kuzmin:PhysPlasma:2002, Tiwari:PRE:2017, Dornheim:PRE:2019, Bonitz2004, Zelener_2018, Dumin:PlasPhys:2022}.

To explore stability and formation of bound states, we use  CMC and PIMC methods; the long-range effects are taken into account via the angular-averaged Ewald potential (AAEP).
We utilize a recently derived expression for the high-temperature density matrix \cite{Demyanov:ContrPlasPhys:2022}, which is a generalized Kelbg \cite{Kelbg:AnnDerPhys:1963} expression for the AAEP. Special attention is paid to the PIMC simulation procedure using the AAEP.
All particles are treated as Boltzmannons due to negligible Fermi statistics effects.
We demonstrate the collapse of a classical TCP even at very low values of the coupling parameter. Then we propose a criterion for monitoring bound states to identify the conditions for classical TCP stability during CMC simulations, despite the non-fulfillment of the condition  \eqref{eq:stab}.
Additionally, we use PIMC simulations to show the formation of bound states as the coupling parameter increases. Finally, we analyze the thermodynamic limit for a weakly coupled TCP and the consistency of both quantum and classical computational approaches. 
%Our results show that taking into account the effects of long-range interaction improves convergence to the thermodynamic limit. 
%This allows us to 
We get the coincidence of the obtained MC energy with the asymptotic Debye--H\"uckel formula at $\Gamma = 0.01$, which validates our generalized Kelbg pseudopotential \cite{Demyanov:ContrPlasPhys:2022} and the simulation method.

The article is organized as follows. Section \ref{sec:tcp} describes the AAEP and plasma parameters. Section~\ref{sec:simmeth} discusses  simulation methods and calculation details related to  the features of the AAEP.
In Section \ref{sec:boundstate}, we consider  the formation of bound states and estimate the number of bound states during MC simulations.
Section~\ref{sec:result} discusses the results for a TCP, in particular, the formation of bound states and their influence on the thermodynamic limit. Finally, we summarize our study in Section~\ref{sec:concl}.

\section{Two-Component Plasma \label{sec:tcp}}
In this section, we describe the Hamiltonian of the system under consideration and relate the thermodynamic parameters to the plasma parameters: the coupling strength $\Gamma$ and the degeneracy parameter $\chi$.

\subsection{Angular--averaged Ewald potential for a TCP}
We consider $N_p + N_e = N = 2N_p = 2N_e$ particles with charges $+e$ and $-e$ for $N_p$ protons and $N_e$ electrons, respectively; here, $e > 0$ is the electron charge. Thus, the system is electroneutral. The particles are placed in a cubic cell of a volume $L^3$ at positions $\textbf{r}_i$, $i = 1,\ldots, N$; we use the same index, $i$, to enumerate all the particles (electrons and protons). The periodic boundary conditions are imposed, so each $i$th particle has an infinite number of images at positions $\textbf{r}_i~+~\bm{\eta}L$, where $\bm{\eta} \in \mathbb{Z}^3$ is the integer vector. The Hamiltonian of the system is:
\begin{equation}
H = K + V_{\text{init}},
\end{equation}
where $K$ and $V_{\text{init}}$ are the kinetic and potential energy, respectively. If we consider a quantum system, the Hamiltonian as well as kinetic and potential energy are operators:
\begin{equation}
\label{eq:kinenergyop}
H\to \hat{H}, K\to\hat{K} = \sum_{i = 1}^N\cfrac{\hat{\textbf{p}}_i^2}{2 m_i},
\end{equation}
where $m_i$ is the mass ($m_e$ and $m_p$ for electrons and protons, respectively), and $\hat{\textbf{p}}_i$ is the momentum operator of the $i$th particle. If the system is classical, the Hamiltonian is a function of momentums and coordinates:
\begin{equation}
H \to H(\textbf{p}_1,\ldots, \textbf{p}_N; \textbf{r}_1,\ldots,\textbf{r}_N), 
\end{equation}
\begin{equation}
K \to K(\textbf{p}_1,\ldots, \textbf{p}_N) = \sum_{i = 1}^N\cfrac{\textbf{p}_i^2}{2 m_i},
\end{equation}
where $\textbf{p}_i$ (without ``hat'') is the momentum, and $\textbf{r}_i$ is the coordinate of the $i$th particle.

In both classical and quantum (coordinate representation) cases, the potential energy has the same form (we use Gaussian units):
\begin{equation}
\label{eq:initPotenergy}
V_{\text{init}} = \cfrac{1}{2} \sideset{}{'}\sum_{\bm{\eta}}\sum_{i,j = 1}^N\cfrac{q_iq_j}{|\textbf{r}_{ij}+L\bm{\eta}|}.
\end{equation}
Here, $q_i = \pm e$ is the charge of the $i$th particle, ${\sum_{i=1}^N q_i = 0}$, and $\textbf{r}_{ij}=\textbf{r}_i~-~\textbf{r}_j$. The prime in summation means that the terms with $\bm{\eta} = \textbf{0}$ are omitted if $i = j$. The sum \eqref{eq:initPotenergy} is conditionally convergent; thus, to obtain the correct result one has to use a special Ewald summation technique  \cite{Ewald:1921, Leeuw:1980}. Hence, the familiar short-range Ewald potential is obtained representing an effective interaction potential between the particles contained only in the main cell. Being angular-dependent, it leads to needless calculations in disordered and isotropic media \cite{Demyanov:JPhysA:2022, DemyanovOCP:PRE:2022}.

Following the E. Yakub and C. Ronchi approach \cite{Yakub:2003, Yakub:2005, Yakub:JPA:2006}, we average this potential over directions and obtain the \emph{angular--averaged Ewald potential} (AAEP) in its shifted form (see Eqs. \text{(41), (56)-(59)} in Ref. \cite{Demyanov:JPhysA:2022}):
\begin{equation}
\label{eq:AAEP}
\varphi(r) = 
\begin{cases}
\cfrac{1}{r} \left[1 + \frac{r}{2 r_m}((r/r_m)^2 - 3)\right], &r\leq r_m,\\
0, &r>r_m.
\end{cases}
\end{equation}
Surprisingly, this potential coincides with the one in the simple ion-sphere model (see Fig.~5 in Ref.~\cite{Ichimaru:1982} and the reasoning above). A detailed analysis of this potential can be found in Sec. 4 of Ref. \cite{Demyanov:JPhysA:2022} (see also Ref. \cite{Fukuda2022} for non-Ewald techniques). Now, the potential energy $V_{\text{init}}$ is replaced with $V$ (see Eq. (59) in \cite{Demyanov:JPhysA:2022}):
\begin{equation}
\label{eq:classical_energy_pot}
V = U_0 + \cfrac{1}{2}\sum_{i = 1}^N\sum_{\substack{j = 1\\ j\neq i}}^{N_{s,i}}q_iq_j\varphi(r_{ij}),\quad U_0 = -\sum\limits_{i=1}^N\cfrac{3q_i^2}{4 r_m}.
\end{equation}
Here, $N_{s,i}$ is the number of ions inside the sphere centered at the $i$th ion, and $r_m = (4\pi/3)^{-1/3}L$ is the radius of the sphere with equivalent volume $L^3$. The reader unfamiliar with the AAEP can find more details in Refs. \cite{Demyanov:JPhysA:2022, DemyanovOCP:PRE:2022, Yakub:2003, Jha:2010}.

\subsection{Plasma parameters}

Thermodynamic properties at an inverse temperature $\beta = (k_BT)^{-1}$ ($k_B$ is the Boltzmann constant) can be specified by the coupling strength $\Gamma$ and the degeneracy parameter $\chi$:
\begin{equation}
\Gamma = e^2\beta/r_a, \quad \chi = n_e\Lambda^3,
\end{equation}
where $r_a = (4\pi N_e/3)^{-1/3}L$ is the mean interparticle distance between the electrons; $n_e = N_e / L^3$ is the \emph{electron} density; $\Lambda = (2\pi \hbar^2 \beta/m_e)^{1/2}$ is the electron thermal de Broglie wavelength. The pair $(r_s, \theta)$ is often used as an alternative to $(\Gamma, \chi)$, where $r_s = r_a / a_B$ is the Brueckner parameter, and $\theta = (\beta E_f)^{-1}$ is the reduced temperature. Here, $E_f = \frac{\hbar^2}{2m_e}(3\pi^2N_e/L^3)^{2/3}$ denotes the Fermi energy and $a_B = \hbar^2 / (m_e e^2)$ is the Bohr radius. If we set some $(\Gamma, \chi)$, the dimensionless inverse temperature and cell volume are defined as follows:
\begin{equation}
\beta E_H = \left(\frac{9\pi}{2}\right)^{1/3}\Gamma^2\chi^{-2/3},\ \left(\frac{L}{a_B}\right)^3 = 6\pi^2 \Gamma^3\chi^{-2}N_e
\end{equation}
as well as $(r_s, \theta)$:
\begin{equation}
r_s = (9\pi/2)^{1/3}\Gamma\chi^{-2/3},\quad \theta = 4/(9\pi)^{1/3}\chi^{-2/3}.
\end{equation}
Here, $E_H = m_e e^4/\hbar^2$ is the Hartree energy.

\section{Simulation Methods \label{sec:simmeth}}
In this section, we describe the  CMC and PIMC technique based upon the AAEP. We  also examine in detail the calculation method using the recently obtained generalized Kelbg pseudopotential \cite{Demyanov:ContrPlasPhys:2022}, which accounts for long-range interaction effects.

\subsection{Classical Monte Carlo}
To simulate a classical TCP, the CMC simulation technique can be used. The simulation procedure is the same as that for a OCP (see Sec. IV in \cite{Brush:JChemPhys:1966}). The average potential energy, $\beta E_{\text{pot}}$, is calculated directly using the formula \eqref{eq:classical_energy_pot}:
\begin{equation}
\label{eq:classicpotenergy}
\beta E_{\text{pot}} =  \beta U_0 + \Bigg\langle\cfrac{1}{2}\sum_{i = 1}^N\sum_{\substack{j = 1\\ j\neq i}}^{N_{s,i}}\beta q_iq_j\varphi(r_{ij}) \Bigg\rangle,
\end{equation}
where $\langle (\ldots)\rangle$ stands for the ensemble averaging; the averaged kinetic energy, $\beta E_{\text{kin}}$, is always exactly equal to $3N/2$.

Since the potential energy, \eqref{eq:classical_energy_pot}, of such a classical system is not bounded from below, charges of opposite sign stick together during the simulation, and the system energy decreases to $-\infty$ (see Fig.~\ref{fig:classical_PIMC_inf}). This problem makes it impossible to directly simulate a classical TCP for a wide range of $\Gamma$ using the CMC method.

Nevertheless, as our calculations show, at very small values of $\Gamma \leq \Gamma_{\text{max}} \ll 1$, a CMC simulation can be performed avoiding the sticking of particles. It is explained by the absence of bound states in a system with $\Gamma \leq \Gamma_{\text{max}}$ during a simulation (more details are in Sec. \ref{subsec:stability}). In this case, the initial configuration should be strongly disordered and the particles of the opposite sign should be far enough apart (a random particle configuration is a good candidate). Otherwise the potential energy may decrease to $-\infty$  even for a very weak coupling, $\Gamma \ll \Gamma_{\text{max}}$ (see Fig.~\ref{fig:classical_PIMC_inf}).

\subsection{Path Integral Monte Carlo via generalized Kelbg pseudopotential}
The thermodynamic properties of a quantum system at a temperature $T$ can be calculated from the density matrix, $\hat\rho(\beta)$, and its coordinate representation, $\rho(\textbf{R}, {\textbf{R}'}; \beta)$:
\begin{equation}
   \hat\rho(\beta) = \exp(-\beta \hat H),\quad  \rho(\textbf{R}, {\textbf{R}'}; \beta) = \langle \textbf{R} |\hat{\rho}(\beta)| {\textbf{R}'}\rangle,
   \label{eq:denmat}
\end{equation}
where $\textbf{R}$ is  the variable for the set of all the coordinates,  $\textbf{R}~\equiv~(\textbf{r}_1, \textbf{r}_2, \dots, \textbf{r}_N)$. Since we consider $\chi \ll 1$, the system obeys the Boltzmann distribution; thus, the permutations over coordinates give an infinitesimal contribution. Such particles are often called \emph{Bolzmannons}. Then the partition function $Q(\beta)$ is expressed as follows:
\begin{equation}
\label{eq:part_func1}
Q(\beta) = \mathrm{Sp} \hat{\rho}(\beta) = \int d\textbf{R} \rho(\textbf{R}, {\textbf{R}}; \beta).
\end{equation}

The PIMC technique is fully based on the (semi)-group property of the exponential function:
\begin{equation}
   \exp(-\beta\hat H) = \prod\limits_{k = 0}^{n}\exp(-\epsilon\hat H),
   \label{eq:group}
\end{equation}
where $\epsilon = \beta / (n + 1)$. 
Using Eq. \eqref{eq:group} and the ``completeness relation'' of the coordinate basis, one gets from Eq.~\eqref{eq:part_func1}:
\begin{equation}
\label{eq:part_func}
Q(\beta) = \int  \prod_{k = 0}^n\rho(\textbf{R}_k, \textbf{R}_{k + 1}; \epsilon) d \textbf{R}_k, 
\end{equation}
where $\textbf{R}_{n + 1} = \textbf{R}_0$, and $\textbf{R}_{k} = (\textbf{r}_{1,k}, \textbf{r}_{2,k}, \dots, \textbf{r}_{N,k})$ is a set of all beads at the $k$th layer.
Now each $i$th particle is considered as a closed discrete path $(\textbf{r}_{i,0}, \textbf{r}_{i,1},\ldots, \textbf{r}_{i,n}, \textbf{r}_{i,0})$, or a number of ``beads'', $\textbf{r}_{i,k}$.

The density matrix satisfies the Bl\"{o}ch equation (see Eq. (2.53) in Ref.~\cite{Feynman:1972:SMS}). Its solution in the first order of a perturbation theory was obtained by Kelbg \cite{Kelbg:AnnDerPhys:1963, Kelbg:78:1963}:
\begin{widetext}
\vspace{-15pt}
\begin{equation}
\label{eq:dmkelbg}
\rho(\textbf{R}, {\textbf{R}'}; r_m, \beta)
 =
\left(\cfrac{\sqrt{m_em_p}}{2\pi\hbar^2\beta}\right)^{3N/2}
e^{-\tfrac{1}{2\hbar^2\beta}\sum\limits_{i=1}^Nm_i(\textbf{r}_i-\textbf{r}'_i)^2}e^{-\beta U_0}\exp\left\{
-
\frac{\beta}{2}\sum_{i=1}^{N}\sum_{\substack{j = 1\\ j\neq i}}^{N_{s,i}}q_iq_j \Phi(\textbf{r}_{ij},\textbf{r}'_{ij}; r_m, \beta)\right\},
\end{equation}
\end{widetext}
where $\Phi(\textbf{r}_{ij},\textbf{r}'_{ij};r_m, \beta)$ denotes the Kelbg functional in the case of the AAEP:
\begin{multline}
\label{eq:kelbgfuncDef}
\Phi(\textbf{r}_{ij},\textbf{r}'_{ij}; r_m, \beta) = {}\\
 \cfrac{1}{8\pi^3}
\int\limits_0^{1}d\alpha
\int {\varphi}(\textbf{t};r_m)
e^{
i \textbf{t}\textbf{d}_{ij}(\alpha)  
}e^{
-\alpha(1-\alpha)
\lambda_{ij}^2(\beta) t^2 
}
 d\textbf{t}.
\end{multline}
Here, $\lambda^2_{ij}(\beta) = \hbar^2\beta/(2\mu_{ij})$, $\mu_{ij}^{-1} = m_i^{-1} + m_j^{-1}$ is the reduced mass, and ${\varphi}(\textbf{t}; r_m)$ is the Fourier transform of the AAEP.
Here and further,
\begin{multline}
\label{eq:sdfdsfvsfs}
\textbf{d}_{ij}(\alpha) = \alpha\textbf{r}_{ij}+(1-\alpha)\textbf{r}'_{ij},  \\
d_{ij}(\alpha) = |\alpha\textbf{r}_{ij}+(1-\alpha)\textbf{r}'_{ij}|,
\end{multline}
and $\textbf{r}'_{ij} = \textbf{r}'_{i}-\textbf{r}'_{j}$. We refer to $\Phi(\textbf{r}_{ij},\textbf{r}'_{ij}; r_m, \beta)$ as the Kelbg-AAE pseudopotential (Kelbg-AAEPP).
The step-by-step derivation of Eq. \eqref{eq:dmkelbg} in the general form can be found elsewhere \cite{Demyanov:Kelbg:2022, Ebeling:CPP:1967}.

We use the Path Integral representation for both electrons and protons. So each electron and proton is represented by a closed path with a characteristic particle size $\Lambda$ and $\Lambda \sqrt{m_e/m_p}$, respectively.

\begin{figure}[h!]
\centering
\includegraphics[width=1\linewidth]{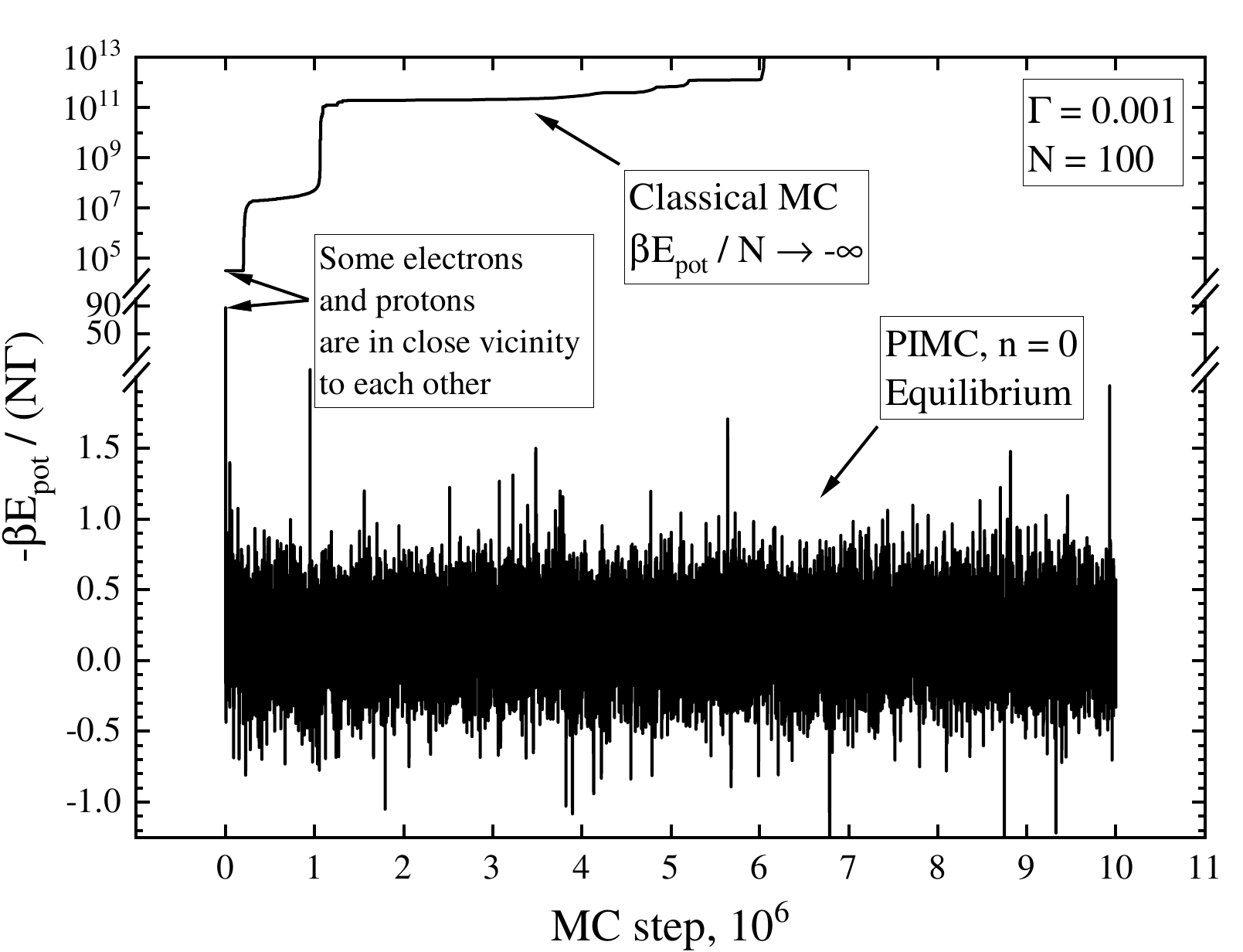}
\caption{Minus potential energy during the simulation by CMC and PIMC ($n = 0$). The initial positions of the particles are the same. In this case, some protons and electrons are in close vicinity of each other. The PIMC simulation quickly approaches the equilibrium section; meanwhile, during the CMC simulation, the potential energy decreases unboundedly.}
\label{fig:classical_PIMC_inf}
\end{figure}

The Kelbg pseudopotential is finite at zero distance, which ensures that the energy of a system is bounded from below; thus, the ``stability of the first kind'' \eqref{eq:stab} is fulfilled \cite{Lieb:2004}. It also eliminates the sticking of particles.  For example (see Fig.~\ref{fig:classical_PIMC_inf}), if the initial configuration consists of protons and electrons, which are in the vicinity of each other, the energy of the system reaches the equilibrium for $\Gamma \ll 1$ during PIMC simulations. In contrast, the CMC simulation results in the decrease of energy  to $-\infty$ in this case.

We rewrite the product of density matrices in Eq. \eqref{eq:part_func} in the following form:
\begin{multline}
\prod_{k = 0}^n\rho(\textbf{R}_k, \textbf{R}_{k + 1}; r_m, \epsilon) \\= \left(\cfrac{\sqrt{m_em_p}}{2\pi\hbar^2\epsilon}\right)^{\frac{3N(n + 1)}{2}}e^{-\beta U_0} e^{-S(\textbf{R}_0,\ldots,\textbf{R}_{n+1};r_m, \epsilon)},
\end{multline}
where
\begin{multline}
\label{eq:action}
S(\textbf{R}_0,\ldots,\textbf{R}_{n+1};r_m, \epsilon) = \sum_{k = 0}^n\Bigg(
\sum_{i = 1}^N\cfrac{m_i(\textbf{r}_{i,k} - \textbf{r}_{i, k+1})^2}{2\hbar^2 \epsilon} \\+ \cfrac{\epsilon}{2}\sum_{i = 1}^N\sum_{\substack{j = 1\\ i\neq j}}^{N_{s,i}}q_iq_j\Phi(\textbf{r}_{ij,k},\textbf{r}_{ij,k+1}; r_m, \epsilon)
\Bigg).
\end{multline}
We refer to $S(\textbf{R}_0,\ldots,\textbf{R}_{n+1};r_m, \epsilon)$ as the \emph{action}; here,  $\textbf{r}_{ij,k} = \textbf{r}_{i,k}-\textbf{r}_{j,k}$.

In the case of AAEP, \eqref{eq:AAEP}, we obtain:
\begin{multline}
\label{eq:PhiAvDefInt}
\Phi(\textbf{r}_{ij},\textbf{r}'_{ij};r_m, \beta) = \Phi_0(\textbf{r}_{ij},\textbf{r}'_{ij};\beta) \\+ \Phi_1(\textbf{r}_{ij},\textbf{r}'_{ij};r_m,\beta),
\end{multline}
where $\Phi_0(\textbf{r}_{ij},\textbf{r}'_{ij};\beta)$ is the familiar Kelbg pseudopotential (see Eq. (3) in \cite{FilinovA:PRE:2004} and Eq. (97) in \cite{Demyanov:Kelbg:2022}):
\begin{equation}
\label{eq:kelbgpseudo}
\Phi_0(\textbf{r}_{ij},\textbf{r}'_{ij};\beta) =
\int\limits_0^{1}
\cfrac{d\alpha}{ d_{ij}(\alpha)} \mathrm{erf}\left(\cfrac{d_{ij}(\alpha) / \lambda_{ij}(\beta)}{2\sqrt{\alpha(1-\alpha)}}\right)
\end{equation}
with diagonal elements:
\begin{multline}
\label{eq:diagkelbgtheta0}
\Phi_0(\textbf{r}_{ij},\textbf{r}_{ij};\beta) \\=  \cfrac{1}{ r_{ij}}\left(1 - e^{-\left(\frac{r_{ij}}{\lambda_{ij}(\beta)}\right)^2} + \frac{\sqrt{\pi}r_{ij}}{\lambda_{ij}(\beta)} \mathrm{erfc}\left(\frac{r_{ij}}{\lambda_{ij}(\beta)}\right)\right).
\end{multline}
The additional term $\Phi_1(\textbf{r}_{ij},\textbf{r}'_{ij};r_m, \beta)$ in Eq. \eqref{eq:PhiAvDefInt}, which accounts for long range interactions, is given in App.~\ref{app:suppl_formulas}, Eq.~\eqref{eq:nondiagAAEPP} (or in Eqs. (29), (30) in Ref.~\cite{Demyanov:ContrPlasPhys:2022}); its diagonal elements are defined by Eq.~\eqref{eq:phi1Def} (or by Eq.~(42) in \cite{Demyanov:ContrPlasPhys:2022}).

The average full energy is the sum of the average kinetic and average potential energy:
\begin{equation}
\label{eq:full_energy_pimc}
\beta E = \beta E_{\text{kin}} + \beta E_{\text{pot}} = -\beta\cfrac{\partial \ln Q(\beta)}{\partial \beta} .
\end{equation}
Differentiating (see Eq. (40) in Ref. \cite{Bohme:PRE:2023}), we obtain:
\begin{equation}
\!\!
\beta E_{\text{kin}} =  \cfrac{3N}{2}(n + 1)- \Bigg\langle\sum_{k = 0}^n\sum_{i = 1}^N\cfrac{m_i(\textbf{r}_{i,k} - \textbf{r}_{i, k+1})^2}{2\hbar^2 \epsilon}\Bigg\rangle,
\end{equation}
\begin{multline}
\label{eq:potenergyav}
\beta E_{\text{pot}} = \cfrac{\epsilon}{2}\Bigg\langle
\sum_{k = 0}^n\sum_{i = 1}^N\sum_{\substack{j = 1\\ i\neq j}}^{N_{s,i}}q_iq_j\Big(\Phi(\textbf{r}_{ij,k},\textbf{r}_{ij,k+1}; r_m, \epsilon) + \\
\epsilon\cfrac{\partial \Phi(\textbf{r}_{ij,k},\textbf{r}_{ij,k+1}; r_m, \epsilon)}{\partial\epsilon}\Big)
\Bigg\rangle + \beta U_0,
\end{multline}
where the ensemble average is the following:
\begin{equation} 
\langle(\ldots)\rangle = \cfrac{1}{Q(\beta)}\int(\ldots)\prod_{k = 0}^n\rho(\textbf{R}_k, \textbf{R}_{k + 1}; r_m, \epsilon) d \textbf{R}_k.
\end{equation}
In Appendix \ref{app:kinetic_energy_classical} we show that $\beta E_{\text{kin}} \xrightarrow{\chi\to 0} 3N/2$. 
The derivative of Kelbg-AAEPP, $\Phi(\textbf{r}_{ij,k},\textbf{r}_{ij,k+1}; r_m, \epsilon)$, over $\epsilon$ is presented in Eq. \eqref{eq:rgoderivbeta} (or in Eq.~(37) in Ref.~\cite{Demyanov:ContrPlasPhys:2022}).

The calculation of kinetic and potential energy are performed in the framework of the ``minimum image convention'' to calculate the interaction between the closest periodic images of beads (see Sec.~III of Ref.~\cite{Brush:JChemPhys:1966}). So each bead, $\textbf{r}_{i,k}$, is an independent ``particle''. We apply this convention to the kinetic energy to allow the path to be outside the main cell; a direct (without the convention) calculation of the difference $|\textbf{r}_{i,k} - \textbf{r}_{i, k+1}|$ may lead to a sharp increase in the action during a simulation. In other words, without the convention, all the trajectories are locked inside the main cell.

\subsection{Simulation parameters}

To obtain the equilibrium thermodynamic properties of a classical TCP with a given $N$ and $\Gamma$, we first make a number of MC steps until the equilibrium is reached; after that $m_{\text{tot}} = 10^7$ MC steps are performed. The statistical error is calculated according to standard block averaging; the equilibrium section is divided in $n_b$ blocks, with each block containing $2\times 10^6$ configurations (i.e., for $m_{\text{tot}}~=~10^7$, $n_b~=~5$). The sampling algorithm, as well as the calculation of thermodynamic properties, are described in App. \ref{app:sampling}.

Note that we consider $\Gamma \ll 1$; in this low-coupling regime, a relatively large number of particles, $N$, should be used since the Debye radius $r_D/a_B = r_s/\sqrt{6\Gamma}$, and the ratio $r_D/L = 1/(2\times 6^{1/6}\pi^{1/3}\sqrt{\Gamma}N_e^{1/3})$ diverges if $\Gamma \to 0$.

During the PIMC simulation, we use dotted particles ($n = 0$) and paths ($n \geq 1$). The degeneracy parameter $\chi$ equals to $10^{-6} \ll 1$ to eliminate the effects of antisymmetrization. The particular values of $\Gamma$, $n$ and $N$ are mentioned in the text, tables, and figures.

Note that the mass ratio, $m_p/m_e$, is equal to $1836$ during the simulation and calculation of action \eqref{eq:action}.

In CMC, \eqref{eq:classical_energy_pot}, and in PIMC for $n = 0$, the calculation of
$\sum_{{j = 1\atop j\neq i}}^{N_{s,i}}$ 
is clearly defined. The calculation algorithm is described in detail in Sec.~4 of Ref.~\cite{Demyanov:JPhysA:2022}. However, an uncertainty arises in calculating the sum
\begin{equation}
\label{eq:hjsfdfdsj}
u^{AAE}(\textbf{r}_i) = \sum\limits_{k = 0}^{n}\sum_{\substack{j = 1\\j\neq i}}^{N_{s, i}}q_j\Phi(\textbf{r}_{ij, k}, \textbf{r}_{ij, k+1};r_m, \epsilon)
\end{equation}
due to the fact that the interaction of four beads, $\textbf{r}_{ij, k}$ and $\textbf{r}_{ij, k+1}$, at once is taken into account. Below we consider how to resolve this uncertainty. 

\subsection{Summation over sphere}
\begin{figure}[h!]
\centering
\includegraphics[width=1\linewidth]{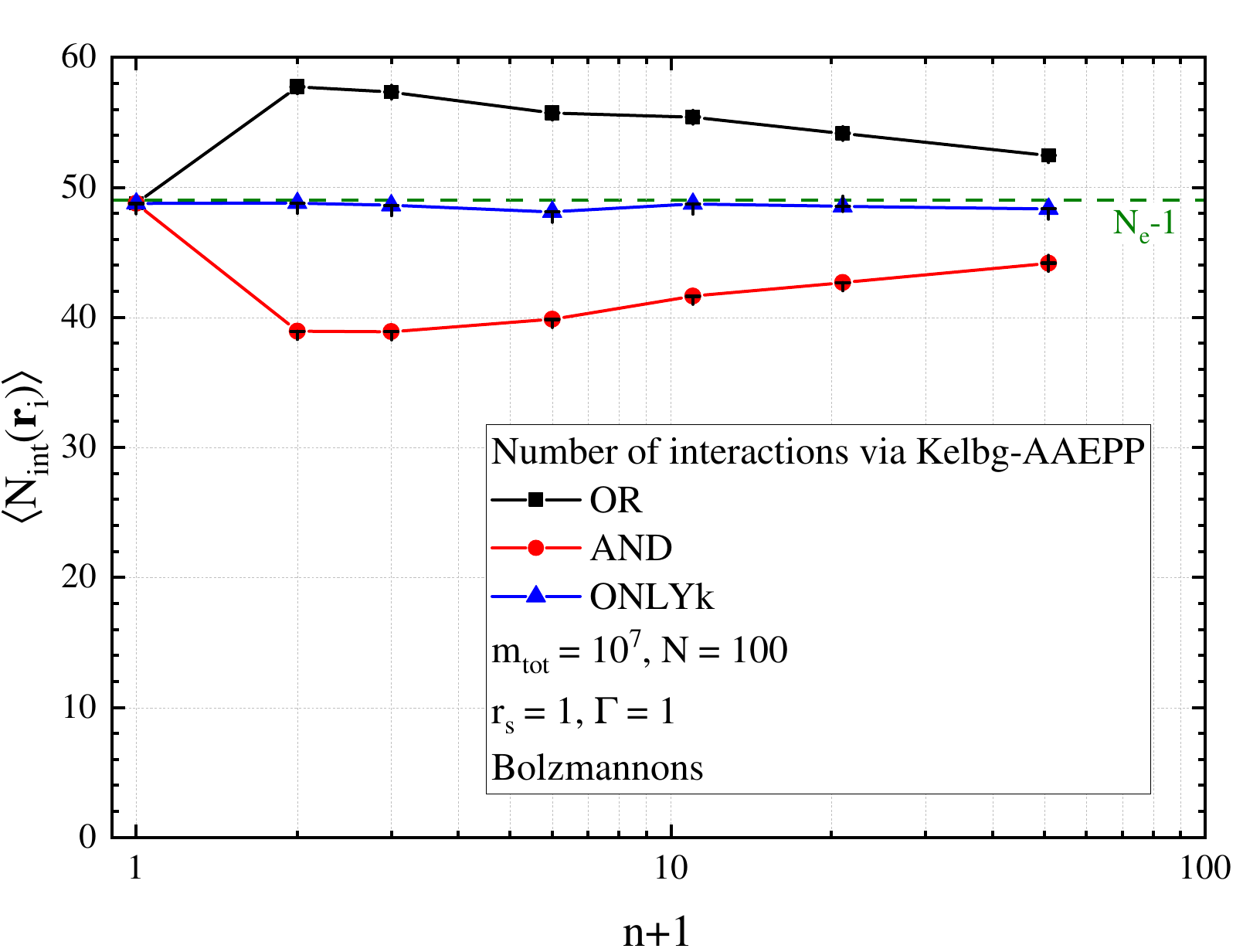}
\caption{Average number of interactions during the PIMC simulation as a function of $n$ for three different methods of accounting for interactions: OR, \eqref{eq:ormethod}; AND, \eqref{eq:andmethod}; ONLYk, \eqref{eq:onlykmethod}.
}
\label{fig:interacnumberorandonly}
\end{figure}
When calculating the ``potential'', $u^\text{AAE}(\textbf{r}_i)$, at some point $\textbf{r}_i$
there is an uncertainty in the contribution of some term $\Phi(\textbf{r}_{ij, k}, \textbf{r}_{ij, k+1};r_m, \epsilon)$ due to the interaction of four beads at once: on the layer $k$ and $k + 1$. If particles are point-like, only the particles within the sphere centered at the $i$th particle will interact with it (see Fig.~3 in Ref. \cite{Demyanov:JPhysA:2022}). If the particles are represented as a discrete path, four relations between the coordinates $\textbf{r}_{i,k}, \textbf{r}_{i,k+1}, \textbf{r}_{j,k}, \textbf{r}_{j,k+1}$ arise:
\begin{enumerate}
\item $|\textbf{r}_{ij,k}|\leq r_m$ and $|\textbf{r}_{ij,k + 1} |\leq r_m$;
\item $|\textbf{r}_{ij,k}|\leq r_m$ and $|\textbf{r}_{ij,k + 1} |> r_m$;
\item $|\textbf{r}_{ij,k}|> r_m$ and $|\textbf{r}_{ij,k + 1}|\leq r_m$;
\item $|\textbf{r}_{ij,k}|> r_m$ and $|\textbf{r}_{ij,k + 1}|> r_m$.
\end{enumerate}

For the first or fourth cases, the calculation procedure is clear: the contribution to the sum \eqref{eq:hjsfdfdsj} is $\Phi(\textbf{r}_{ij, k}, \textbf{r}_{ij, k+1};r_m, \epsilon) \neq 0$ (the beads interact) or this contribution is zero (the beads do not interact), respectively. However, the second and third cases present uncertainties: there is an interaction on the layer $k$ but not on the layer $k+1$ (and vice versa). To check the calculation method, we implemented three methods: ``OR'', ``AND'', ``ONLYk''. For convenience, $\Phi$ denotes a contribution to the sum \eqref{eq:hjsfdfdsj}.

``OR'': If $|\textbf{r}_{ij,k}|\leq r_m$ or  $|\textbf{r}_{ij,k + 1}|\leq r_m$, the contribution is nonzero:
\begin{equation}
\label{eq:ormethod}
\Phi
=
\begin{cases}
\Phi(\textbf{r}_{ij, k}, \textbf{r}_{ij, k+1};r_m, \epsilon),  \substack{ |\textbf{r}_{ij,k}|\leq r_m \\ \text{ or } |\textbf{r}_{ij,k + 1}|\leq r_m}, \\
0, \text{ else}.
\end{cases}
\end{equation}

``AND'': The contribution is nonzero if the interaction occurs on both layers immediately:
\begin{equation}
\label{eq:andmethod}
\Phi
=
\begin{cases}
\Phi(\textbf{r}_{ij, k}, \textbf{r}_{ij, k+1};r_m, \epsilon), \substack {|\textbf{r}_{ij,k}|\leq r_m  \\\text{ and } |\textbf{r}_{ij,k + 1}|\leq r_m}, \\
0, \text{ else}.
\end{cases}
\end{equation}

``ONLYk'': The contribution is nonzero if the interaction occurs only on the first layer $k$:
\begin{equation}
\label{eq:onlykmethod}
\Phi
=
\begin{cases}
\Phi(\textbf{r}_{ij, k}, \textbf{r}_{ij, k+1};r_m, \epsilon),\quad   |\textbf{r}_{ij,k}|\leq r_m, \\
0, \text{ else}.
\end{cases}
\end{equation}
The latter approach arises from the reasoning in App. \ref{app:pot_energy_nondiag}.

Let us now examine the average number of interactions
\begin{equation}
\langle N_{\text{int}}(\textbf{r}_s)\rangle = \left\langle\cfrac{1}{n + 1}\sum\limits_{k = 0}^{n}\sum_{\substack{j_e = 1\\j_e\neq s}}^{N_{s, se}}\frac{\Phi(\textbf{r}_{sj_e, k}, \textbf{r}_{sj_e, k+1};r_m, \epsilon) }{ \Phi(\textbf{r}_{sj_e, k}, \textbf{r}_{sj_e, k+1};r_m, \epsilon)}\right\rangle
\end{equation}
between electrons along the PIMC simulation for some \emph{fixed proton positions} vs. the number of partitions, $n$, for $\Gamma = r_s = 1$ and $N_e = 50$. Here, $s$ is the electron number chosen for the displacement at a current MC step, $j_e$ enumerates electrons, and $N_{s,se}$ is the number of electron beads in the sphere centered at the $s$th electron. The average number of interactions should be close to the number of neighbors in the cubic cell, $N_e - 1$. The result of the calculation is shown in Fig.~\ref{fig:interacnumberorandonly}.

In this case, the parameter $\chi \approx 3.76$, so the exchange effects for electrons should be taken into account. We carry out the calculation without antisymmetrization only to demonstrate the difference in the number of interactions.

We see that the three methods give the same results for $n = 0$. When increasing $n$, the ``OR'' method \eqref{eq:ormethod} overestimates the number of interactions compared to $N_e - 1$, while the ``AND'' one \eqref{eq:andmethod} underestimates it. In the third approach \eqref{eq:onlykmethod}, the number of interactions changes weakly with increasing $n$, and close to the average value $N_e - 1$. We also see that with increasing $n$, the average number of interactions tends to $N_e - 1$ in all cases, since $\textbf{r}_{ij, k + 1} \to \textbf{r}_{ij, k}$ at $n\to\infty$.

Thus, we use further the ``ONLYk'' method as the most reliable and weakly dependent on the number of partitions, $n$.

\section{Definition of a bound state \label{sec:boundstate}}

Since antisymmetry effects are negligible at $\chi \ll 1$, one can easily separate the kinetic and potential energies, as in Eq. \eqref{eq:full_energy_pimc}. The full Hamiltonian is $\hat{H} = U_0 + \hat{\mathcal{H}}$, where $\hat{\mathcal{H}} = \hat{K} + U(\textbf{R})$ and $U(\textbf{R}) = V - U_0$ stands for the pair interactions (see Eq. \eqref{eq:classical_energy_pot}). The Hamiltonian $\hat{\mathcal{H}}$ can be rewritten as a sum of single-particle Hamiltonians $\hat{\mathcal{H}} = \sum_{i = 1}^N\hat{h}_i$, where
$2\hat{h_i} = \hat{\textbf{p}}^2_i / m_i  + q_i\sum_{j = 1\atop j\neq i}^{N_{s,i}}q_j\varphi(r_{ij})$. Similarly, it is possible to decompose the energy by the sum of single--particle contributions for a certain particle configuration:
\begin{equation}
\beta \mathcal{E} = \sum_{i = 1}^N \beta\varepsilon_i,
\end{equation}
\begin{equation}
\label{eq:full_energy_part_shift}
\beta \varepsilon_i = \cfrac{3}{2}(n + 1) - \sum_{k = 0}^n\cfrac{m_i(\textbf{r}_{i,k} - \textbf{r}_{i, k+1})^2}{2\hbar^2 \epsilon} + \beta\tilde{U}_{i, pot}.
\end{equation}
In the CMC, the potential energy of the $i$th particle is expressed via the AAEP:
\begin{equation}
\beta\tilde{U}_{i, pot} = \cfrac{\beta}{2}\sum_{\substack{j = 1\\ j\neq i}}^{N_{s,i}}q_iq_j\varphi(r_{ij}).
\end{equation}
In PIMC, the same value is expressed via the Kelbg-AAEPP:
\begin{multline}
\beta\tilde{U}_{i, pot} = 
\cfrac{\epsilon}{2}
\sum_{k = 0}^n\sum_{\substack{j = 1\\ i\neq j}}^{N_{s,i}}q_iq_j\Big(\Phi(\textbf{r}_{ij,k},\textbf{r}_{ij,k+1}; r_m, \epsilon) \\+
\epsilon\cfrac{\partial \Phi(\textbf{r}_{ij,k},\textbf{r}_{ij,k+1}; r_m, \epsilon)}{\partial\epsilon}\Big).
\end{multline}
The feature is that both AAEP, $\varphi(r_{ij})$ and Kelbg-AAEPP, $\Phi(\textbf{r}_{ij,k},\textbf{r}_{ij,k+1}; r_m, \epsilon)$, are shifted  to vanish at large interparticle distances. Thus, it is natural to \emph{define} the bound state for a particle as follows.

If for a PIMC configuration $(\textbf{R}_0, \textbf{R}_1, \ldots, \textbf{R}_n)$ or CMC configuration $\textbf{R}$, the energy $\beta\varepsilon_i$, \eqref{eq:full_energy_part_shift}, of the $i$th  particle is negative, then the $i$th particle is in the bound state; if this energy is positive, the $i$th particle is unbounded.

\begin{figure}[h!]
\centering
\includegraphics[width=1\linewidth]{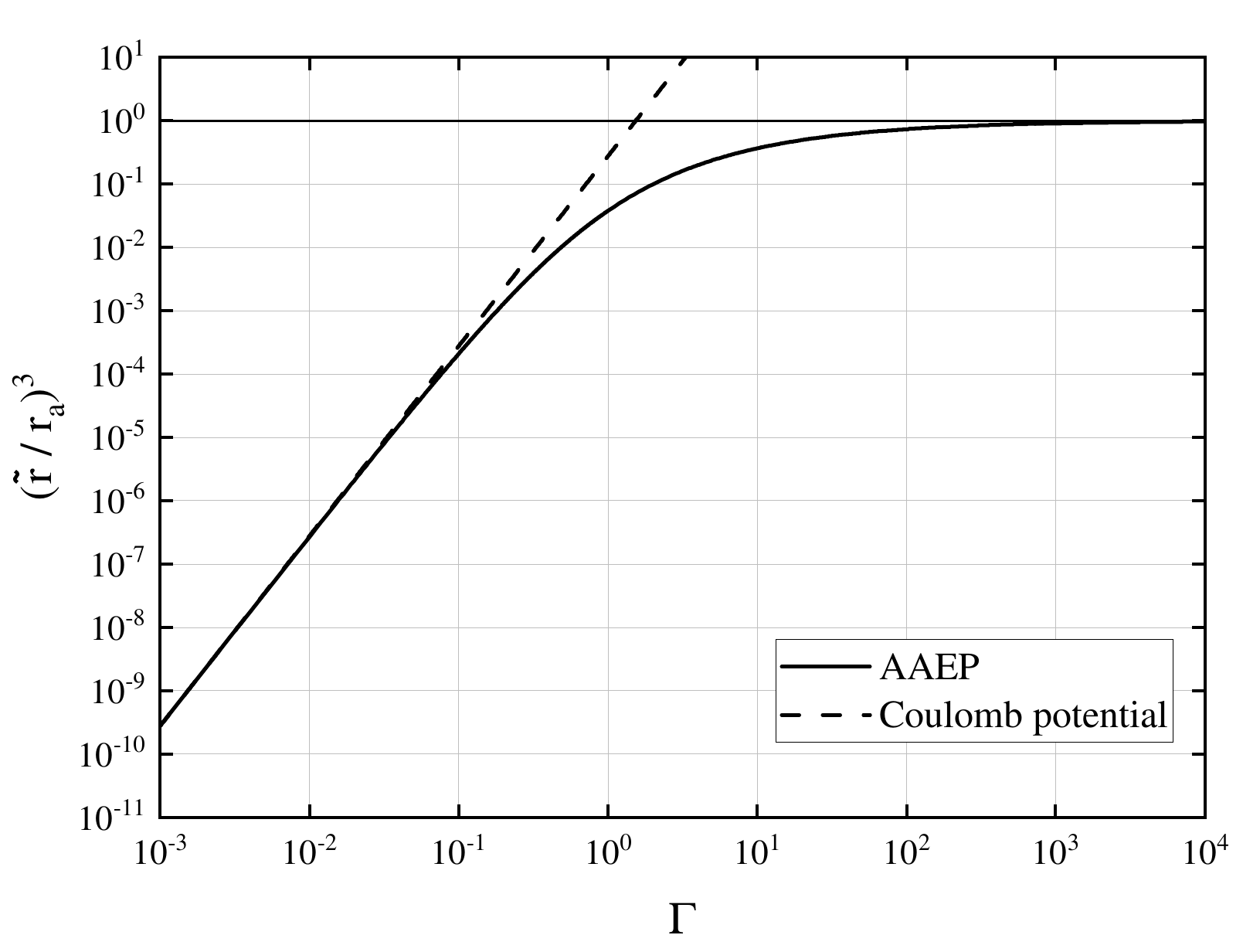}
\caption{Probability of transition into a bound state for different values of $\Gamma$ when a particle is displaced to a random position inside the computational cell during MC simulations. Here, $\tilde r$ is a solution of Eq. \eqref{eq:bound_sphere_rad}. Solid line---the AAEP, dashed line---the Coulomb potential. At $\Gamma \ll 1$, both potentials lead to identical probability. At stronger coupling, the result for the Coulomb one diverges; in contrast, the probability tends to $1$ for the AAEP in the limit $\Gamma\to\infty$. }
\label{fig:probabboundstates}
\end{figure}

Now we  \emph{estimate} the probability of transition of some particle into a bound state when it is randomly displaced in an MC simulation.
To do this, we consider only two point particles in the cell ($N_e = N_p = 1$), which interact via the AAEP. The following equation defines the radius of a sphere, $\tilde{r}$:
\begin{equation}
\label{eq:bound_sphere_rad}
\beta\varepsilon_1 = 0 \Rightarrow \cfrac{3}{2} - \cfrac{\beta e^2}{2a_B}\varphi(\tilde{r}/a_B) = 0.
\end{equation}
So, if the distance between two particles is less than $\tilde{r}$, the energies $\beta \varepsilon_1$ and $\beta \varepsilon_2$ of the electron and proton are negative: the bound state is formed. Otherwise, the state is unbound. Solving the cubic equation \eqref{eq:bound_sphere_rad}, one can find a cumbersome expression for $\tilde{r}(\Gamma, \chi)$. We write it only in the limit $\Gamma \to 0$:
\begin{equation}
\label{eq:radius_probab_bound}
\tilde{r}(\Gamma, \chi) / a_B = \cfrac{\pi \Gamma^2}{2\chi^{2/3}} + O(\Gamma^3).
\end{equation}
Here, we used the equality $\beta e^2/a_B = (9\pi/2)^{1/3}\Gamma^2\chi^{-2/3}$.
The volume of the sphere is $4 \pi \tilde{r}^3(\Gamma, \chi)/3$:
\begin{equation}
 \cfrac{4 \pi\tilde{r}^3(\Gamma, \chi)}{3 a^3_B} = \cfrac{\pi^4\Gamma^6}{6 \chi^2} + O(\Gamma^7).
\end{equation}
If a particle is displaced randomly, the probability of getting into the sphere (i.e., into the bound state) is:
\begin{equation}
\label{eq:prob_free}
\cfrac{4\pi \tilde{r}^3/3}{L^3} =\cfrac{\tilde{r}^3}{r_a^3} =  \cfrac{\pi^2 \Gamma^3}{36} + O(\Gamma ^4).
\end{equation}
Note that the ratio in \eqref{eq:prob_free} is \emph{independent of} $\chi$; it depends only on $\Gamma$. 

Equation  \eqref{eq:bound_sphere_rad} can be solved for the Coulomb potential. This results in Eqs. \eqref{eq:radius_probab_bound}--\eqref{eq:prob_free} without any corrections ``$O(\ldots)$''.
However, for the AAEP, the ratio $\tilde{r}^3/r_a^3(\Gamma) \leq 1$ for any $\Gamma$ has a meaning of probability, whereas it tends to infinity at $\Gamma \to \infty$ for the Coulomb potential.

The probability $\tilde{r}^3/r_a^3(\Gamma)$ is presented for both potentials as a function of $\Gamma$ in Fig.~\ref{fig:probabboundstates}. This function allows estimating the number of transitions to a bound state during the simulation as
\begin{equation}
\label{eq:tansit_bound_number}
(\tilde{r}/r_a)^3\times m_{\text{tot}},
\end{equation}
where $m_{\text{tot}}$ denotes the full number of MC steps.

\section{Results and discussion \label{sec:result}}
In this section, we determine the range of $\Gamma$ in which bound states are absent and propose a technique for tracking the appearance and decay of bound states during a simulation. We also calculate the thermodynamic energy limit at $\Gamma~=~0.01$ and analyze the behavior of the statistical error in the presence of bound states at $\Gamma~=~0.05$.

\subsection{$\Gamma$-region of unbound states and a condition of stability for CMC \label{subsec:stability}}

To begin, we compare the results obtained using CMC and PIMC with $n=0$, $n=10$. The AAEP and Kelbg-AAEPP are utilized in MC and PIMC, respectively. The following simulation parameters were chosen for both methods: $0.001\leq \Gamma \leq 0.1$ and $N = 100$. Note that in all cases, the Debye radius, $ r_D/a_B =  r_s/\sqrt{6\Gamma}$, is either larger than the cell size, $L$, or of similar magnitude. The $N$-dependence and thermodynamic limit are considered in Sec. \ref{subsec:therm_limit}.
\begin{table}[!ht]
    \centering
	\caption{Dimensionless potential energy per particle, $-\beta E_{\text{pot}}/(N\Gamma)$, for $N~=~100$ and various $\Gamma$. The energy is calculated using CMC and PIMC for $n = 0$ and $n = 10$. The digits in brackets correspond to one standard deviation. The bound states are not observed for $\Gamma \leq \Gamma_{\text{max}} = 0.01$. The thermodynamic limit for $\Gamma \leq \Gamma_{\text{max}}$ is considered in Sec. \ref{subsec:therm_limit}.}
	\label{tab:N100}
\begin{ruledtabular}
    \begin{tabular}{c|ccc}
        $\Gamma$ & CMC & PIMC, $n = 0$ & PIMC, $n = 10$ \\ \hline
        0.001  & 0.2463(12) & 0.24628(48) & 0.2468(14) \\ 
        0.002  & 0.2484(8)  & 0.2483(12) & 0.2491(15) \\ 
        0.005  & 0.2552(9) & 0.2559(14) & 0.2557(8) \\ 
        0.01  & 0.2661(10)  & 0.26686(65) & 0.2654(11) \\  \hline
        0.02  & 0.23(2) & 0.2881(17) & 0.2886(6)  \\ 
        0.03  & 0.45(26) & 0.33(4) & 0.312(3) \\ 
        0.04  & 0.48(20) & 0.35(4) & 0.41(7) \\ 
        0.05  & 0.51(27) & 0.7(8) & 0.363(10) \\ 
        0.06  & 1.9(6) & 0.67(18) & $2\pm 2$ \\ 
        0.1 & $9\pm 5$ & $8\pm 2$ & $3\pm 3$ \\
    \end{tabular}
\end{ruledtabular}
\end{table}

\begin{figure}[h!]
\centering
\includegraphics[width=1\linewidth]{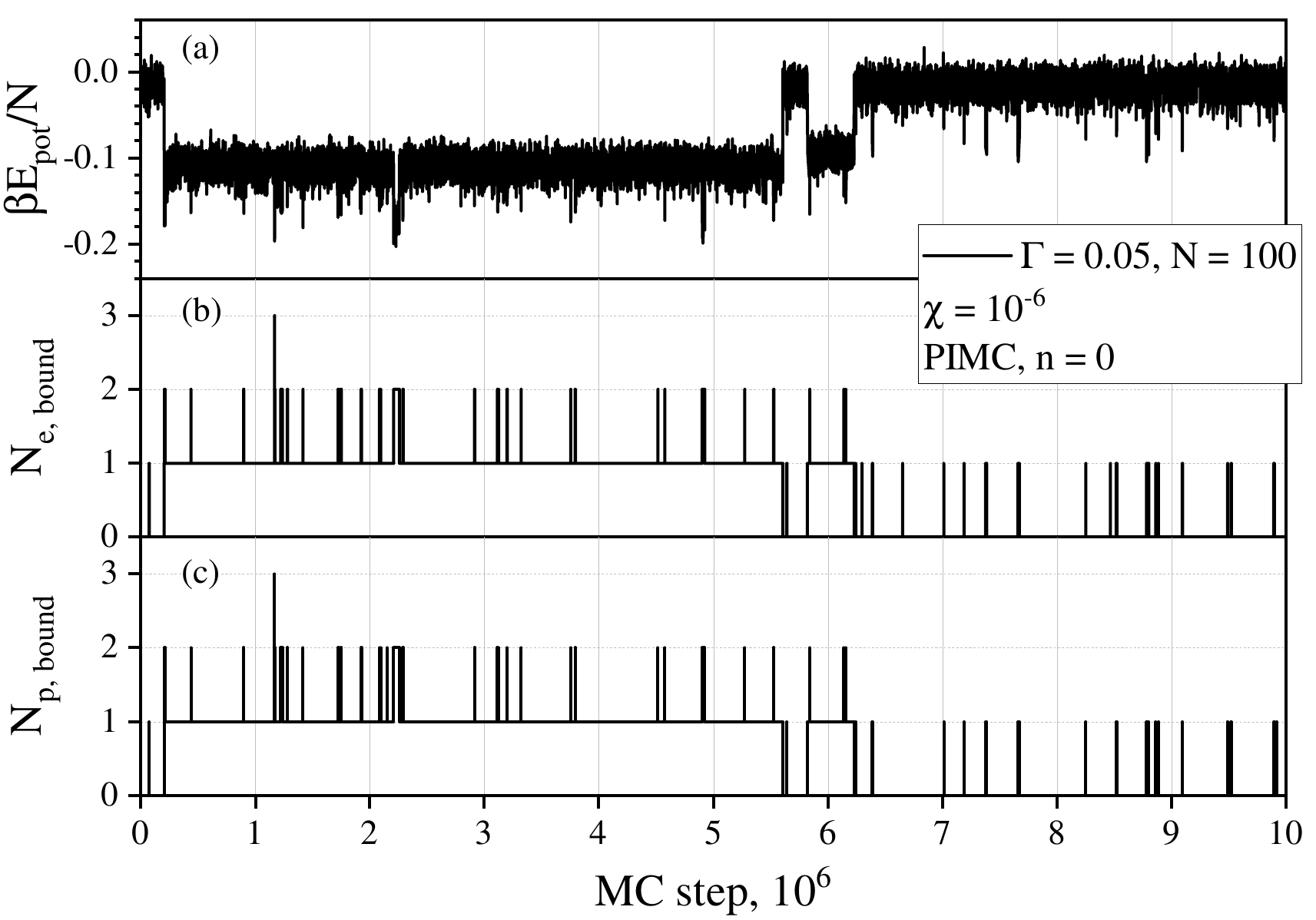}
\caption{Dimensionless potential energy per particle and the number of bound states vs. MC step in the PIMC simulation for $N = 100$ particles and $\Gamma = 0.05$. Each sharp decrease in energy is accompanied by the appearance of particles with negative energy, \eqref{eq:full_energy_part_shift}. Also, each sharp increase in energy, on the contrary, occurs when some particles jump to an unbound state.}
\label{fig:bound_states}
\end{figure}

Our simulations indicate that for $\Gamma \leq \Gamma_{\text{max}} = 0.01$, the results obtained using the three methods are in close agreement. However, at $\Gamma > \Gamma_{\text{max}}$ energy jumps are observed as a result of bound states formation (see Fig.~\ref{fig:bound_states}(a) for $\Gamma = 0.05$).  This leads to an increase in the relative statistical error of the averaged potential energy, $\beta E_{\text{pot}}$.

In both classical and path integral MC simulations, we do not observe particle sticking for small values of $\Gamma \leq \Gamma_{\text{max}} = 0.01$. This can be explained by the low probability of the bound state formation. For example, this probability is approximately $\sim 10^{-7}$ at $\Gamma = 0.01$  (as shown in Fig.~\ref{fig:probabboundstates}). Thus, for $m_{\text{tot}} = 10^7$ MC steps, only a small number of bound states would be expected to form. As the value of $\Gamma$ decreases, the probability of a bound state formation decreases proportional to $\Gamma^3$. Hence, in the MC simulation, it is nearly impossible to observe the formation of a bound state at $\Gamma \leq \Gamma_{\text{max}} = 0.01$ for $m_{\text{tot}} \sim 10^7$. This means that the CMC method can be used in the low coupling regime, $\Gamma \leq 0.01$, to obtain a reliable thermodynamic limit, as the probability of particle sticking is negligible for a sufficiently large number of MC steps. Therefore, we can say that a classical TCP can be considered stable in the $\Gamma \leq 0.01$ region, even though the energy of such a system is not bounded from below.

Next, we perform a PIMC simulation at $\Gamma = 0.05$. We monitor the number of electrons and protons that have negative energies, as defined by equation \eqref{eq:full_energy_part_shift}. The results are displayed in Fig.~\ref{fig:bound_states}. During the simulation, we observe that bound electron states formed $39\pm 33$ times in $m_{\text{tot}}/n_b$ steps: totally $193$ times. Using Eq.~\eqref{eq:tansit_bound_number} for $m_{\text{tot}}\to m_{\text{tot}}/n_b$, we predict $59$ occurrences, which agrees with the PIMC result within the error bars.

\subsection{Thermodynamic limit at $\Gamma = 0.01$\label{subsec:therm_limit}}
\begin{figure}[h!]
\centering
\includegraphics[width=1\linewidth]{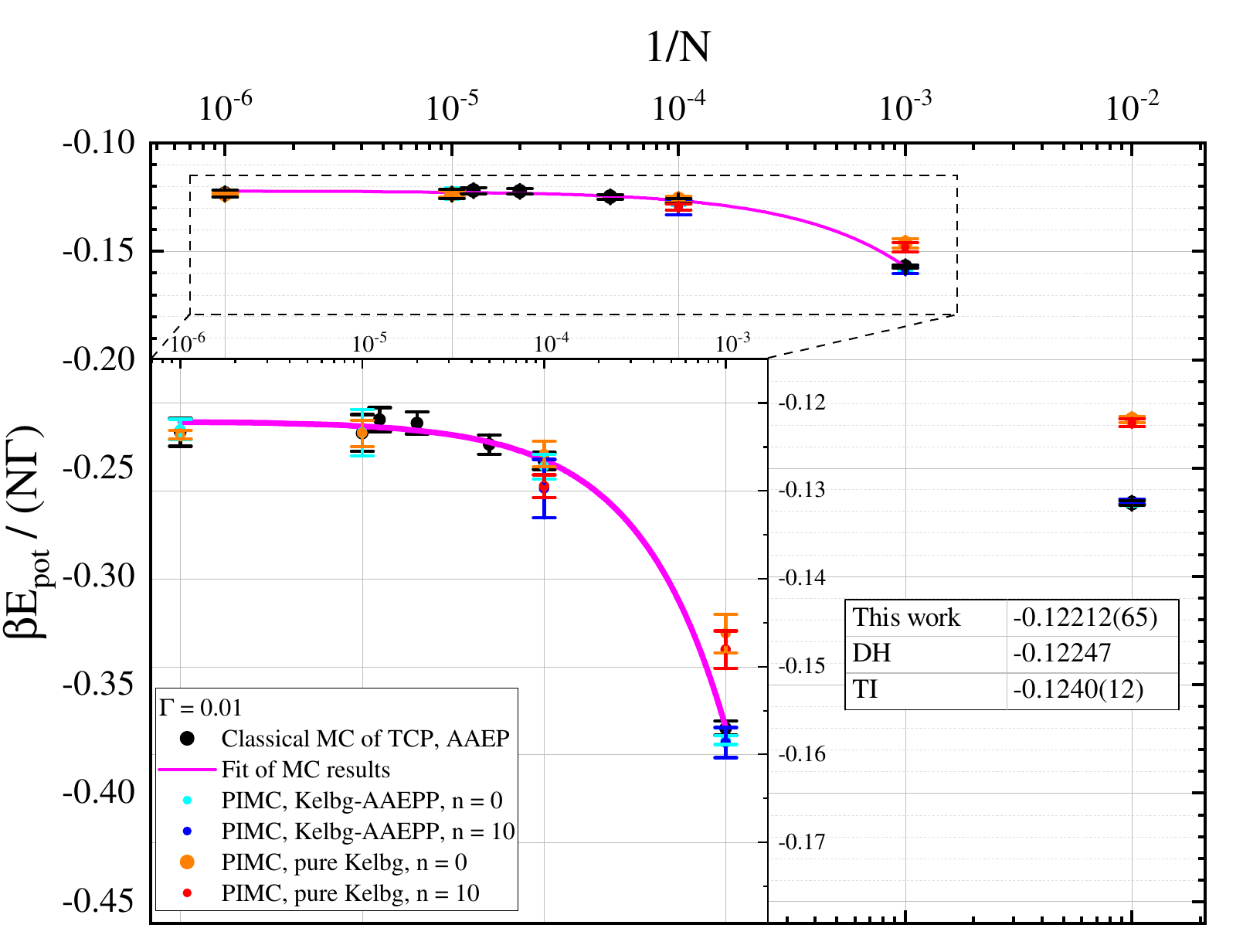}
\caption{Dimensionless potential energy per particle for $\Gamma~=~0.01$ and different $N$. The calculations were carried out using CMC and PIMC (for $n = 0$ and $n = 10$) with and without long--range interaction effects (the latter corresponds to the ``pure Kelbg'' pseudopotential). The thermodynamic limit is calculated using the results of the CMC simulation.}
\label{fig:thermlimit_g0.01}
\end{figure}

We calculate the thermodynamic limit by using a method that has been successful for an OCP (see Ref.~\cite{DemyanovOCP:PRE:2022}). The simulation is carried out via both CMC and PIMC. In the PIMC method, we use both the usual Kelbg pseudopotential (neglecting long-range interaction effects) and Kelbg-AAEPP. The results for $\Gamma = 0.01$ are shown in Fig.~\ref{fig:thermlimit_g0.01}.

In the low $\Gamma$-regime, we can consider point particles, meaning that the results for $n = 0$ and $n = 10$ are nearly the same for both pseudopotentials.  
Furthermore, the result of the CMC simulation is the same as that of the PIMC one with the Kelbg-AAEPP. 
Also, the usual Kelbg pseudopotential shows similar $N$-convergence as the Kelbg-AAEPP, which is explained by the strong disorder of the weakly-coupled plasma.

We use Eq. \eqref{eq:fitFunc} to calculate the thermodynamic limit $\beta E_{\text{pot}}/(N \Gamma) = -0.12212(65)$ for $\Gamma = 0.01$; the data of the CMC is used. We find that the obtained limit coincides with the Debye--H\"{u}ckel (DH) \cite{Debye:1923} approximation within the computational accuracy, $\beta E^{\text{DH}}_{\text{pot}}/(N \Gamma) =\sqrt{1.5\Gamma} = -0.12247$. Another theoretical result, the approximation of Tanaka and Ichimaru (TI, see Eq. (9) in Ref. \cite{Tanaka:PRA:1985} or Eq. (3.142) in \cite{Ichimaru:PhysRep:1987}) also gives a close value, $\beta E^{\text{TI}}_{\text{pot}}/(N \Gamma) = -0.1240(12)$, within the stated accuracy of TI approximation as $1\%$.

So in the range of $\Gamma \leq 0.01$, the DH approximation is valid. However, as $\Gamma$ increases, the presence of bound states causes convergence issues in PIMC simulations. We will illustrate this in detail for $\Gamma = 0.05$.

\begin{table*}[ht!]
\centering
\caption{MC results for the average potential energy \eqref{eq:potenergyav}, $-\beta E_{\text{pot}}/(N\Gamma)$. Five different methods are used for the number of particles from $N = 10^2$ to $10^6$ and $\Gamma = 0.01$; $m_{\text{tot}} = 10^7$, $n_b = 5$. The digits in parentheses correspond to one standard deviation.
}
\label{tab:addlabel1}
\begin{ruledtabular}
\begin{tabular}{c|c|cccccccc} 
Method          &     $n$    & $10^2$       & $10^3$        & $10^4$        & $2\times10^4$        & $5\times 10^4$   & $8\times 10^4$        & $10^5$      & $10^6$  \bigstrut[t] \\ 
\hline
Classical MC         &    ---          & 0.2661(10)	& 0.15702(81)          & 0.12663(97)	 & 0.1248(11)	 & 0.1223(13)  & 0.1219(13) & 0.1235(21)	 & 0.1234(16)	 \bigstrut[t]\\
\hline
Kelbg-AAEPP  &   $0$   & 0.26686(65)	& 0.15835(48)	     & 0.1273(14)		 & ---	 & ---  & --- & 0.1234(27)	 & 0.1230(12)	 \bigstrut[t]\\
\hline
Kelbg-AAEPP    &  $10$   & 0.2654(11)	& 0.1587(17)	       & 0.1298(33)		 & ---	 & --- & ---& ---	 & ---	 \bigstrut[t]\\
\hline
pure Kelbg                     &        0      &  0.2275(15)	 & 0.1463(22)	& 0.1258(15)  & ---  & ---  & ---  & 0.1235(15)	& 0.1236(45)	\bigstrut[t] \\
\hline
pure Kelbg                   &           10         &  0.2291(18)	  & 0.1481(22)         & 0.1295(13)	   & ---   &  --- & ---   & ---        & ---  	\bigstrut[t]  \\
\end{tabular}
\end{ruledtabular}
\end{table*}

\subsection{Statistical error for $\Gamma = 0.05$}
\begin{figure*}[ht!]
(a)~\includegraphics[width=0.46\linewidth]{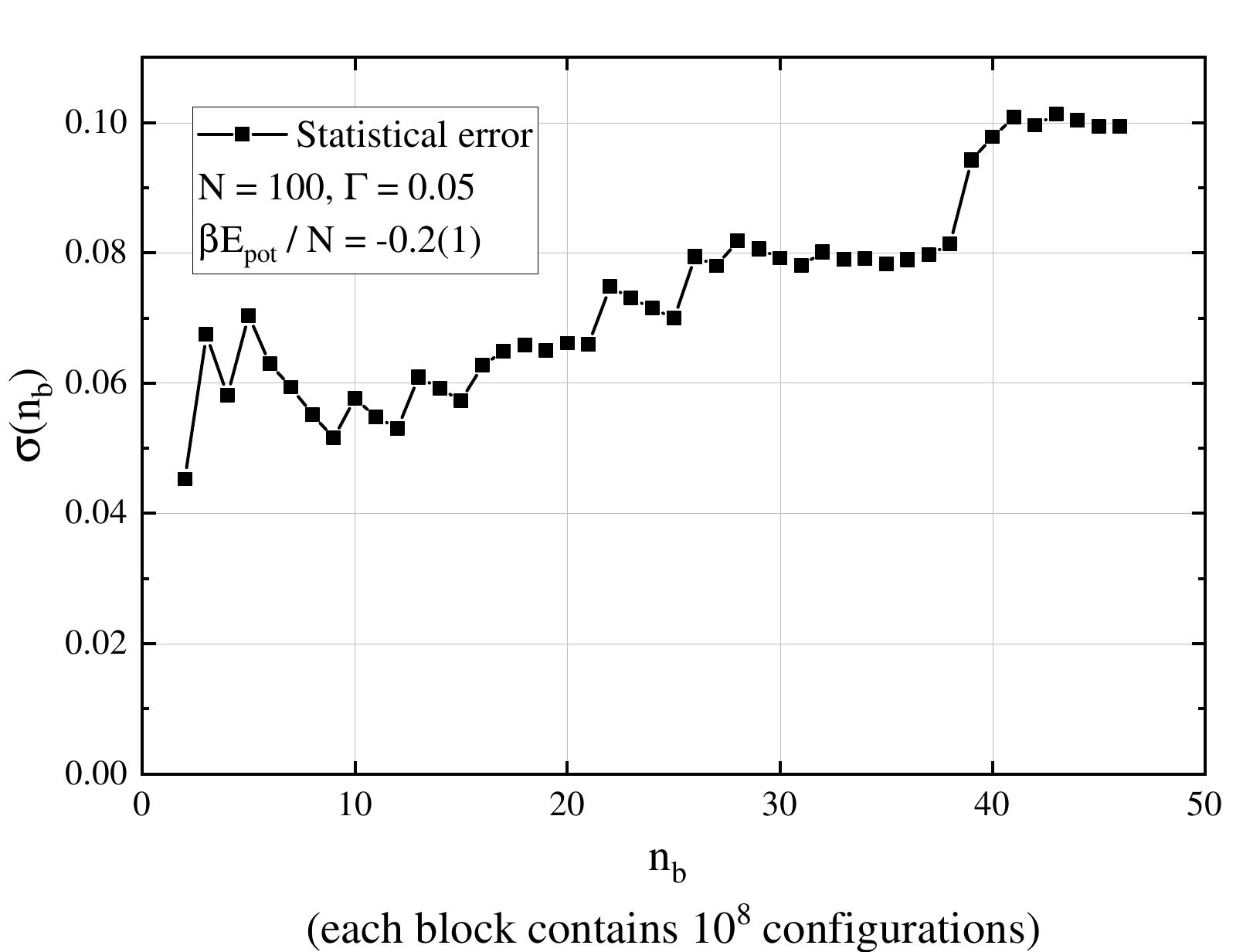}
(b)~\includegraphics[width=0.46\linewidth]{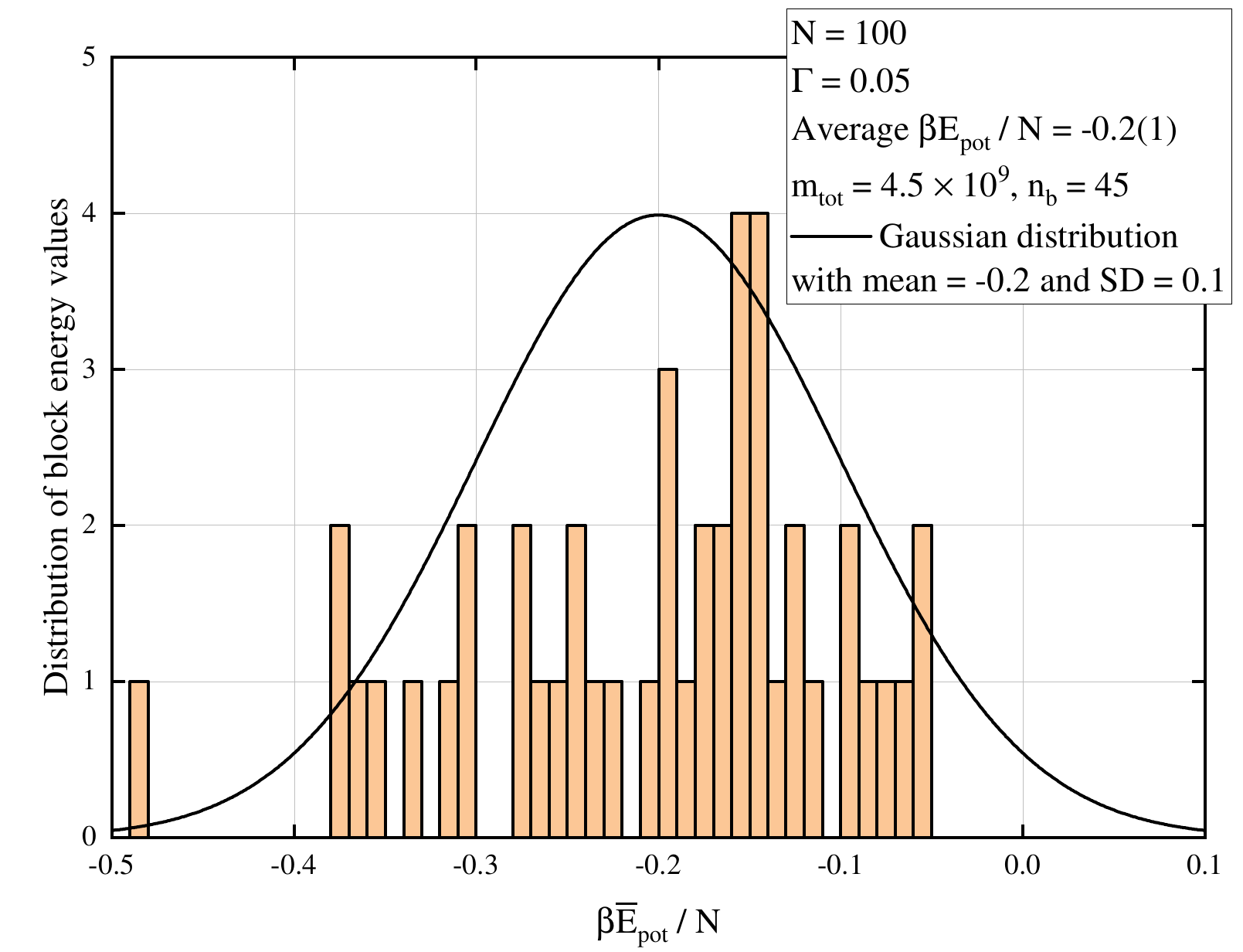}
\caption{Dependence of statistical error of $\beta E_{\text{pot}}/N$  on the number of blocks, $n_b$, for $\Gamma = 0.05$ and $N = 10^2$. (a) The statistical error, \eqref{eq:statError}, as a function of $n_b$; each block contains $m_{\text{tot}}/n_b = 10^8$ configurations. Contrary to our expectation, the statistical error grows as $n_b$ increases. (b) The distribution of block energy values, $\beta \bar{E}_{\text{pot}} / N$ (see Eq. \eqref{eq:statError} and reasoning above); totally $n_b = 45$ blocks are used. This histogram has a non-Gaussian form.
 }
\label{Fig:nblockdependence}
\end{figure*}

As shown above, as the parameter $\Gamma$ increases, bound states are observed. If such a state forms, the energy per particle drops sharply; at the same time, when such a bound state decays, the energy (per particle) jumps (see Fig.~\ref{fig:bound_states}). This leads to a significant increase in the statistical error of the average potential energy.

Since about $300$ such jumps for $\Gamma = 0.05$ occur during a PIMC simulation of $m_{\text{tot}} = 10^7$ steps, it is necessary to increase the block size: we use now $m_{\text{tot}}/n_b = 10^8$. We carried out a long simulation of $m_{\text{tot}} = 4.5\times 10^9$ PIMC steps with $n = 0$; this simulation is divided into $n_b = 45$ blocks. In total, we obtained $45$ block energy values, $\beta \bar{E}_{\text{pot}}(l)/N$, $l = 1,\ldots,45$, for $\Gamma = 0.05$.

It turned out that even such a long simulation still have a huge statistical error. We obtained the average potential energy $\beta E_{\text{pot}}/N = -0.2\pm 0.1$; thus, the relative error was $50\%$.

To investigate the reasons for this, we calculated the energy error as a function of the number of blocks (see Fig.~\ref{Fig:nblockdependence}(a)); the number of configurations in each block remained constant $10^8$ (in other words, with a fixed $m_{\text{tot}}/n_b$ ratio, $m_{\text{tot}}$ increased). We see that the statistical error not only fails to decrease, but in fact, it increases as the number of blocks increases.

In Fig.~\ref{Fig:nblockdependence}(b), one can see the distribution of block energies; this distribution is non-Gaussian.

Thus, PIMC sampling, in the presence of a small number of bound states, requires a huge computational cost. One of the solutions to this problem could be the Wang--Landau (WL) algorithm \cite{Wang:PRL:2001} to directly compute the density of states. The sampling in the energy space inherent to the WL method should be more efficient in the presence of bound states  to reduce the statistical error.

Another possible solution could be the Path Integral Molecular Dynamics (PIMD) technique, which can be used for Bolzmannons at any $\chi > 0$. By including the movement of all beads simultaneously, PIMD can more efficiently sample the configurations to reduce statistical error. Therefore, we hope that PIMD can be used to improve the accuracy of the PIMC simulation in the presence of a small number of bound states.

In the case of nondegenerate TCP (when $(-\ln \chi) \gg 1$), the CMC and PIMC simulations produced identical results both at $n = 0$ and $n = 10$ (see Tab. \ref{tab:N100}). Therefore, the classical MD approach with the diagonal Kelbg-AAE pseudopotential can be employed to simulate the nondegenerate TCP. In the process of MD calculation, each particle is shifted according to  the forces acting on it, resulting in enhanced algorithm efficiency.

\section{Conclusion \label{sec:concl}}
The problem of stability of matter was theoretically solved more than 60 years ago. However, this outstanding achievement had very little effect on computer simulations of Coulomb systems. This is especially true for a TCP. Since consistent account of quantum effects and the Fermi statistics requires an enormous amount of computations, a (near-)classical TCP is usually replaced by a stable system using a number of time-tested techniques. In particular, pseudopotentials or a truncated Coulomb potential are often used to limit the attractive Coulomb potential from below and stabilize a Coulomb system with unlike charges. Such procedures of artificial stabilization definitely influence the thermodynamic properties of a TCP. Therefore, it is important to develop a consistent account of quantum and long-range effects as well as  Fermi statistics in atomistic modeling.

In this paper, we studied the thermodynamic properties of a non-degenerate weakly coupled TCP using the CMC and PIMC approaches. The main peculiarity of this work is the use of AAEP in both classical and quantum simulations. As the degeneracy parameter is very small ($10^{-6}$), we disregard the antisymmetrization effects and consider all the particles as Boltzmannons. We thoroughly investigate the calculation of potential energy in PIMC using the Kelbg-AAE pseudopotential and offer a special procedure ``ONLYk'',  validated both theoretically and numerically. 
We considered a hydrogen plasma, but the entire proposed calculation scheme is also valid for an electrically neutral system of unequal charges.
Particular attention is paid to the formation of bound states as the coupling parameter $\Gamma$ increases. We estimate the number of transitions to a bound state and show that this parameter is crucial for simulations. At $\Gamma \le 0.01$, the probability of a bound state formation is very small, so both CMC and PIMC give very close results coinciding with the Debye--H\"uckel approximation. Moreover, we calculate the thermodynamic limit at $\Gamma=0.01$.
%and show that neglecting the long-range interaction effects slows down convergence to the limit. 
As the probability of a bound state formation rises proportional to $\Gamma^3$, the number of bound states becomes significant already at $\Gamma = 0.05$. For such a plasma, the convergence of energy cannot be achieved using conventional algorithms.
We hope that alternative simulation methods, such as PIMD and classical MD, could significantly improve the convergence. We plan to continue our research of a nondegenerate TCP using PIMD or MD with the Kelbg-AAEPP in the forthcoming papers to increase the coupling parameter.

\begin{acknowledgments}
We thank I.V. Morozov, E.M. Apfelbaum,  A.S. Larkin, and V.S. Filinov for fruitful discussions during the preparation of this paper. We thank E. M. Bazanova for proofreading the initial version of the manuscript.
The authors acknowledge the JIHT RAS Supercomputer Centre, the Joint Supercomputer Centre of the Russian Academy of Sciences, and the Shared Resource Centre ``Far Easten Computing Resource'' IACP FEB RAS for providing computing time.
\end{acknowledgments}

%\vspace{1cm}
\appendix
\section{PIMC sampling and averaging \label{app:sampling}}

For an initial configuration $(\textbf{R}_0, \textbf{R}_1, \ldots, \textbf{R}_n)$, we calculate the initial kinetic and potential energy and action. Then a particle $i$ is randomly chosen
and displaced. The displacement occurs as follows: first, the particle is displaced (as a whole) by a random vector (that is, all beads $\textbf{r}_{i, k}$ are displaced by the same vector). Then, each component of each bead, $\textbf{r}_{i,k}$, $k\in [0, n]$, is shifted by a random value $\delta \in (0, \sqrt{\hbar^2\epsilon/m_i})$. Finally, we obtain a new closed path $(\textbf{r}'_{i, 0}, \textbf{r}'_{i, 1},\ldots,\textbf{r}'_{i, n}, \textbf{r}'_{i, 0})$. Then the changes in action $\Delta S$ and energy $\beta\Delta E_{\text{kin}}, \beta\Delta E_{\text{pot}}$ are calculated. This configuration is accepted with the probability $\min(1, e^{-\Delta S})$. Then the procedure is repeated until the equilibrium section is observed. The following $m_{\text{tot}}$ MC steps are performed to collect statistics.

The initial configuration with $n \geq 1$, from which the simulation starts, is prepared using a similar simulation without interaction. This yields an initial configuration $(\textbf{R}_0, \textbf{R}_1, \ldots, \textbf{R}_n)$ in which the path of each particle, $(\textbf{r}_{i, 0}, \textbf{r}_{i, 1},\ldots,\textbf{r}_{i,n}, \textbf{r}_{i,0})$, is sampled from the distribution
\begin{equation}
\left(\frac{\sqrt{m_em_p}}{2\pi\hbar^2\epsilon}\right)^{\frac{3N(n + 1)}{2}} e^{-\sum_{k = 0}^n
\sum_{i = 1}^N\frac{m_i(\textbf{r}_{i,k} - \textbf{r}_{i, k+1})^2}{2\hbar^2 \epsilon}}.
\end{equation}

The sequence of energies on the equilibrium stage of simulation is divided into $n_b$ blocks; averaging the values over each block generates $n_b$ energy values, $\beta\bar{E}_{\text{pot}}(l)/N$, where $l = 1,\ldots,n_b$ (see Fig.~2 of Ref. \cite{DemyanovOCP:PRE:2022}). The statistical error is calculated as the root of the variance of these $n_b$ values:
\begin{equation}
\label{eq:statError}
\sigma = \sqrt{\cfrac{1}{n_b - 1}\sum_{l=1}^{n_b}\left(\cfrac{\beta\bar{E}_{\text{pot}}(l)}{N} - \cfrac{\beta E_{\text{pot}}}{N}\right)^2}.
\end{equation}
Here, $\beta E_{\text{pot}}$ refers to the average over all $m_{\text{tot}}$ configurations.

To calculate the thermodynamic limit, we fit a curve over all the calculated energy values for different $N$ as a function of $1/N$:
\begin{equation}
\label{eq:fitFunc}
\cfrac{\beta E_{\text{pot}}}{N}(1/N) = \cfrac{\beta E_{\text{pot}}}{N}(0) + b\left(1/N\right)^{\gamma},
\end{equation}
where $b$, $\gamma$, and $\frac{\beta E_{\text{pot}}}{N}(0)$ are the fitting parameters; the last one is
the energy value in the thermodynamic limit.

\begin{widetext}
\section{Supplementary formulas \label{app:suppl_formulas}}
In this Appendix, the following notations are used (see also Eq. \eqref{eq:sdfdsfvsfs}):
\begin{equation}
x_{ij}(\alpha) = d_{ij}(\alpha) / \lambda, \quad x_m = r_m/\lambda, \quad \textbf{x}_{ij} = \textbf{r}_{ij} / \lambda, \quad x_{ij} = r_{ij} / \lambda, \quad \lambda^2 = \hbar^2\beta/(2\mu),
\end{equation}
where $\mu$ is the reduced mass. The additional term $\Phi_1(\textbf{r}_{ij},\textbf{r}'_{ij};r_m, \beta)$ that accounts for long range interactions is the following (see Eq. (29) in \cite{Demyanov:ContrPlasPhys:2022}):
\begin{equation}
\label{eq:nondiagAAEPP}
\Phi_1(\textbf{r}_{ij},\textbf{r}'_{ij};r_m, \beta) = 
\cfrac{1}{\pi}
\int\limits_0^{1}
\cfrac{d\alpha}{d_{ij}(\alpha)} \times  I\left(\frac{d_{ij}(\alpha)}{\lambda}, \frac{r_m}{\lambda}, \alpha\right),
\end{equation}
where (see Eqs. (32)--(34) in \cite{Demyanov:ContrPlasPhys:2022}):
\begin{equation}
I(x_{ij}(\alpha), x_m, \alpha)  = 
\frac{1}{4 x_m^3}\left[2 \sqrt{\pi } \sqrt{(1-\alpha) \alpha} 
\left( 
f_1 (x_{ij}(\alpha))-
f_1 (-x_{ij}(\alpha))\right)+
f_2(x_{ij}(\alpha))-
f_2(-x_{ij}(\alpha)) \right],
\end{equation}
\begin{equation}
f_1 (x_{ij}(\alpha)) \equiv  
\left[4 (1-\alpha) \alpha-(2 x_m-x_{ij}(\alpha)) (x_m+x_{ij}(\alpha))\right]e^{-\frac{(x_m+x_{ij}(\alpha))^2}{4 (1-\alpha) \alpha}},
\end{equation}
\begin{equation}
f_2(x_{ij}(\alpha)) \equiv \pi  \left[-3 x_{ij}(\alpha) \left(2 (\alpha-1) \alpha+x_m^2\right)-2 x_m^3+x^3_{ij}(\alpha)\right]\text{erf}\left(\frac{x_m+x_{ij}(\alpha)}{2 \sqrt{(1-\alpha) \alpha}}\right).
\end{equation}

The derivative of the Kelbg-AAEPP over inverse temperature is the following:
\begin{equation}
\label{eq:rgoderivbeta}
\beta\cfrac{\partial \Phi(\textbf{r}_{ij},\textbf{r}'_{ij}; r_m, \beta)}{\partial \beta}
= -\cfrac{1}{2\lambda \sqrt{\pi }}
\int\limits_0^{1}
\frac{ e^{-\frac{d^2_{ij}(\alpha)}{4 (1-\alpha) \alpha\lambda^2}}}{\sqrt{(1-\alpha) \alpha}}d\alpha
-
\cfrac{1}{\lambda\pi}
\int\limits_0^{1}
I_{\text{der}}\left(\frac{d_{ij}(\alpha)}{\lambda}, \frac{r_m}{\lambda}, \alpha\right)d\alpha,
\end{equation}
where $I_{\text{der}}(x_{ij}(\alpha), x_m, \alpha)$ is:
\begin{multline}
I_{\text{der}}(x_{ij}(\alpha), x_m, \alpha) =
\frac{3 \sqrt{\pi } ((1-\alpha ) \alpha )^{3/2}}{x_{ij}(\alpha) x_m^3} \left(e^{-\frac{(x_m-x_{ij}(\alpha))^2}{4 (1-\alpha) \alpha
   }}-e^{-\frac{(x_{ij}(\alpha)+x_m)^2}{4 (1-\alpha) \alpha }}\right)\\-\frac{3 \pi  \alpha(1-\alpha)}{2
   x_m^3} \left(\text{erf}\left(\frac{x_m-x_{ij}(\alpha)}{2 \sqrt{(1-\alpha) \alpha}}\right)+\text{erf}\left(\frac{x_m+x_{ij}(\alpha)}{2 \sqrt{(1-\alpha) \alpha}}\right)\right).
\end{multline}
All integrals over $\alpha$ in Eqs. \eqref{eq:kelbgpseudo}, \eqref{eq:nondiagAAEPP} and  \eqref{eq:rgoderivbeta} are computed using the GSL library \cite{galassi2021scientific}.

The diagonal Kelbg-AAEPP can be obtained in a direct analytical form (see Eqs. (42)-(46) in \cite{Demyanov:ContrPlasPhys:2022}):
\begin{equation}
\label{eq:phi1Def}
\Phi_1(\textbf{r},\textbf{r};r_m, \beta) \equiv \Phi_1(r; r_m,\beta) =  \cfrac{4}{r\pi} 
\left[
I_{\text{all}}\left(\frac{r}{\lambda}, \frac{r_m}{\lambda}\right) -I_{\text{all}}\left(-\frac{r}{\lambda}, \frac{r_m}{\lambda}\right) -\frac{\pi ^{3/2}}{4} \times \frac{r}{\lambda}
\right],
\end{equation}
where
\begin{equation}
I_{\text{all}}(x, x_m) = I_{\text{exp}}(x, x_m) + I_{\text{erf}}(x, x_m) + I_{\text{mod}}(x, x_m),
\end{equation}
\begin{equation}
\label{eq:expcontr}
I_{\text{exp}}(x, x_m) = \frac{\pi  e^{-(x_m+x)^2} (x_m+x)}{128 x_m^3 |x_m+x|} \left(2 x^2 x_m-2 x^3+10 x x_m^2-5 x+6 x_m^3+3
   x_m\right),
\end{equation}
\begin{equation}
I_{\text{erf}}(x, x_m) = \frac{\pi ^{3/2}}{256 x_m^3} \left(4 (x_m+x) \left(\left(x^2+3\right) x_m-x \left(x^2+3\right)+5 x x_m^2+3
   x_m^3\right)-3\right) \text{erf}\left(|x_m+x|\right),
\end{equation}
\begin{equation}
\label{eq:modcontr}
I_{\text{mod}}(x, x_m) = \frac{\pi  |x_m+x|}{16 x_m^3 (x_m+x)} \left(x^3-3 x x_m^2+x-2 x_m^3\right).
\end{equation}
The modules in Eqs. \eqref{eq:expcontr}--\eqref{eq:modcontr} play a significant role in Eq. \eqref{eq:phi1Def} in the term $I_{\text{all}}(-r / \lambda, r_m/ \lambda)$.

\section{Partition function in the limit $\chi\to 0$ \label{app:kinetic_energy_classical}}
In this section, we consider the partition function in the classical limit $\chi\to 0$. Let us rewrite $Q(\beta)$ \eqref{eq:part_func1} as follows:
\begin{multline}
Q(\beta) = \left(\cfrac{\sqrt{m_em_p}}{2\pi\hbar^2\epsilon}\right)^{\frac{3N(n + 1)}{2}}e^{-\beta U_0} \int\limits_{V^N} d \textbf{R}_0 
\times\\
\int\limits_{\mathbb{R}^{3Nn}} e^{-\sum\limits_{k = 0}^n
\sum\limits_{i = 1}^N\tfrac{m_i(\textbf{r}_{i,k} - \textbf{r}_{i, k+1})^2}{2\hbar^2 \epsilon}} \exp\left\{-\tfrac{\beta}{2(n+1)}\sum_{k = 0}^n\sum_{i = 1}^N\sum_{\substack{j = 1\\ i\neq j}}^{N_{s,i}}q_iq_j\Phi(\textbf{r}_{ij,k},\textbf{r}_{ij,k+1}; r_m, \epsilon)\right\} \prod_{k = 1}^n d \textbf{R}_k.
\end{multline}
\end{widetext}
The characteristic size of each electron and proton is of the order of  $\lambda(\beta) = \Lambda/\sqrt{2\pi}$ and $\sqrt{m_e/m_p}\Lambda/\sqrt{2\pi}$, respectively. If $\chi\to 0$, then the particle size is negligible compared to the interparticle distance $r_a$ ($\Lambda/r_a = (9\pi/2)^{-1/6}\chi^{1/3}\xrightarrow{\chi\to 0} 0$). So, we can replace all the interactions in the potential energy by the interaction of the same beads, $\textbf{r}_{ij,k}\approx \textbf{r}_{ij,0}$:
\begin{multline}
\tfrac{\beta}{2(n+1)}\sum_{k = 0}^n\Phi(\textbf{r}_{ij,k},\textbf{r}_{ij,k+1}; r_m, \epsilon)
\\
\approx
\tfrac{\beta}{2(n+1)}\sum_{k = 0}^n\Phi(\textbf{r}_{ij,0},\textbf{r}_{ij,0}; r_m,  \epsilon)
\\=
\tfrac{\beta}{2}\Phi(\textbf{r}_{ij,0},\textbf{r}_{ij,0}; r_m,  \epsilon).
\end{multline}

The distance between the beads $|\textbf{r}_{ij,k}|$ is of the order of $r_s$; then $|\textbf{r}_{ij,k}|/\lambda \sim r_s/\lambda = (9\pi/2)^{1/6}\chi^{-1/3} \xrightarrow{\chi\to 0} \infty$. If we consider the expression for the diagonal Kelbg pseudopotential \eqref{eq:diagkelbgtheta0}, then in the classical limits, it tends to a usual interaction potential: $\Phi(\textbf{r}_{ij},\textbf{r}_{ij}; r_m,  \epsilon) \xrightarrow{\chi\to 0}\varphi(|\textbf{r}_{ij}|)$. The latter can be also validated from the following reasoning: all quantum effects arise on the thermal de Broglie wavelength; if the interparticle distance is larger than their wavelength, the quantum effects are negligible.

Altogether we can replace the sum over $k$ in the potential energy by the sum of identical terms that do not depend on  temperature:
\begin{equation}
\tfrac{\beta}{2(n+1)}\sum_{k = 0}^n\Phi(\textbf{r}_{ij,k},\textbf{r}_{ij,k+1}; r_m, \epsilon)
\approx \tfrac{\beta}{2}\varphi(r_{ij,0}).
\end{equation}
Thus, the partition function in the limit $\chi\to 0$ has the following form:
\begin{multline}
Q(\beta) 
\approx \left(\cfrac{\sqrt{m_em_p}}{2\pi\hbar^2\beta}\right)^{\frac{3N}{2}} Q_K(\beta)
\\
\times
e^{-\beta U_0} \int\limits_{V^N}d \textbf{R}_0  \exp\left\{-\tfrac{\beta}{2}\sum\limits_{i = 1}^N\sum\limits_{\substack{j = 1\\ i\neq j}}^{N_{s,i}}q_iq_j\varphi(r_{ij,0})\right\} 
,
\end{multline}
where $Q_K(\beta)$ is the kinetic ideal gas contribution (see Eq. \eqref{eq:dsfjhsdfse}). In the following subsection, we will show that  $Q_K(\beta) = 1$, so the partition function then takes the form typical of a classical system:
\begin{equation}
Q(\beta) = \left(\tfrac{\sqrt{m_em_p}}{2\pi\hbar^2\beta}\right)^{\frac{3N}{2}}  \int\limits_{V^N} e^{-\beta U_0-\tfrac{\beta}{2}\sum\limits_{i = 1}^N\sum\limits_{{j = 1\atop i\neq j}}^{N_{s,i}}q_iq_j\varphi(r_{ij,0})} d \textbf{R}_0.
\end{equation}
If we now calculate the energy:
\begin{equation}
\beta E =  \cfrac{3N}{2}+\beta U_0+
\left\langle\tfrac{1}{2}\sum_{i = 1}^N\sum_{\substack{j = 1\\ i\neq j}}^{N_{s,i}}q_iq_j\varphi(r_{ij,0})\right\rangle,
\end{equation}
we get that $\beta E_{\text{kin}} = 3N/2$, and $\beta E_\text{pot}$ is the same as in Eq. \eqref{eq:classicpotenergy}.

\begin{widetext}
\subsubsection*{Kinetic contribution in the partition function of ideal gas in Path Integral}
In this subsection, we calculate the ideal gas kinetic energy contribution of the partition function, $Q_K(\beta)$:
\begin{equation}
\label{eq:dsfjhsdfse}
Q_K(\beta) = (n+1)^{\frac{3N}{2}}\left(\cfrac{\sqrt{m_em_p}}{2\pi\hbar^2\epsilon}\right)^{\frac{3Nn}{2}} \int\limits_{\mathbb{R}^{3Nn}}  e^{-\sum\limits_{k = 0}^n
\sum\limits_{i = 1}^N\tfrac{m_i(\textbf{r}_{i,k} - \textbf{r}_{i, k+1})^2}{2\hbar^2 \epsilon}} \prod_{k = 1}^n d \textbf{R}_k = 
\prod_{i = 1}^N
I_i,
\end{equation}
where 
\begin{equation}
I_i = (n+1)^{\frac{3}{2}}\left(\cfrac{m_i}{2\pi\hbar^2\epsilon}\right)^{\frac{3n}{2}} 
\int\limits_{\mathbb{R}^{3n}}  e^{-\sum\limits_{k = 0}^n
\tfrac{m_i(\textbf{r}_{i,k} - \textbf{r}_{i, k+1})^2}{2\hbar^2 \epsilon}}d\textbf{R}^{(i)}, \quad \textbf{R}^{(i)} = (\textbf{r}_{i, 1}, \textbf{r}_{i, 2}, \ldots, \textbf{r}_{i, n}).
\end{equation}
Here, $\textbf{R}^{(i)}$ is the set of all beads of $i$th particle. All the integrals $I_i$ have the same form, $I_i = I$. Thus, $Q_K(\beta)$ is the product of $N$ identical integrals: $Q_K(\beta) = I^N$.
So our goal is to calculate the following integral:
\begin{equation}
\label{eq:sfhsdfef}
I = (n+1)^{\frac{3}{2}}\left(\cfrac{m}{2\pi\hbar^2\epsilon}\right)^{\frac{3n}{2}} 
\int\limits_{\mathbb{R}^{3n}}  e^{-\sum\limits_{k = 1}^{n+1}
\tfrac{m(\textbf{r}_{k} - \textbf{r}_{k-1})^2}{2\hbar^2 \epsilon}}d\textbf{r}_1d\textbf{r}_2\cdots d\textbf{r}_n,
\end{equation}
with a condition $\textbf{r}_{n+1} = \textbf{r}_0$. Here, $m$ is the mass of a particle. Let us make a substitution of the variables:
\begin{equation}
\textbf{h}_k = (\textbf{r}_k - \textbf{r}_{k - 1})\sqrt{\cfrac{m}{2\hbar^2 \epsilon}}, 
\quad
\textbf{r}_{n} = \textbf{r}_{0} + \sqrt{\cfrac{2\hbar^2 \epsilon}{m}}\sum_{k = 1}^n\textbf{h}_{k},
\quad 
\textbf{h}^2_{n+1} = \left(\sum_{k = 1}^n\textbf{h}_{k}\right)^2,
\quad d\textbf{r}_k = d(\textbf{r}_k -\textbf{r}_{k-1}) = \left(\cfrac{2\hbar^2 \epsilon}{m}\right)^{\frac{3}{2}} d \textbf{h}_k.
\end{equation}
Then integral \eqref{eq:sfhsdfef} is transformed to the following expression:
\begin{equation}
I = (n+1)^{\frac{3}{2}}\pi^{-\frac{3n}{2}} 
\int\limits_{\mathbb{R}^{3n}}  e^{-\sum\limits_{k = 1}^{n}
\textbf{h}^2_k-\left(\sum_{k = 1}^n\textbf{h}_{k}\right)^2}d\textbf{h}_1d\textbf{h}_2\cdots d\textbf{h}_n.
\end{equation}
To calculate this integral, we introduce the following variables:
\begin{equation}
\textbf{w}_n = \textbf{h}_1+\textbf{h}_2+\ldots+\textbf{h}_{n - 1}, \quad q_n^2 = \textbf{h}^2_1+\textbf{h}^2_2+\ldots+\textbf{h}^2_{n - 1}, \quad w_n = |\textbf{w}_n|
\end{equation}
Now we integrate over the last variable, $\textbf{h}_n$:
\begin{equation}
\label{eq:dsffewfdsf}
I = (n+1)^{\frac{3}{2}}\pi^{-\frac{3n}{2}} 
\int\limits_{\mathbb{R}^{3(n-1)}} 
d\textbf{h}_1d\textbf{h}_2\cdots d\textbf{h}_{n-1}\int\limits_{\mathbb{R}^3}d\textbf{h}_n
  e^{-q_n^2}e^{-
\textbf{h}^2_n-\left(\textbf{w}_n+\textbf{h}_{n}\right)^2},
\end{equation}
We calculate this integral in the spherical coordinates:
\begin{equation}
\int\limits_{\mathbb{R}^3}d\textbf{h}_n
  e^{-q_n^2}e^{-
\textbf{h}^2_n-\left(\textbf{w}_n+\textbf{h}_{n}\right)^2}
=
2\pi e^{-q_n^2}e^{-w_n^2}\int\limits_{0}^{\infty} dh_n h_n^2\int\limits_{-1}^1du
  e^{-2\left(h_n^2 + w_n h_n u\right)} 
=
\left(\cfrac{\pi}{2}\right)^{3/2}e^{-q_{n-1}^2-\textbf{h}^2_{n-1}-\frac{(\textbf{w}_{n-1}+\textbf{h}_{n-1})^2}{2}}
\end{equation}
and obtain
\begin{equation}
I = (n+1)^{\frac{3}{2}}\pi^{-\frac{3n}{2}}(\pi/2)^{3/2} 
\int\limits_{\mathbb{R}^{3(n-1)}} 
d\textbf{h}_1d\textbf{h}_2\cdots d\textbf{h}_{n-1}
e^{-q_{n-1}^2-\textbf{h}^2_{n-1}-(\textbf{w}_{n-1}+\textbf{h}_{n-1})^2/2}
\end{equation}
One can see that this is almost the same integral as in Eq. \eqref{eq:dsffewfdsf}. 
Let us integrate $m$ times:
\begin{equation}
I = (n+1)^{\frac{3}{2}}\pi^{-\frac{3n}{2}}\cfrac{\pi^{3m/2}}{(m+1)^{3/2}}
\int\limits_{\mathbb{R}^{3(n-m)}} 
d\textbf{h}_1d\textbf{h}_2\cdots d\textbf{h}_{n-m}
e^{-q_{n-m}^2-\textbf{h}^2_{n-m}-(\textbf{w}_{n-m}+\textbf{h}_{n-m})^2/(m+1)}.
\end{equation}
If $m = n - 1$, then we obtain the simple integral (since $\textbf{w}_1 = \textbf{0}$ and $q_1^2 = 0$):
\begin{equation}
I = (n+1)^{\frac{3}{2}}\pi^{-\frac{3n}{2}}\cfrac{\pi^{3(n-1)/2}}{n^{3/2}}
\int\limits_{\mathbb{R}^{3}} 
d\textbf{h}_1
e^{-\textbf{h}^2_1-\textbf{h}_1^2/n} = 1.
\end{equation}
Thus, $Q_K(\beta) = I^N = 1$.

\section{Reasoning for ONLYk method \eqref{eq:onlykmethod} \label{app:pot_energy_nondiag}}
In the first order of the perturbation theory, the density matrix has the following form (see Eq. (30) in Ref. \cite{Demyanov:Kelbg:2022} or Eq. (13) in Ref. \cite{Kelbg:AnnDerPhys:1963}):
\begin{equation}
\label{eq:matriPlot1}
\hat{\rho}(\beta) = e^{-\beta\hat{V}}e^{-\beta\hat{K}} + e^{-\beta\hat{V}}e^{-\beta\hat{K}}\int\limits_0^{\beta}\beta_1 \cfrac{d }{d  \beta_1}\left(e^{\beta_1\hat{K}}\hat{V}e^{-\beta_1\hat{K}}\right) d\beta_1.
\end{equation}
The operators $\hat{K}$ and $\hat{V}$ are the same as those defined in Eqs. \eqref{eq:kinenergyop} and \eqref{eq:classical_energy_pot}, respectively.

The next step in the derivation of Kelbg functional \eqref{eq:kelbgfuncDef} is to rewrite the potential energy in the Fourier form:
\begin{equation}
\label{eq:sfhsdfe}
\langle \textbf{R}_k|\hat{V} = U_0\langle \textbf{R}_k| +
\cfrac{1}{16\pi^3}\sum_{i \neq j}
q_iq_j\int d\textbf{t} v(\textbf{t})e^{i\textbf{t}\cdot\textbf{r}_{ij, k}}\langle \textbf{R}_k|.
\end{equation}
We should remember that if $|\textbf{r}_{ij, k}| > r_m$, then $\hat{V} = U_0$.
By performing some transformations specified in \cite{Demyanov:Kelbg:2022}, we write Eq.~(49) in Ref. \cite{Kelbg:AnnDerPhys:1963}, acting the density matrix $\hat{\rho}(\beta)$ on the vector $\langle \textbf{R}_k|$:
\begin{multline}
\label{eq:ghewjd}
\langle \textbf{R}_k|\hat{\rho}(\beta) = 
\langle \textbf{R}_k|e^{-\beta\hat{V}}e^{-\beta\hat{K}} + 
\langle\textbf{R}_k|e^{-\beta\hat{V}}e^{-\beta\hat{K}}\int\limits_0^{\beta}\beta_1 \cfrac{d }{d  \beta_1}\left(e^{\beta_1\hat{K}}\hat{V}e^{-\beta_1\hat{K}}\right) d\beta_1
=
\langle \textbf{R}_k|\left(e^{-\beta\hat{V}}e^{-\beta\hat{K}}\right) \\
{} +\cfrac{1}{16\pi^3}
\sum_{i\neq j}
q_iq_j 
\int d\textbf{t} v(\textbf{t}) e^{i\textbf{t}\cdot\textbf{r}_{ij,k}}
\langle \textbf{R}_k|
e^{-\beta\hat{V}}\int\limits_0^{\beta}\beta_1 \cfrac{d }{d  \beta_1}
\left(
e^{\tfrac{\hbar (\beta_1-\beta)}{m}\,  \textbf{t} (\hat{\textbf{p}}_i - \hat{\textbf{p}}_j) +  \tfrac{\hbar^2(\beta_1-\beta)}{m}\,t^2}
\right) 
e^{-\beta\hat{K}}d\beta_1.
\end{multline}
Thus, if $|\textbf{r}_{ij, k}| > r_m$, the second term in Eq. \eqref{eq:sfhsdfe} and, hence, in Eq. \eqref{eq:ghewjd}, is zero. This condition is true regardless of the further projection of $\langle \textbf{R}_k|\hat{\rho}(\beta)$ onto $|\textbf{R}_{k+1}\rangle$. In other words, we can rewrite the quantity $\langle\textbf{R}_k|\hat{\rho}(\beta)$ from Eq. \eqref{eq:ghewjd} as follows (using Eq. \eqref{eq:sfhsdfe}):
\begin{equation}
\label{eq:dsfsfefd}
\langle \textbf{R}_k|\hat{\rho}(\beta) =\langle \textbf{R}_k|e^{-\beta\hat{V}}e^{-\beta\hat{K}} + 
\begin{cases}
\langle \textbf{R}_k|e^{-\beta\hat{V}}e^{-\beta\hat{K}}\int\limits_0^{\beta}\beta_1 \cfrac{d }{d  \beta_1}\left(e^{\beta_1\hat{K}}\hat{V}e^{-\beta_1\hat{K}}\right) d\beta_1, &\text{ if } |\textbf{r}_{ij, k}| \leq r_m,\\
0, &\text{ if } |\textbf{r}_{ij, k}| > r_m.
\end{cases}
\end{equation}

After projecting  $\langle \textbf{R}_k|\hat{\rho}(\beta)$ onto  $|\textbf{R}_{k+1}\rangle$, we obtain:
\begin{equation}
\label{eq:sadzxfe}
\langle \textbf{R}_k|\hat{\rho}(\beta) |\textbf{R}_{k+1}\rangle
=
\begin{cases}
\text{Eq. \eqref{eq:dmkelbg}}, & \text{ if } |\textbf{r}_{ij, k}| \leq r_m,\\
 e^{-\beta U_0}\langle\textbf{R}_k|e^{-\beta\hat{K}} |\textbf{R}_{k+1}\rangle, & \text{ if } |\textbf{r}_{ij, k}| > r_m.
\end{cases}
\end{equation}
The transformation from  the top line of Eq.~\eqref{eq:dsfsfefd} to the top line of Eq.~\eqref{eq:sadzxfe} is fully explained in Ref. \cite{Demyanov:Kelbg:2022}.

The bottom line of Eq. \eqref{eq:sadzxfe} corresponds to Eq. \eqref{eq:dmkelbg} if $\Phi(\textbf{r}_{ij},\textbf{r}'_{ij}; r_m, \beta) = 0$. So, if $|\textbf{r}_{ij, k}| \leq r_m$, the contribution to the sum \eqref{eq:hjsfdfdsj} is non-zero, and if  $|\textbf{r}_{ij, k}| > r_m$, the contribution is zero. This is how the ONLYk method in Eq. \eqref{eq:onlykmethod} is derived.

\end{widetext}
%\bibliography{refs}

\begin{thebibliography}{107}%
\makeatletter
\providecommand \@ifxundefined [1]{%
 \@ifx{#1\undefined}
}%
\providecommand \@ifnum [1]{%
 \ifnum #1\expandafter \@firstoftwo
 \else \expandafter \@secondoftwo
 \fi
}%
\providecommand \@ifx [1]{%
 \ifx #1\expandafter \@firstoftwo
 \else \expandafter \@secondoftwo
 \fi
}%
\providecommand \natexlab [1]{#1}%
\providecommand \enquote  [1]{``#1''}%
\providecommand \bibnamefont  [1]{#1}%
\providecommand \bibfnamefont [1]{#1}%
\providecommand \citenamefont [1]{#1}%
\providecommand \href@noop [0]{\@secondoftwo}%
\providecommand \href [0]{\begingroup \@sanitize@url \@href}%
\providecommand \@href[1]{\@@startlink{#1}\@@href}%
\providecommand \@@href[1]{\endgroup#1\@@endlink}%
\providecommand \@sanitize@url [0]{\catcode `\\12\catcode `\$12\catcode
  `\&12\catcode `\#12\catcode `\^12\catcode `\_12\catcode `\%12\relax}%
\providecommand \@@startlink[1]{}%
\providecommand \@@endlink[0]{}%
\providecommand \url  [0]{\begingroup\@sanitize@url \@url }%
\providecommand \@url [1]{\endgroup\@href {#1}{\urlprefix }}%
\providecommand \urlprefix  [0]{URL }%
\providecommand \Eprint [0]{\href }%
\providecommand \doibase [0]{https://doi.org/}%
\providecommand \selectlanguage [0]{\@gobble}%
\providecommand \bibinfo  [0]{\@secondoftwo}%
\providecommand \bibfield  [0]{\@secondoftwo}%
\providecommand \translation [1]{[#1]}%
\providecommand \BibitemOpen [0]{}%
\providecommand \bibitemStop [0]{}%
\providecommand \bibitemNoStop [0]{.\EOS\space}%
\providecommand \EOS [0]{\spacefactor3000\relax}%
\providecommand \BibitemShut  [1]{\csname bibitem#1\endcsname}%
\let\auto@bib@innerbib\@empty
%</preamble>
\bibitem [{\citenamefont {Baus}\ and\ \citenamefont
  {Hansen}(1980)}]{Baus:PR:1980}%
  \BibitemOpen
  \bibfield  {author} {\bibinfo {author} {\bibfnamefont {M.}~\bibnamefont
  {Baus}}\ and\ \bibinfo {author} {\bibfnamefont {J.-P.}\ \bibnamefont
  {Hansen}},\ }\href
  {https://doi.org/https://doi.org/10.1016/0370-1573(80)90022-8} {\bibfield
  {journal} {\bibinfo  {journal} {Physics Reports}\ }\textbf {\bibinfo {volume}
  {59}},\ \bibinfo {pages} {1} (\bibinfo {year} {1980})}\BibitemShut {NoStop}%
\bibitem [{\citenamefont {Earnshaw}(1848)}]{earnshaw1848nature}%
  \BibitemOpen
  \bibfield  {author} {\bibinfo {author} {\bibfnamefont {S.}~\bibnamefont
  {Earnshaw}},\ }\href
  {https://archive.org/details/transactionsofca07camb/page/96/mode/2up}
  {\bibfield  {journal} {\bibinfo  {journal} {Transactions of the Cambridge
  Philosophical Society}\ }\textbf {\bibinfo {volume} {7}},\ \bibinfo {pages}
  {97} (\bibinfo {year} {1848})}\BibitemShut {NoStop}%
\bibitem [{\citenamefont {Dyson}\ and\ \citenamefont
  {Lenard}(1967)}]{Dyson_Lenard:JMP:1967}%
  \BibitemOpen
  \bibfield  {author} {\bibinfo {author} {\bibfnamefont {F.~J.}\ \bibnamefont
  {Dyson}}\ and\ \bibinfo {author} {\bibfnamefont {A.}~\bibnamefont {Lenard}},\
  }\href@noop {} {\bibfield  {journal} {\bibinfo  {journal} {Journal of
  Mathematical Physics}\ }\textbf {\bibinfo {volume} {8}},\ \bibinfo {pages}
  {423–434} (\bibinfo {year} {1967})}\BibitemShut {NoStop}%
\bibitem [{\citenamefont {Lenard}\ and\ \citenamefont
  {Dyson}(1968)}]{Lenard_Dyson:JMP:1968}%
  \BibitemOpen
  \bibfield  {author} {\bibinfo {author} {\bibfnamefont {A.}~\bibnamefont
  {Lenard}}\ and\ \bibinfo {author} {\bibfnamefont {F.~J.}\ \bibnamefont
  {Dyson}},\ }\href {https://doi.org/10.1063/1.1664631} {\bibfield  {journal}
  {\bibinfo  {journal} {Journal of Mathematical Physics}\ }\textbf {\bibinfo
  {volume} {9}},\ \bibinfo {pages} {698–711} (\bibinfo {year}
  {1968})}\BibitemShut {NoStop}%
\bibitem [{\citenamefont {Lieb}\ and\ \citenamefont
  {Thirring}(1975)}]{Lieb:PRL:1975}%
  \BibitemOpen
  \bibfield  {author} {\bibinfo {author} {\bibfnamefont {E.~H.}\ \bibnamefont
  {Lieb}}\ and\ \bibinfo {author} {\bibfnamefont {W.~E.}\ \bibnamefont
  {Thirring}},\ }\href {https://doi.org/10.1103/PhysRevLett.35.687} {\bibfield
  {journal} {\bibinfo  {journal} {Phys. Rev. Lett.}\ }\textbf {\bibinfo
  {volume} {35}},\ \bibinfo {pages} {687} (\bibinfo {year} {1975})}\BibitemShut
  {NoStop}%
\bibitem [{\citenamefont {Dyson}(1967)}]{Dyson:JMP:1967}%
  \BibitemOpen
  \bibfield  {author} {\bibinfo {author} {\bibfnamefont {F.~J.}\ \bibnamefont
  {Dyson}},\ }\href {https://doi.org/10.1063/1.1705389} {\bibfield  {journal}
  {\bibinfo  {journal} {Journal of Mathematical Physics}\ }\textbf {\bibinfo
  {volume} {8}},\ \bibinfo {pages} {1538–1545} (\bibinfo {year}
  {1967})}\BibitemShut {NoStop}%
\bibitem [{\citenamefont {Lieb}\ and\ \citenamefont
  {Narnhofer}(1975)}]{Lieb1975}%
  \BibitemOpen
  \bibfield  {author} {\bibinfo {author} {\bibfnamefont {E.~H.}\ \bibnamefont
  {Lieb}}\ and\ \bibinfo {author} {\bibfnamefont {H.}~\bibnamefont
  {Narnhofer}},\ }\href {https://doi.org/10.1007/BF01012066} {\bibfield
  {journal} {\bibinfo  {journal} {Journal of Statistical Physics}\ }\textbf
  {\bibinfo {volume} {12}},\ \bibinfo {pages} {291} (\bibinfo {year}
  {1975})}\BibitemShut {NoStop}%
\bibitem [{\citenamefont {Lieb}\ and\ \citenamefont
  {Lebowitz}(1972)}]{Lieb:AM:1972}%
  \BibitemOpen
  \bibfield  {author} {\bibinfo {author} {\bibfnamefont {E.~H.}\ \bibnamefont
  {Lieb}}\ and\ \bibinfo {author} {\bibfnamefont {J.~L.}\ \bibnamefont
  {Lebowitz}},\ }\href
  {https://doi.org/https://doi.org/10.1016/0001-8708(72)90023-0} {\bibfield
  {journal} {\bibinfo  {journal} {Advances in Mathematics}\ }\textbf {\bibinfo
  {volume} {9}},\ \bibinfo {pages} {316} (\bibinfo {year} {1972})}\BibitemShut
  {NoStop}%
\bibitem [{\citenamefont {Hansen}(1987)}]{Hansen1987}%
  \BibitemOpen
  \bibfield  {author} {\bibinfo {author} {\bibfnamefont {J.-P.}\ \bibnamefont
  {Hansen}},\ }\bibinfo {title} {{Two-Component Plasmas in Two and Three
  Dimensions}},\ in\ \href {https://doi.org/10.1007/978-1-4613-1891-0_11}
  {\emph {\bibinfo {booktitle} {Strongly Coupled Plasma Physics}}},\ \bibinfo
  {editor} {edited by\ \bibinfo {editor} {\bibfnamefont {F.~J.}\ \bibnamefont
  {Rogers}}\ and\ \bibinfo {editor} {\bibfnamefont {H.~E.}\ \bibnamefont
  {Dewitt}}}\ (\bibinfo  {publisher} {Springer US},\ \bibinfo {address}
  {Boston, MA},\ \bibinfo {year} {1987})\ pp.\ \bibinfo {pages}
  {111--122}\BibitemShut {NoStop}%
\bibitem [{\citenamefont {Hansen}\ and\ \citenamefont
  {McDonald}(1978)}]{PhysRevLett.41.1379}%
  \BibitemOpen
  \bibfield  {author} {\bibinfo {author} {\bibfnamefont {J.~P.}\ \bibnamefont
  {Hansen}}\ and\ \bibinfo {author} {\bibfnamefont {I.~R.}\ \bibnamefont
  {McDonald}},\ }\href {https://doi.org/10.1103/PhysRevLett.41.1379} {\bibfield
   {journal} {\bibinfo  {journal} {Phys. Rev. Lett.}\ }\textbf {\bibinfo
  {volume} {41}},\ \bibinfo {pages} {1379} (\bibinfo {year}
  {1978})}\BibitemShut {NoStop}%
\bibitem [{\citenamefont {Hansen}\ and\ \citenamefont
  {McDonald}(1981)}]{PhysRevA.23.2041}%
  \BibitemOpen
  \bibfield  {author} {\bibinfo {author} {\bibfnamefont {J.~P.}\ \bibnamefont
  {Hansen}}\ and\ \bibinfo {author} {\bibfnamefont {I.~R.}\ \bibnamefont
  {McDonald}},\ }\href {https://doi.org/10.1103/PhysRevA.23.2041} {\bibfield
  {journal} {\bibinfo  {journal} {Phys. Rev. A}\ }\textbf {\bibinfo {volume}
  {23}},\ \bibinfo {pages} {2041} (\bibinfo {year} {1981})}\BibitemShut
  {NoStop}%
\bibitem [{\citenamefont {Deutsch}(1977)}]{DEUTSCH1977317}%
  \BibitemOpen
  \bibfield  {author} {\bibinfo {author} {\bibfnamefont {C.}~\bibnamefont
  {Deutsch}},\ }\href
  {https://doi.org/https://doi.org/10.1016/0375-9601(77)90111-6} {\bibfield
  {journal} {\bibinfo  {journal} {Physics Letters A}\ }\textbf {\bibinfo
  {volume} {60}},\ \bibinfo {pages} {317} (\bibinfo {year} {1977})}\BibitemShut
  {NoStop}%
\bibitem [{\citenamefont {Tiwari}\ \emph {et~al.}(2017)\citenamefont {Tiwari},
  \citenamefont {Shaffer},\ and\ \citenamefont {Baalrud}}]{Tiwari:PRE:2017}%
  \BibitemOpen
  \bibfield  {author} {\bibinfo {author} {\bibfnamefont {S.~K.}\ \bibnamefont
  {Tiwari}}, \bibinfo {author} {\bibfnamefont {N.~R.}\ \bibnamefont
  {Shaffer}},\ and\ \bibinfo {author} {\bibfnamefont {S.~D.}\ \bibnamefont
  {Baalrud}},\ }\href {https://doi.org/10.1103/PhysRevE.95.043204} {\bibfield
  {journal} {\bibinfo  {journal} {Phys. Rev. E}\ }\textbf {\bibinfo {volume}
  {95}},\ \bibinfo {pages} {043204} (\bibinfo {year} {2017})}\BibitemShut
  {NoStop}%
\bibitem [{\citenamefont {Zelener}(1977)}]{Zelener1977equation}%
  \BibitemOpen
  \bibfield  {author} {\bibinfo {author} {\bibfnamefont {B.~V.}\ \bibnamefont
  {Zelener}},\ }\href@noop {} {\bibfield  {journal} {\bibinfo  {journal} {TVT}\
  }\textbf {\bibinfo {volume} {15}},\ \bibinfo {pages} {893} (\bibinfo {year}
  {1977})}\BibitemShut {NoStop}%
\bibitem [{\citenamefont {Zelener}\ \emph {et~al.}(2016)\citenamefont
  {Zelener}, \citenamefont {Zelener},\ and\ \citenamefont
  {Butlitsky}}]{Zelener_2016}%
  \BibitemOpen
  \bibfield  {author} {\bibinfo {author} {\bibfnamefont {B.~V.}\ \bibnamefont
  {Zelener}}, \bibinfo {author} {\bibfnamefont {B.~B.}\ \bibnamefont
  {Zelener}},\ and\ \bibinfo {author} {\bibfnamefont {M.~A.}\ \bibnamefont
  {Butlitsky}},\ }\href {https://doi.org/10.1088/1742-6596/774/1/012158}
  {\bibfield  {journal} {\bibinfo  {journal} {Journal of Physics: Conference
  Series}\ }\textbf {\bibinfo {volume} {774}},\ \bibinfo {pages} {012158}
  (\bibinfo {year} {2016})}\BibitemShut {NoStop}%
\bibitem [{\citenamefont {Norman}\ and\ \citenamefont
  {Valuev}(1979)}]{Norman_1979}%
  \BibitemOpen
  \bibfield  {author} {\bibinfo {author} {\bibfnamefont {G.~E.}\ \bibnamefont
  {Norman}}\ and\ \bibinfo {author} {\bibfnamefont {A.~A.}\ \bibnamefont
  {Valuev}},\ }\href {https://doi.org/10.1088/0032-1028/21/6/002} {\bibfield
  {journal} {\bibinfo  {journal} {Plasma Physics}\ }\textbf {\bibinfo {volume}
  {21}},\ \bibinfo {pages} {531} (\bibinfo {year} {1979})}\BibitemShut
  {NoStop}%
\bibitem [{\citenamefont {Kuzmin}\ and\ \citenamefont
  {O’Neil}(2002)}]{Kuzmin:PhysPlasma:2002}%
  \BibitemOpen
  \bibfield  {author} {\bibinfo {author} {\bibfnamefont {S.~G.}\ \bibnamefont
  {Kuzmin}}\ and\ \bibinfo {author} {\bibfnamefont {T.~M.}\ \bibnamefont
  {O’Neil}},\ }\href {https://doi.org/10.1063/1.1497166} {\bibfield
  {journal} {\bibinfo  {journal} {Physics of Plasmas}\ }\textbf {\bibinfo
  {volume} {9}},\ \bibinfo {pages} {3743} (\bibinfo {year} {2002})},\ \Eprint
  {https://arxiv.org/abs/https://doi.org/10.1063/1.1497166}
  {https://doi.org/10.1063/1.1497166} \BibitemShut {NoStop}%
\bibitem [{\citenamefont {Maiorov}\ \emph {et~al.}(1991)\citenamefont
  {Maiorov}, \citenamefont {Tkachev},\ and\ \citenamefont
  {Yakovlenko}}]{Maiorov1991}%
  \BibitemOpen
  \bibfield  {author} {\bibinfo {author} {\bibfnamefont {S.~A.}\ \bibnamefont
  {Maiorov}}, \bibinfo {author} {\bibfnamefont {A.~N.}\ \bibnamefont
  {Tkachev}},\ and\ \bibinfo {author} {\bibfnamefont {S.~I.}\ \bibnamefont
  {Yakovlenko}},\ }\href {https://doi.org/10.1007/BF00895472} {\bibfield
  {journal} {\bibinfo  {journal} {Soviet Physics Journal}\ }\textbf {\bibinfo
  {volume} {34}},\ \bibinfo {pages} {951} (\bibinfo {year} {1991})}\BibitemShut
  {NoStop}%
\bibitem [{\citenamefont {Maiorov}\ \emph {et~al.}(1995)\citenamefont
  {Maiorov}, \citenamefont {Tkachev},\ and\ \citenamefont
  {Yakovlenko}}]{Maiorov_1995}%
  \BibitemOpen
  \bibfield  {author} {\bibinfo {author} {\bibfnamefont {S.~A.}\ \bibnamefont
  {Maiorov}}, \bibinfo {author} {\bibfnamefont {A.~N.}\ \bibnamefont
  {Tkachev}},\ and\ \bibinfo {author} {\bibfnamefont {S.~I.}\ \bibnamefont
  {Yakovlenko}},\ }\href {https://doi.org/10.1088/0031-8949/51/4/012}
  {\bibfield  {journal} {\bibinfo  {journal} {Physica Scripta}\ }\textbf
  {\bibinfo {volume} {51}},\ \bibinfo {pages} {498} (\bibinfo {year}
  {1995})}\BibitemShut {NoStop}%
\bibitem [{\citenamefont {Gabdullin}\ \emph {et~al.}(2016)\citenamefont
  {Gabdullin}, \citenamefont {Ramazanov}, \citenamefont {Ismagambetova},\ and\
  \citenamefont {Karimova}}]{gabdullin2016thermodynamic}%
  \BibitemOpen
  \bibfield  {author} {\bibinfo {author} {\bibfnamefont {M.}~\bibnamefont
  {Gabdullin}}, \bibinfo {author} {\bibfnamefont {T.}~\bibnamefont
  {Ramazanov}}, \bibinfo {author} {\bibfnamefont {T.}~\bibnamefont
  {Ismagambetova}},\ and\ \bibinfo {author} {\bibfnamefont {A.}~\bibnamefont
  {Karimova}},\ }in\ \href {https://doi.org/10.12955/cbup.v4.860} {\emph
  {\bibinfo {booktitle} {CBU International Conference Proceedings}}},\
  Vol.~\bibinfo {volume} {4}\ (\bibinfo {year} {2016})\ pp.\ \bibinfo {pages}
  {826--831}\BibitemShut {NoStop}%
\bibitem [{\citenamefont {Butlitsky}\ \emph {et~al.}(2008)\citenamefont
  {Butlitsky}, \citenamefont {Zelener}, \citenamefont {Zelener},\ and\
  \citenamefont {Manykin}}]{Butlitsky2008}%
  \BibitemOpen
  \bibfield  {author} {\bibinfo {author} {\bibfnamefont {M.~A.}\ \bibnamefont
  {Butlitsky}}, \bibinfo {author} {\bibfnamefont {B.~B.}\ \bibnamefont
  {Zelener}}, \bibinfo {author} {\bibfnamefont {B.~V.}\ \bibnamefont
  {Zelener}},\ and\ \bibinfo {author} {\bibfnamefont {E.~A.}\ \bibnamefont
  {Manykin}},\ }\href {https://doi.org/10.1134/S0965542508010119} {\bibfield
  {journal} {\bibinfo  {journal} {Computational Mathematics and Mathematical
  Physics}\ }\textbf {\bibinfo {volume} {48}},\ \bibinfo {pages} {147}
  (\bibinfo {year} {2008})}\BibitemShut {NoStop}%
\bibitem [{\citenamefont {Bonitz}\ \emph {et~al.}(2004)\citenamefont {Bonitz},
  \citenamefont {Zelener}, \citenamefont {Zelener}, \citenamefont {Manykin},
  \citenamefont {Filinov},\ and\ \citenamefont {Fortov}}]{Bonitz2004}%
  \BibitemOpen
  \bibfield  {author} {\bibinfo {author} {\bibfnamefont {M.}~\bibnamefont
  {Bonitz}}, \bibinfo {author} {\bibfnamefont {B.~B.}\ \bibnamefont {Zelener}},
  \bibinfo {author} {\bibfnamefont {B.~V.}\ \bibnamefont {Zelener}}, \bibinfo
  {author} {\bibfnamefont {E.~A.}\ \bibnamefont {Manykin}}, \bibinfo {author}
  {\bibfnamefont {V.~S.}\ \bibnamefont {Filinov}},\ and\ \bibinfo {author}
  {\bibfnamefont {V.~E.}\ \bibnamefont {Fortov}},\ }\href
  {https://doi.org/10.1134/1.1757672} {\bibfield  {journal} {\bibinfo
  {journal} {Journal of Experimental and Theoretical Physics}\ }\textbf
  {\bibinfo {volume} {98}},\ \bibinfo {pages} {719} (\bibinfo {year}
  {2004})}\BibitemShut {NoStop}%
\bibitem [{\citenamefont {Kelbg}(1963{\natexlab{a}})}]{Kelbg:AnnDerPhys:1963}%
  \BibitemOpen
  \bibfield  {author} {\bibinfo {author} {\bibfnamefont {G.}~\bibnamefont
  {Kelbg}},\ }\href {https://doi.org/https://doi.org/10.1002/andp.19634670308}
  {\bibfield  {journal} {\bibinfo  {journal} {Annalen der Physik}\ }\textbf
  {\bibinfo {volume} {467}},\ \bibinfo {pages} {219} (\bibinfo {year}
  {1963}{\natexlab{a}})}\BibitemShut {NoStop}%
\bibitem [{\citenamefont {Bonitz}\ \emph {et~al.}(2023)\citenamefont {Bonitz},
  \citenamefont {Ebeling}, \citenamefont {Filinov}, \citenamefont {Kraeft},
  \citenamefont {Redmer},\ and\ \citenamefont
  {Röpke}}]{Bonitz:ContrPlasPhys:2023}%
  \BibitemOpen
  \bibfield  {author} {\bibinfo {author} {\bibfnamefont {M.}~\bibnamefont
  {Bonitz}}, \bibinfo {author} {\bibfnamefont {W.}~\bibnamefont {Ebeling}},
  \bibinfo {author} {\bibfnamefont {A.}~\bibnamefont {Filinov}}, \bibinfo
  {author} {\bibfnamefont {W.}~\bibnamefont {Kraeft}}, \bibinfo {author}
  {\bibfnamefont {R.}~\bibnamefont {Redmer}},\ and\ \bibinfo {author}
  {\bibfnamefont {G.}~\bibnamefont {Röpke}},\ }\href
  {https://doi.org/https://doi.org/10.1002/ctpp.202300029} {\bibfield
  {journal} {\bibinfo  {journal} {Contributions to Plasma Physics}\ }\textbf
  {\bibinfo {volume} {63}},\ \bibinfo {pages} {e202300029} (\bibinfo {year}
  {2023})}\BibitemShut {NoStop}%
\bibitem [{\citenamefont {Ebeling}\ \emph {et~al.}(1999)\citenamefont
  {Ebeling}, \citenamefont {Norman}, \citenamefont {Valuev},\ and\
  \citenamefont {Valuev}}]{Ebeling:CPP:1999}%
  \BibitemOpen
  \bibfield  {author} {\bibinfo {author} {\bibfnamefont {W.}~\bibnamefont
  {Ebeling}}, \bibinfo {author} {\bibfnamefont {G.~E.}\ \bibnamefont {Norman}},
  \bibinfo {author} {\bibfnamefont {A.~A.}\ \bibnamefont {Valuev}},\ and\
  \bibinfo {author} {\bibfnamefont {I.~A.}\ \bibnamefont {Valuev}},\ }\href
  {https://doi.org/https://doi.org/10.1002/ctpp.2150390115} {\bibfield
  {journal} {\bibinfo  {journal} {Contributions to Plasma Physics}\ }\textbf
  {\bibinfo {volume} {39}},\ \bibinfo {pages} {61} (\bibinfo {year}
  {1999})}\BibitemShut {NoStop}%
\bibitem [{\citenamefont {Lavrinenko}\ \emph {et~al.}(2018)\citenamefont
  {Lavrinenko}, \citenamefont {Morozov},\ and\ \citenamefont
  {Valuev}}]{Lavrinenko:2018}%
  \BibitemOpen
  \bibfield  {author} {\bibinfo {author} {\bibfnamefont {Y.~S.}\ \bibnamefont
  {Lavrinenko}}, \bibinfo {author} {\bibfnamefont {I.~V.}\ \bibnamefont
  {Morozov}},\ and\ \bibinfo {author} {\bibfnamefont {I.~A.}\ \bibnamefont
  {Valuev}},\ }\href {https://doi.org/10.1088/1742-6596/946/1/012097}
  {\bibfield  {journal} {\bibinfo  {journal} {Journal of Physics: Conference
  Series}\ }\textbf {\bibinfo {volume} {946}},\ \bibinfo {pages} {012097}
  (\bibinfo {year} {2018})}\BibitemShut {NoStop}%
\bibitem [{\citenamefont {Filinov}\ \emph {et~al.}(2020)\citenamefont
  {Filinov}, \citenamefont {Larkin},\ and\ \citenamefont
  {Levashov}}]{Filinov:PRE:2020}%
  \BibitemOpen
  \bibfield  {author} {\bibinfo {author} {\bibfnamefont {V.~S.}\ \bibnamefont
  {Filinov}}, \bibinfo {author} {\bibfnamefont {A.~S.}\ \bibnamefont
  {Larkin}},\ and\ \bibinfo {author} {\bibfnamefont {P.~R.}\ \bibnamefont
  {Levashov}},\ }\href {https://doi.org/10.1103/PhysRevE.102.033203} {\bibfield
   {journal} {\bibinfo  {journal} {Phys. Rev. E}\ }\textbf {\bibinfo {volume}
  {102}},\ \bibinfo {pages} {033203} (\bibinfo {year} {2020})}\BibitemShut
  {NoStop}%
\bibitem [{\citenamefont {Filinov}\ \emph
  {et~al.}(2001{\natexlab{a}})\citenamefont {Filinov}, \citenamefont {Bonitz},
  \citenamefont {Kremp}, \citenamefont {Kraeft}, \citenamefont {Ebeling},
  \citenamefont {Levashov},\ and\ \citenamefont {Fortov}}]{Filinov:CPP:2001}%
  \BibitemOpen
  \bibfield  {author} {\bibinfo {author} {\bibfnamefont {V.}~\bibnamefont
  {Filinov}}, \bibinfo {author} {\bibfnamefont {M.}~\bibnamefont {Bonitz}},
  \bibinfo {author} {\bibfnamefont {D.}~\bibnamefont {Kremp}}, \bibinfo
  {author} {\bibfnamefont {W.-D.}\ \bibnamefont {Kraeft}}, \bibinfo {author}
  {\bibfnamefont {W.}~\bibnamefont {Ebeling}}, \bibinfo {author} {\bibfnamefont
  {P.}~\bibnamefont {Levashov}},\ and\ \bibinfo {author} {\bibfnamefont
  {V.}~\bibnamefont {Fortov}},\ }\href
  {https://doi.org/https://doi.org/10.1002/1521-3986(200103)41:2/3<135::AID-CTPP135>3.0.CO;2-C}
  {\bibfield  {journal} {\bibinfo  {journal} {Contributions to Plasma Physics}\
  }\textbf {\bibinfo {volume} {41}},\ \bibinfo {pages} {135} (\bibinfo {year}
  {2001}{\natexlab{a}})}\BibitemShut {NoStop}%
\bibitem [{\citenamefont {Deutsch}\ \emph {et~al.}(1978)\citenamefont
  {Deutsch}, \citenamefont {Gombert},\ and\ \citenamefont
  {Minoo}}]{DEUTSCH1978381}%
  \BibitemOpen
  \bibfield  {author} {\bibinfo {author} {\bibfnamefont {C.}~\bibnamefont
  {Deutsch}}, \bibinfo {author} {\bibfnamefont {M.}~\bibnamefont {Gombert}},\
  and\ \bibinfo {author} {\bibfnamefont {H.}~\bibnamefont {Minoo}},\ }\href
  {https://doi.org/https://doi.org/10.1016/0375-9601(78)90066-X} {\bibfield
  {journal} {\bibinfo  {journal} {Physics Letters A}\ }\textbf {\bibinfo
  {volume} {66}},\ \bibinfo {pages} {381} (\bibinfo {year} {1978})}\BibitemShut
  {NoStop}%
\bibitem [{\citenamefont {Benedict}\ \emph {et~al.}(2012)\citenamefont
  {Benedict}, \citenamefont {Surh}, \citenamefont {Castor}, \citenamefont
  {Khairallah}, \citenamefont {Whitley}, \citenamefont {Richards},
  \citenamefont {Glosli}, \citenamefont {Murillo}, \citenamefont {Scullard},
  \citenamefont {Grabowski}, \citenamefont {Michta},\ and\ \citenamefont
  {Graziani}}]{Benedict:PRE:2012}%
  \BibitemOpen
  \bibfield  {author} {\bibinfo {author} {\bibfnamefont {L.~X.}\ \bibnamefont
  {Benedict}}, \bibinfo {author} {\bibfnamefont {M.~P.}\ \bibnamefont {Surh}},
  \bibinfo {author} {\bibfnamefont {J.~I.}\ \bibnamefont {Castor}}, \bibinfo
  {author} {\bibfnamefont {S.~A.}\ \bibnamefont {Khairallah}}, \bibinfo
  {author} {\bibfnamefont {H.~D.}\ \bibnamefont {Whitley}}, \bibinfo {author}
  {\bibfnamefont {D.~F.}\ \bibnamefont {Richards}}, \bibinfo {author}
  {\bibfnamefont {J.~N.}\ \bibnamefont {Glosli}}, \bibinfo {author}
  {\bibfnamefont {M.~S.}\ \bibnamefont {Murillo}}, \bibinfo {author}
  {\bibfnamefont {C.~R.}\ \bibnamefont {Scullard}}, \bibinfo {author}
  {\bibfnamefont {P.~E.}\ \bibnamefont {Grabowski}}, \bibinfo {author}
  {\bibfnamefont {D.}~\bibnamefont {Michta}},\ and\ \bibinfo {author}
  {\bibfnamefont {F.~R.}\ \bibnamefont {Graziani}},\ }\href
  {https://doi.org/10.1103/PhysRevE.86.046406} {\bibfield  {journal} {\bibinfo
  {journal} {Phys. Rev. E}\ }\textbf {\bibinfo {volume} {86}},\ \bibinfo
  {pages} {046406} (\bibinfo {year} {2012})}\BibitemShut {NoStop}%
\bibitem [{\citenamefont {Kurilenkov}\ and\ \citenamefont
  {Valuev}(1984)}]{Kurilenkov:1984}%
  \BibitemOpen
  \bibfield  {author} {\bibinfo {author} {\bibfnamefont {Y.~K.}\ \bibnamefont
  {Kurilenkov}}\ and\ \bibinfo {author} {\bibfnamefont {A.~A.}\ \bibnamefont
  {Valuev}},\ }\href {https://doi.org/https://doi.org/10.1002/ctpp.19840240304}
  {\bibfield  {journal} {\bibinfo  {journal} {Beiträge aus der Plasmaphysik}\
  }\textbf {\bibinfo {volume} {24}},\ \bibinfo {pages} {161} (\bibinfo {year}
  {1984})}\BibitemShut {NoStop}%
\bibitem [{\citenamefont {Khomkin}\ and\ \citenamefont
  {Shumikhin}(2020)}]{Khomkin2020}%
  \BibitemOpen
  \bibfield  {author} {\bibinfo {author} {\bibfnamefont {A.~L.}\ \bibnamefont
  {Khomkin}}\ and\ \bibinfo {author} {\bibfnamefont {A.~S.}\ \bibnamefont
  {Shumikhin}},\ }\href {https://doi.org/10.1134/S0018151X20030098} {\bibfield
  {journal} {\bibinfo  {journal} {High Temperature}\ }\textbf {\bibinfo
  {volume} {58}},\ \bibinfo {pages} {305} (\bibinfo {year} {2020})}\BibitemShut
  {NoStop}%
\bibitem [{\citenamefont {Zelener}\ \emph {et~al.}(2018)\citenamefont
  {Zelener}, \citenamefont {Zelener}, \citenamefont {Manykin}, \citenamefont
  {Bronin}, \citenamefont {Bobrov},\ and\ \citenamefont
  {Khikhlukha}}]{Zelener_2018}%
  \BibitemOpen
  \bibfield  {author} {\bibinfo {author} {\bibfnamefont {B.~B.}\ \bibnamefont
  {Zelener}}, \bibinfo {author} {\bibfnamefont {B.~V.}\ \bibnamefont
  {Zelener}}, \bibinfo {author} {\bibfnamefont {E.~A.}\ \bibnamefont
  {Manykin}}, \bibinfo {author} {\bibfnamefont {S.~Y.}\ \bibnamefont {Bronin}},
  \bibinfo {author} {\bibfnamefont {A.~A.}\ \bibnamefont {Bobrov}},\ and\
  \bibinfo {author} {\bibfnamefont {D.~R.}\ \bibnamefont {Khikhlukha}},\ }\href
  {https://doi.org/10.1088/1742-6596/946/1/012126} {\bibfield  {journal}
  {\bibinfo  {journal} {Journal of Physics: Conference Series}\ }\textbf
  {\bibinfo {volume} {946}},\ \bibinfo {pages} {012126} (\bibinfo {year}
  {2018})}\BibitemShut {NoStop}%
\bibitem [{\citenamefont {Morozov}\ \emph {et~al.}(2005)\citenamefont
  {Morozov}, \citenamefont {Reinholz}, \citenamefont {R\"opke}, \citenamefont
  {Wierling},\ and\ \citenamefont {Zwicknagel}}]{Morozov:PRE:2005}%
  \BibitemOpen
  \bibfield  {author} {\bibinfo {author} {\bibfnamefont {I.}~\bibnamefont
  {Morozov}}, \bibinfo {author} {\bibfnamefont {H.}~\bibnamefont {Reinholz}},
  \bibinfo {author} {\bibfnamefont {G.}~\bibnamefont {R\"opke}}, \bibinfo
  {author} {\bibfnamefont {A.}~\bibnamefont {Wierling}},\ and\ \bibinfo
  {author} {\bibfnamefont {G.}~\bibnamefont {Zwicknagel}},\ }\href
  {https://doi.org/10.1103/PhysRevE.71.066408} {\bibfield  {journal} {\bibinfo
  {journal} {Phys. Rev. E}\ }\textbf {\bibinfo {volume} {71}},\ \bibinfo
  {pages} {066408} (\bibinfo {year} {2005})}\BibitemShut {NoStop}%
\bibitem [{\citenamefont {Dumin}\ and\ \citenamefont
  {Lukashenko}(2022)}]{Dumin:PlasPhys:2022}%
  \BibitemOpen
  \bibfield  {author} {\bibinfo {author} {\bibfnamefont {Y.~V.}\ \bibnamefont
  {Dumin}}\ and\ \bibinfo {author} {\bibfnamefont {A.~T.}\ \bibnamefont
  {Lukashenko}},\ }\href {https://doi.org/10.1063/5.0093840} {\bibfield
  {journal} {\bibinfo  {journal} {Physics of Plasmas}\ }\textbf {\bibinfo
  {volume} {29}},\ \bibinfo {pages} {113506} (\bibinfo {year}
  {2022})}\BibitemShut {NoStop}%
\bibitem [{\citenamefont {Barker}(1965)}]{Barker1965}%
  \BibitemOpen
  \bibfield  {author} {\bibinfo {author} {\bibfnamefont {A.~A.}\ \bibnamefont
  {Barker}},\ }\href {https://doi.org/10.1071/PH650119} {\bibfield  {journal}
  {\bibinfo  {journal} {Australian Journal of Physics}\ }\textbf {\bibinfo
  {volume} {18}},\ \bibinfo {pages} {119} (\bibinfo {year} {1965})}\BibitemShut
  {NoStop}%
\bibitem [{\citenamefont {Barker}(1968)}]{Barker:PR:1968}%
  \BibitemOpen
  \bibfield  {author} {\bibinfo {author} {\bibfnamefont {A.~A.}\ \bibnamefont
  {Barker}},\ }\href {https://doi.org/10.1103/PhysRev.171.186} {\bibfield
  {journal} {\bibinfo  {journal} {Phys. Rev.}\ }\textbf {\bibinfo {volume}
  {171}},\ \bibinfo {pages} {186} (\bibinfo {year} {1968})}\BibitemShut
  {NoStop}%
\bibitem [{\citenamefont {Zelener}\ \emph {et~al.}(1972)\citenamefont
  {Zelener}, \citenamefont {Norman},\ and\ \citenamefont
  {Filinov}}]{ZelNorFil72}%
  \BibitemOpen
  \bibfield  {author} {\bibinfo {author} {\bibfnamefont {B.~V.}\ \bibnamefont
  {Zelener}}, \bibinfo {author} {\bibfnamefont {H.~E.}\ \bibnamefont
  {Norman}},\ and\ \bibinfo {author} {\bibfnamefont {V.~S.}\ \bibnamefont
  {Filinov}},\ }\href {http://mi.mathnet.ru/tvt10306} {\bibfield  {journal}
  {\bibinfo  {journal} {TVT}\ }\textbf {\bibinfo {volume} {10}},\ \bibinfo
  {pages} {1160} (\bibinfo {year} {1972})}\BibitemShut {NoStop}%
\bibitem [{\citenamefont {Valuev}\ \emph {et~al.}(1974)\citenamefont {Valuev},
  \citenamefont {Norman},\ and\ \citenamefont {Filinov}}]{ValNorFil74}%
  \BibitemOpen
  \bibfield  {author} {\bibinfo {author} {\bibfnamefont {A.}~\bibnamefont
  {Valuev}}, \bibinfo {author} {\bibfnamefont {H.~E.}\ \bibnamefont {Norman}},\
  and\ \bibinfo {author} {\bibfnamefont {V.~S.}\ \bibnamefont {Filinov}},\
  }\href {http://mi.mathnet.ru/tvt7436} {\bibfield  {journal} {\bibinfo
  {journal} {TVT}\ }\textbf {\bibinfo {volume} {12}},\ \bibinfo {pages} {931}
  (\bibinfo {year} {1974})}\BibitemShut {NoStop}%
\bibitem [{\citenamefont {Hansen}\ and\ \citenamefont
  {Viot}(1985)}]{Hansen1985}%
  \BibitemOpen
  \bibfield  {author} {\bibinfo {author} {\bibfnamefont {J.~P.}\ \bibnamefont
  {Hansen}}\ and\ \bibinfo {author} {\bibfnamefont {P.}~\bibnamefont {Viot}},\
  }\href {https://doi.org/10.1007/BF01010417} {\bibfield  {journal} {\bibinfo
  {journal} {Journal of Statistical Physics}\ }\textbf {\bibinfo {volume}
  {38}},\ \bibinfo {pages} {823} (\bibinfo {year} {1985})}\BibitemShut
  {NoStop}%
\bibitem [{\citenamefont {Redmer}\ and\ \citenamefont
  {R\"opke}(2010)}]{Redmer:CPP:2010}%
  \BibitemOpen
  \bibfield  {author} {\bibinfo {author} {\bibfnamefont {R.}~\bibnamefont
  {Redmer}}\ and\ \bibinfo {author} {\bibfnamefont {G.}~\bibnamefont
  {R\"opke}},\ }\href {https://doi.org/https://doi.org/10.1002/ctpp.201000079}
  {\bibfield  {journal} {\bibinfo  {journal} {Contributions to Plasma Physics}\
  }\textbf {\bibinfo {volume} {50}},\ \bibinfo {pages} {970} (\bibinfo {year}
  {2010})}\BibitemShut {NoStop}%
\bibitem [{\citenamefont {Holst}\ \emph {et~al.}(2008)\citenamefont {Holst},
  \citenamefont {Redmer},\ and\ \citenamefont {Desjarlais}}]{Holst:PRB:2008}%
  \BibitemOpen
  \bibfield  {author} {\bibinfo {author} {\bibfnamefont {B.}~\bibnamefont
  {Holst}}, \bibinfo {author} {\bibfnamefont {R.}~\bibnamefont {Redmer}},\ and\
  \bibinfo {author} {\bibfnamefont {M.~P.}\ \bibnamefont {Desjarlais}},\ }\href
  {https://doi.org/10.1103/PhysRevB.77.184201} {\bibfield  {journal} {\bibinfo
  {journal} {Phys. Rev. B}\ }\textbf {\bibinfo {volume} {77}},\ \bibinfo
  {pages} {184201} (\bibinfo {year} {2008})}\BibitemShut {NoStop}%
\bibitem [{\citenamefont {Norman}\ \emph {et~al.}(2019)\citenamefont {Norman},
  \citenamefont {Saitov},\ and\ \citenamefont
  {Sartan}}]{Norman:ContrPlasPhys:2019}%
  \BibitemOpen
  \bibfield  {author} {\bibinfo {author} {\bibfnamefont {G.~E.}\ \bibnamefont
  {Norman}}, \bibinfo {author} {\bibfnamefont {I.~M.}\ \bibnamefont {Saitov}},\
  and\ \bibinfo {author} {\bibfnamefont {R.~A.}\ \bibnamefont {Sartan}},\
  }\href@noop {} {\bibfield  {journal} {\bibinfo  {journal} {Contributions to
  Plasma Physics}\ }\textbf {\bibinfo {volume} {59}},\ \bibinfo {pages}
  {e201800173} (\bibinfo {year} {2019})}\BibitemShut {NoStop}%
\bibitem [{\citenamefont {Knyazev}\ and\ \citenamefont
  {Levashov}(2016)}]{Knyazev:PhysPlas:2016}%
  \BibitemOpen
  \bibfield  {author} {\bibinfo {author} {\bibfnamefont {D.~V.}\ \bibnamefont
  {Knyazev}}\ and\ \bibinfo {author} {\bibfnamefont {P.~R.}\ \bibnamefont
  {Levashov}},\ }\href {https://doi.org/10.1063/1.4966565} {\bibfield
  {journal} {\bibinfo  {journal} {Physics of Plasmas}\ }\textbf {\bibinfo
  {volume} {23}},\ \bibinfo {pages} {102708} (\bibinfo {year}
  {2016})}\BibitemShut {NoStop}%
\bibitem [{\citenamefont {Zwicknagel}\ \emph {et~al.}(1993)\citenamefont
  {Zwicknagel}, \citenamefont {Klakow}, \citenamefont {Reinhard},\ and\
  \citenamefont {Toepffer}}]{Zwicknagel:CPP:1993}%
  \BibitemOpen
  \bibfield  {author} {\bibinfo {author} {\bibfnamefont {G.}~\bibnamefont
  {Zwicknagel}}, \bibinfo {author} {\bibfnamefont {D.}~\bibnamefont {Klakow}},
  \bibinfo {author} {\bibfnamefont {P.-G.}\ \bibnamefont {Reinhard}},\ and\
  \bibinfo {author} {\bibfnamefont {C.}~\bibnamefont {Toepffer}},\ }\href
  {https://doi.org/https://doi.org/10.1002/ctpp.2150330510} {\bibfield
  {journal} {\bibinfo  {journal} {Contributions to Plasma Physics}\ }\textbf
  {\bibinfo {volume} {33}},\ \bibinfo {pages} {395} (\bibinfo {year}
  {1993})}\BibitemShut {NoStop}%
\bibitem [{\citenamefont {Lavrinenko}\ \emph {et~al.}(2019)\citenamefont
  {Lavrinenko}, \citenamefont {Morozov},\ and\ \citenamefont
  {Valuev}}]{Lavrinenko:CPP:2019}%
  \BibitemOpen
  \bibfield  {author} {\bibinfo {author} {\bibfnamefont {Y.~S.}\ \bibnamefont
  {Lavrinenko}}, \bibinfo {author} {\bibfnamefont {I.~V.}\ \bibnamefont
  {Morozov}},\ and\ \bibinfo {author} {\bibfnamefont {I.~A.}\ \bibnamefont
  {Valuev}},\ }\href {https://doi.org/https://doi.org/10.1002/ctpp.201800179}
  {\bibfield  {journal} {\bibinfo  {journal} {Contributions to Plasma Physics}\
  }\textbf {\bibinfo {volume} {59}},\ \bibinfo {pages} {e201800179} (\bibinfo
  {year} {2019})}\BibitemShut {NoStop}%
\bibitem [{\citenamefont {Lavrinenko}\ \emph {et~al.}(2021)\citenamefont
  {Lavrinenko}, \citenamefont {Levashov}, \citenamefont {Minakov},
  \citenamefont {Morozov},\ and\ \citenamefont {Valuev}}]{Lavrinenko:PRE:2021}%
  \BibitemOpen
  \bibfield  {author} {\bibinfo {author} {\bibfnamefont {Y.}~\bibnamefont
  {Lavrinenko}}, \bibinfo {author} {\bibfnamefont {P.~R.}\ \bibnamefont
  {Levashov}}, \bibinfo {author} {\bibfnamefont {D.~V.}\ \bibnamefont
  {Minakov}}, \bibinfo {author} {\bibfnamefont {I.~V.}\ \bibnamefont
  {Morozov}},\ and\ \bibinfo {author} {\bibfnamefont {I.~A.}\ \bibnamefont
  {Valuev}},\ }\href {https://doi.org/10.1103/PhysRevE.104.045304} {\bibfield
  {journal} {\bibinfo  {journal} {Phys. Rev. E}\ }\textbf {\bibinfo {volume}
  {104}},\ \bibinfo {pages} {045304} (\bibinfo {year} {2021})}\BibitemShut
  {NoStop}%
\bibitem [{\citenamefont {Filinov}\ \emph
  {et~al.}(2001{\natexlab{b}})\citenamefont {Filinov}, \citenamefont {Bonitz},
  \citenamefont {Ebeling},\ and\ \citenamefont {Fortov}}]{Filinov_2001}%
  \BibitemOpen
  \bibfield  {author} {\bibinfo {author} {\bibfnamefont {V.~S.}\ \bibnamefont
  {Filinov}}, \bibinfo {author} {\bibfnamefont {M.}~\bibnamefont {Bonitz}},
  \bibinfo {author} {\bibfnamefont {W.}~\bibnamefont {Ebeling}},\ and\ \bibinfo
  {author} {\bibfnamefont {V.~E.}\ \bibnamefont {Fortov}},\ }\href
  {https://doi.org/10.1088/0741-3335/43/6/301} {\bibfield  {journal} {\bibinfo
  {journal} {Plasma Physics and Controlled Fusion}\ }\textbf {\bibinfo {volume}
  {43}},\ \bibinfo {pages} {743} (\bibinfo {year}
  {2001}{\natexlab{b}})}\BibitemShut {NoStop}%
\bibitem [{\citenamefont {Militzer}\ and\ \citenamefont
  {Ceperley}(2001)}]{Militzer:PRE:2001}%
  \BibitemOpen
  \bibfield  {author} {\bibinfo {author} {\bibfnamefont {B.}~\bibnamefont
  {Militzer}}\ and\ \bibinfo {author} {\bibfnamefont {D.~M.}\ \bibnamefont
  {Ceperley}},\ }\href {https://doi.org/10.1103/PhysRevE.63.066404} {\bibfield
  {journal} {\bibinfo  {journal} {Phys. Rev. E}\ }\textbf {\bibinfo {volume}
  {63}},\ \bibinfo {pages} {066404} (\bibinfo {year} {2001})}\BibitemShut
  {NoStop}%
\bibitem [{\citenamefont {Bezkrovniy}\ \emph {et~al.}(2004)\citenamefont
  {Bezkrovniy}, \citenamefont {Filinov}, \citenamefont {Kremp}, \citenamefont
  {Bonitz}, \citenamefont {Schlanges}, \citenamefont {Kraeft}, \citenamefont
  {Levashov},\ and\ \citenamefont {Fortov}}]{Bezkrovniy:PRE:2004}%
  \BibitemOpen
  \bibfield  {author} {\bibinfo {author} {\bibfnamefont {V.}~\bibnamefont
  {Bezkrovniy}}, \bibinfo {author} {\bibfnamefont {V.~S.}\ \bibnamefont
  {Filinov}}, \bibinfo {author} {\bibfnamefont {D.}~\bibnamefont {Kremp}},
  \bibinfo {author} {\bibfnamefont {M.}~\bibnamefont {Bonitz}}, \bibinfo
  {author} {\bibfnamefont {M.}~\bibnamefont {Schlanges}}, \bibinfo {author}
  {\bibfnamefont {W.~D.}\ \bibnamefont {Kraeft}}, \bibinfo {author}
  {\bibfnamefont {P.~R.}\ \bibnamefont {Levashov}},\ and\ \bibinfo {author}
  {\bibfnamefont {V.~E.}\ \bibnamefont {Fortov}},\ }\href
  {https://doi.org/10.1103/PhysRevE.70.057401} {\bibfield  {journal} {\bibinfo
  {journal} {Phys. Rev. E}\ }\textbf {\bibinfo {volume} {70}},\ \bibinfo
  {pages} {057401} (\bibinfo {year} {2004})}\BibitemShut {NoStop}%
\bibitem [{\citenamefont {Alexandru}\ \emph {et~al.}(2022)\citenamefont
  {Alexandru}, \citenamefont {Ba\ifmmode~\mbox{\c{s}}\else \c{s}\fi{}ar},
  \citenamefont {Bedaque},\ and\ \citenamefont
  {Warrington}}]{Alexandru:RevModPhys:2022}%
  \BibitemOpen
  \bibfield  {author} {\bibinfo {author} {\bibfnamefont {A.}~\bibnamefont
  {Alexandru}}, \bibinfo {author} {\bibfnamefont {G.~m.~c.}\ \bibnamefont
  {Ba\ifmmode~\mbox{\c{s}}\else \c{s}\fi{}ar}}, \bibinfo {author}
  {\bibfnamefont {P.~F.}\ \bibnamefont {Bedaque}},\ and\ \bibinfo {author}
  {\bibfnamefont {N.~C.}\ \bibnamefont {Warrington}},\ }\href
  {https://doi.org/10.1103/RevModPhys.94.015006} {\bibfield  {journal}
  {\bibinfo  {journal} {Rev. Mod. Phys.}\ }\textbf {\bibinfo {volume} {94}},\
  \bibinfo {pages} {015006} (\bibinfo {year} {2022})}\BibitemShut {NoStop}%
\bibitem [{\citenamefont {Filinov}\ \emph {et~al.}(2022)\citenamefont
  {Filinov}, \citenamefont {Levashov},\ and\ \citenamefont
  {Larkin}}]{Filinov:PhysPlas:2022}%
  \BibitemOpen
  \bibfield  {author} {\bibinfo {author} {\bibfnamefont {V.~S.}\ \bibnamefont
  {Filinov}}, \bibinfo {author} {\bibfnamefont {P.~R.}\ \bibnamefont
  {Levashov}},\ and\ \bibinfo {author} {\bibfnamefont {A.~S.}\ \bibnamefont
  {Larkin}},\ }\href {https://doi.org/10.1063/5.0089836} {\bibfield  {journal}
  {\bibinfo  {journal} {Physics of Plasmas}\ }\textbf {\bibinfo {volume}
  {29}},\ \bibinfo {pages} {052106} (\bibinfo {year} {2022})}\BibitemShut
  {NoStop}%
\bibitem [{\citenamefont {Larkin}\ \emph {et~al.}(2021)\citenamefont {Larkin},
  \citenamefont {Filinov},\ and\ \citenamefont
  {Levashov}}]{Larkin:PhysPlas:2021}%
  \BibitemOpen
  \bibfield  {author} {\bibinfo {author} {\bibfnamefont {A.~S.}\ \bibnamefont
  {Larkin}}, \bibinfo {author} {\bibfnamefont {V.~S.}\ \bibnamefont
  {Filinov}},\ and\ \bibinfo {author} {\bibfnamefont {P.~R.}\ \bibnamefont
  {Levashov}},\ }\href {https://doi.org/10.1063/5.0072354} {\bibfield
  {journal} {\bibinfo  {journal} {Physics of Plasmas}\ }\textbf {\bibinfo
  {volume} {28}},\ \bibinfo {pages} {122712} (\bibinfo {year}
  {2021})}\BibitemShut {NoStop}%
\bibitem [{\citenamefont {Dornheim}\ \emph {et~al.}(2018)\citenamefont
  {Dornheim}, \citenamefont {Groth},\ and\ \citenamefont
  {Bonitz}}]{DORNHEIM20181}%
  \BibitemOpen
  \bibfield  {author} {\bibinfo {author} {\bibfnamefont {T.}~\bibnamefont
  {Dornheim}}, \bibinfo {author} {\bibfnamefont {S.}~\bibnamefont {Groth}},\
  and\ \bibinfo {author} {\bibfnamefont {M.}~\bibnamefont {Bonitz}},\ }\href
  {https://doi.org/https://doi.org/10.1016/j.physrep.2018.04.001} {\bibfield
  {journal} {\bibinfo  {journal} {Physics Reports}\ }\textbf {\bibinfo {volume}
  {744}},\ \bibinfo {pages} {1} (\bibinfo {year} {2018})}\BibitemShut {NoStop}%
\bibitem [{\citenamefont {Ceperley}(1991)}]{Ceperley1991}%
  \BibitemOpen
  \bibfield  {author} {\bibinfo {author} {\bibfnamefont {D.~M.}\ \bibnamefont
  {Ceperley}},\ }\href {https://doi.org/10.1007/BF01030009} {\bibfield
  {journal} {\bibinfo  {journal} {Journal of Statistical Physics}\ }\textbf
  {\bibinfo {volume} {63}},\ \bibinfo {pages} {1237} (\bibinfo {year}
  {1991})}\BibitemShut {NoStop}%
\bibitem [{\citenamefont {Ichimaru}\ \emph {et~al.}(1985)\citenamefont
  {Ichimaru}, \citenamefont {Mitake}, \citenamefont {Tanaka},\ and\
  \citenamefont {Yan}}]{Ichimaru:PhysRevA:1985}%
  \BibitemOpen
  \bibfield  {author} {\bibinfo {author} {\bibfnamefont {S.}~\bibnamefont
  {Ichimaru}}, \bibinfo {author} {\bibfnamefont {S.}~\bibnamefont {Mitake}},
  \bibinfo {author} {\bibfnamefont {S.}~\bibnamefont {Tanaka}},\ and\ \bibinfo
  {author} {\bibfnamefont {X.-Z.}\ \bibnamefont {Yan}},\ }\href
  {https://doi.org/10.1103/PhysRevA.32.1768} {\bibfield  {journal} {\bibinfo
  {journal} {Phys. Rev. A}\ }\textbf {\bibinfo {volume} {32}},\ \bibinfo
  {pages} {1768} (\bibinfo {year} {1985})}\BibitemShut {NoStop}%
\bibitem [{\citenamefont {Ramazanov}\ \emph {et~al.}(2014)\citenamefont
  {Ramazanov}, \citenamefont {Moldabekov}, \citenamefont {Gabdullin},\ and\
  \citenamefont {Ismagambetova}}]{Ramazanov:PhysPlasmas:2014}%
  \BibitemOpen
  \bibfield  {author} {\bibinfo {author} {\bibfnamefont {T.~S.}\ \bibnamefont
  {Ramazanov}}, \bibinfo {author} {\bibfnamefont {Z.~A.}\ \bibnamefont
  {Moldabekov}}, \bibinfo {author} {\bibfnamefont {M.~T.}\ \bibnamefont
  {Gabdullin}},\ and\ \bibinfo {author} {\bibfnamefont {T.~N.}\ \bibnamefont
  {Ismagambetova}},\ }\href {https://doi.org/10.1063/1.4862549} {\bibfield
  {journal} {\bibinfo  {journal} {Physics of Plasmas}\ }\textbf {\bibinfo
  {volume} {21}},\ \bibinfo {pages} {012706} (\bibinfo {year}
  {2014})}\BibitemShut {NoStop}%
\bibitem [{\citenamefont {Tanaka}\ and\ \citenamefont
  {Ichimaru}(1985)}]{Tanaka:PRA:1985}%
  \BibitemOpen
  \bibfield  {author} {\bibinfo {author} {\bibfnamefont {S.}~\bibnamefont
  {Tanaka}}\ and\ \bibinfo {author} {\bibfnamefont {S.}~\bibnamefont
  {Ichimaru}},\ }\href {https://doi.org/10.1103/PhysRevA.32.3756} {\bibfield
  {journal} {\bibinfo  {journal} {Phys. Rev. A}\ }\textbf {\bibinfo {volume}
  {32}},\ \bibinfo {pages} {3756} (\bibinfo {year} {1985})}\BibitemShut
  {NoStop}%
\bibitem [{\citenamefont {Ichimaru}\ \emph {et~al.}(1987)\citenamefont
  {Ichimaru}, \citenamefont {Iyetomi},\ and\ \citenamefont
  {Tanaka}}]{Ichimaru:PhysRep:1987}%
  \BibitemOpen
  \bibfield  {author} {\bibinfo {author} {\bibfnamefont {S.}~\bibnamefont
  {Ichimaru}}, \bibinfo {author} {\bibfnamefont {H.}~\bibnamefont {Iyetomi}},\
  and\ \bibinfo {author} {\bibfnamefont {S.}~\bibnamefont {Tanaka}},\ }\href
  {https://doi.org/https://doi.org/10.1016/0370-1573(87)90125-6} {\bibfield
  {journal} {\bibinfo  {journal} {Physics Reports}\ }\textbf {\bibinfo {volume}
  {149}},\ \bibinfo {pages} {91} (\bibinfo {year} {1987})}\BibitemShut
  {NoStop}%
\bibitem [{\citenamefont {Chabrier}\ and\ \citenamefont
  {Potekhin}(1998)}]{ChabrierPotekhin:PRE:1998}%
  \BibitemOpen
  \bibfield  {author} {\bibinfo {author} {\bibfnamefont {G.}~\bibnamefont
  {Chabrier}}\ and\ \bibinfo {author} {\bibfnamefont {A.~Y.}\ \bibnamefont
  {Potekhin}},\ }\href {https://doi.org/10.1103/PhysRevE.58.4941} {\bibfield
  {journal} {\bibinfo  {journal} {Phys. Rev. E}\ }\textbf {\bibinfo {volume}
  {58}},\ \bibinfo {pages} {4941} (\bibinfo {year} {1998})}\BibitemShut
  {NoStop}%
\bibitem [{\citenamefont {Potekhin}\ and\ \citenamefont
  {Chabrier}(2000)}]{Potekhin:PRE:2000}%
  \BibitemOpen
  \bibfield  {author} {\bibinfo {author} {\bibfnamefont {A.~Y.}\ \bibnamefont
  {Potekhin}}\ and\ \bibinfo {author} {\bibfnamefont {G.}~\bibnamefont
  {Chabrier}},\ }\href {https://doi.org/10.1103/PhysRevE.62.8554} {\bibfield
  {journal} {\bibinfo  {journal} {Phys. Rev. E}\ }\textbf {\bibinfo {volume}
  {62}},\ \bibinfo {pages} {8554} (\bibinfo {year} {2000})}\BibitemShut
  {NoStop}%
\bibitem [{\citenamefont {Potekhin}\ \emph
  {et~al.}(2009{\natexlab{a}})\citenamefont {Potekhin}, \citenamefont
  {Chabrier},\ and\ \citenamefont {Rogers}}]{Potekhin:PRE:2009}%
  \BibitemOpen
  \bibfield  {author} {\bibinfo {author} {\bibfnamefont {A.~Y.}\ \bibnamefont
  {Potekhin}}, \bibinfo {author} {\bibfnamefont {G.}~\bibnamefont {Chabrier}},\
  and\ \bibinfo {author} {\bibfnamefont {F.~J.}\ \bibnamefont {Rogers}},\
  }\href {https://doi.org/10.1103/PhysRevE.79.016411} {\bibfield  {journal}
  {\bibinfo  {journal} {Phys. Rev. E}\ }\textbf {\bibinfo {volume} {79}},\
  \bibinfo {pages} {016411} (\bibinfo {year} {2009}{\natexlab{a}})}\BibitemShut
  {NoStop}%
\bibitem [{\citenamefont {Potekhin}\ \emph
  {et~al.}(2009{\natexlab{b}})\citenamefont {Potekhin}, \citenamefont
  {Chabrier}, \citenamefont {Chugunov}, \citenamefont {DeWitt},\ and\
  \citenamefont {Rogers}}]{PotekhinAddendum:PRE:2009}%
  \BibitemOpen
  \bibfield  {author} {\bibinfo {author} {\bibfnamefont {A.~Y.}\ \bibnamefont
  {Potekhin}}, \bibinfo {author} {\bibfnamefont {G.}~\bibnamefont {Chabrier}},
  \bibinfo {author} {\bibfnamefont {A.~I.}\ \bibnamefont {Chugunov}}, \bibinfo
  {author} {\bibfnamefont {H.~E.}\ \bibnamefont {DeWitt}},\ and\ \bibinfo
  {author} {\bibfnamefont {F.~J.}\ \bibnamefont {Rogers}},\ }\href
  {https://doi.org/10.1103/PhysRevE.80.047401} {\bibfield  {journal} {\bibinfo
  {journal} {Phys. Rev. E}\ }\textbf {\bibinfo {volume} {80}},\ \bibinfo
  {pages} {047401} (\bibinfo {year} {2009}{\natexlab{b}})}\BibitemShut
  {NoStop}%
\bibitem [{\citenamefont {Baiko}\ and\ \citenamefont
  {Chugunov}(2021)}]{BaikoChugunov:2021}%
  \BibitemOpen
  \bibfield  {author} {\bibinfo {author} {\bibfnamefont {D.~A.}\ \bibnamefont
  {Baiko}}\ and\ \bibinfo {author} {\bibfnamefont {A.~I.}\ \bibnamefont
  {Chugunov}},\ }\href {https://doi.org/10.1093/mnras/stab3613} {\bibfield
  {journal} {\bibinfo  {journal} {Monthly Notices of the Royal Astronomical
  Society}\ }\textbf {\bibinfo {volume} {510}},\ \bibinfo {pages} {2628}
  (\bibinfo {year} {2021})}\BibitemShut {NoStop}%
\bibitem [{\citenamefont {Morozov}(2006)}]{MorozovIV:2006}%
  \BibitemOpen
  \bibfield  {author} {\bibinfo {author} {\bibfnamefont {I.~V.}\ \bibnamefont
  {Morozov}},\ }\href@noop {} {\bibfield  {journal} {\bibinfo  {journal}
  {Physics of Extreme States of Matter}\ ,\ \bibinfo {pages} {219}} (\bibinfo
  {year} {2006})}\BibitemShut {NoStop}%
\bibitem [{\citenamefont {Lankin}\ and\ \citenamefont
  {Norman}(2008)}]{Lankin2008}%
  \BibitemOpen
  \bibfield  {author} {\bibinfo {author} {\bibfnamefont {A.~V.}\ \bibnamefont
  {Lankin}}\ and\ \bibinfo {author} {\bibfnamefont {G.~E.}\ \bibnamefont
  {Norman}},\ }\href {https://doi.org/10.1134/s10740-008-2002-1} {\bibfield
  {journal} {\bibinfo  {journal} {High Temperature}\ }\textbf {\bibinfo
  {volume} {46}},\ \bibinfo {pages} {148} (\bibinfo {year} {2008})}\BibitemShut
  {NoStop}%
\bibitem [{\citenamefont {Ebeling}\ \emph {et~al.}(1976)\citenamefont
  {Ebeling}, \citenamefont {Kraeft},\ and\ \citenamefont
  {Kremp}}]{Ebeling:book:1976}%
  \BibitemOpen
  \bibfield  {author} {\bibinfo {author} {\bibfnamefont {W.}~\bibnamefont
  {Ebeling}}, \bibinfo {author} {\bibfnamefont {W.}~\bibnamefont {Kraeft}},\
  and\ \bibinfo {author} {\bibfnamefont {D.}~\bibnamefont {Kremp}},\
  }\href@noop {} {\emph {\bibinfo {title} {Theory of bound states and
  ionization equilibrium in plasmas and solids}}}\ (\bibinfo  {publisher}
  {Akademie‐Verlag},\ \bibinfo {year} {1976})\BibitemShut {NoStop}%
\bibitem [{\citenamefont {Lankin}\ and\ \citenamefont
  {Norman}(2009)}]{Lankin_2009}%
  \BibitemOpen
  \bibfield  {author} {\bibinfo {author} {\bibfnamefont {A.~V.}\ \bibnamefont
  {Lankin}}\ and\ \bibinfo {author} {\bibfnamefont {G.~E.}\ \bibnamefont
  {Norman}},\ }\href {https://doi.org/10.1088/1751-8113/42/21/214032}
  {\bibfield  {journal} {\bibinfo  {journal} {Journal of Physics A:
  Mathematical and Theoretical}\ }\textbf {\bibinfo {volume} {42}},\ \bibinfo
  {pages} {214032} (\bibinfo {year} {2009})}\BibitemShut {NoStop}%
\bibitem [{\citenamefont {Starostin}\ and\ \citenamefont
  {Roerich}(2006)}]{Starostin_2006}%
  \BibitemOpen
  \bibfield  {author} {\bibinfo {author} {\bibfnamefont {A.~N.}\ \bibnamefont
  {Starostin}}\ and\ \bibinfo {author} {\bibfnamefont {V.~C.}\ \bibnamefont
  {Roerich}},\ }\href {https://doi.org/10.1088/0305-4470/39/17/S18} {\bibfield
  {journal} {\bibinfo  {journal} {Journal of Physics A: Mathematical and
  General}\ }\textbf {\bibinfo {volume} {39}},\ \bibinfo {pages} {4431}
  (\bibinfo {year} {2006})}\BibitemShut {NoStop}%
\bibitem [{\citenamefont {Starostin}\ \emph {et~al.}(2009)\citenamefont
  {Starostin}, \citenamefont {Roerich}, \citenamefont {Gryaznov}, \citenamefont
  {Fortov},\ and\ \citenamefont {Iosilevskiy}}]{Starostin_2009}%
  \BibitemOpen
  \bibfield  {author} {\bibinfo {author} {\bibfnamefont {A.~N.}\ \bibnamefont
  {Starostin}}, \bibinfo {author} {\bibfnamefont {V.~C.}\ \bibnamefont
  {Roerich}}, \bibinfo {author} {\bibfnamefont {V.~K.}\ \bibnamefont
  {Gryaznov}}, \bibinfo {author} {\bibfnamefont {V.~E.}\ \bibnamefont
  {Fortov}},\ and\ \bibinfo {author} {\bibfnamefont {I.~L.}\ \bibnamefont
  {Iosilevskiy}},\ }\href {https://doi.org/10.1088/1751-8113/42/21/214009}
  {\bibfield  {journal} {\bibinfo  {journal} {Journal of Physics A:
  Mathematical and Theoretical}\ }\textbf {\bibinfo {volume} {42}},\ \bibinfo
  {pages} {214009} (\bibinfo {year} {2009})}\BibitemShut {NoStop}%
\bibitem [{\citenamefont {Eastwood}\ and\ \citenamefont
  {Hockney}(1974)}]{Eastwood:JCP:1974}%
  \BibitemOpen
  \bibfield  {author} {\bibinfo {author} {\bibfnamefont {J.~W.}\ \bibnamefont
  {Eastwood}}\ and\ \bibinfo {author} {\bibfnamefont {R.~W.}\ \bibnamefont
  {Hockney}},\ }\href@noop {} {\bibfield  {journal} {\bibinfo  {journal}
  {Journal of Computational Physics}\ }\textbf {\bibinfo {volume} {16}},\
  \bibinfo {pages} {342} (\bibinfo {year} {1974})}\BibitemShut {NoStop}%
\bibitem [{\citenamefont {Greengard}\ and\ \citenamefont
  {Rokhlin}(1987)}]{Greengard:JCP:1987}%
  \BibitemOpen
  \bibfield  {author} {\bibinfo {author} {\bibfnamefont {L.}~\bibnamefont
  {Greengard}}\ and\ \bibinfo {author} {\bibfnamefont {V.}~\bibnamefont
  {Rokhlin}},\ }\href@noop {} {\bibfield  {journal} {\bibinfo  {journal}
  {Journal of computational physics}\ }\textbf {\bibinfo {volume} {73}},\
  \bibinfo {pages} {325} (\bibinfo {year} {1987})}\BibitemShut {NoStop}%
\bibitem [{\citenamefont {Essmann}\ \emph {et~al.}(1995)\citenamefont
  {Essmann}, \citenamefont {Perera}, \citenamefont {Berkowitz}, \citenamefont
  {Darden}, \citenamefont {Lee},\ and\ \citenamefont
  {Pedersen}}]{Essmann:1995}%
  \BibitemOpen
  \bibfield  {author} {\bibinfo {author} {\bibfnamefont {U.}~\bibnamefont
  {Essmann}}, \bibinfo {author} {\bibfnamefont {L.}~\bibnamefont {Perera}},
  \bibinfo {author} {\bibfnamefont {M.~L.}\ \bibnamefont {Berkowitz}}, \bibinfo
  {author} {\bibfnamefont {T.}~\bibnamefont {Darden}}, \bibinfo {author}
  {\bibfnamefont {H.}~\bibnamefont {Lee}},\ and\ \bibinfo {author}
  {\bibfnamefont {L.~G.}\ \bibnamefont {Pedersen}},\ }\href
  {https://doi.org/10.1063/1.470117} {\bibfield  {journal} {\bibinfo  {journal}
  {The Journal of Chemical Physics}\ }\textbf {\bibinfo {volume} {103}},\
  \bibinfo {pages} {8577} (\bibinfo {year} {1995})}\BibitemShut {NoStop}%
\bibitem [{\citenamefont {Lavrinenko}\ \emph {et~al.}(2016)\citenamefont
  {Lavrinenko}, \citenamefont {Morozov},\ and\ \citenamefont
  {Valuev}}]{Lavrinenko:2016}%
  \BibitemOpen
  \bibfield  {author} {\bibinfo {author} {\bibfnamefont {Y.~S.}\ \bibnamefont
  {Lavrinenko}}, \bibinfo {author} {\bibfnamefont {I.}~\bibnamefont
  {Morozov}},\ and\ \bibinfo {author} {\bibfnamefont {I.~A.}\ \bibnamefont
  {Valuev}},\ }\href {https://doi.org/https://doi.org/10.1002/ctpp.201500139}
  {\bibfield  {journal} {\bibinfo  {journal} {Contributions to Plasma Physics}\
  }\textbf {\bibinfo {volume} {56}},\ \bibinfo {pages} {448} (\bibinfo {year}
  {2016})}\BibitemShut {NoStop}%
\bibitem [{\citenamefont {Brush}\ \emph {et~al.}(1966)\citenamefont {Brush},
  \citenamefont {Sahlin},\ and\ \citenamefont {Teller}}]{Brush:JChemPhys:1966}%
  \BibitemOpen
  \bibfield  {author} {\bibinfo {author} {\bibfnamefont {S.~G.}\ \bibnamefont
  {Brush}}, \bibinfo {author} {\bibfnamefont {H.~L.}\ \bibnamefont {Sahlin}},\
  and\ \bibinfo {author} {\bibfnamefont {E.}~\bibnamefont {Teller}},\ }\href
  {https://doi.org/10.1063/1.1727895} {\bibfield  {journal} {\bibinfo
  {journal} {The Journal of Chemical Physics}\ }\textbf {\bibinfo {volume}
  {45}},\ \bibinfo {pages} {2102} (\bibinfo {year} {1966})}\BibitemShut
  {NoStop}%
\bibitem [{\citenamefont {Hansen}(1973)}]{Hansen:PRA:1973}%
  \BibitemOpen
  \bibfield  {author} {\bibinfo {author} {\bibfnamefont {J.~P.}\ \bibnamefont
  {Hansen}},\ }\href {https://doi.org/10.1103/PhysRevA.8.3096} {\bibfield
  {journal} {\bibinfo  {journal} {Phys. Rev. A}\ }\textbf {\bibinfo {volume}
  {8}},\ \bibinfo {pages} {3096} (\bibinfo {year} {1973})}\BibitemShut
  {NoStop}%
\bibitem [{\citenamefont {Lucco~Castello}\ and\ \citenamefont
  {Tolias}(2022)}]{Lucco:PRE:2022}%
  \BibitemOpen
  \bibfield  {author} {\bibinfo {author} {\bibfnamefont {F.}~\bibnamefont
  {Lucco~Castello}}\ and\ \bibinfo {author} {\bibfnamefont {P.}~\bibnamefont
  {Tolias}},\ }\href {https://doi.org/10.1103/PhysRevE.105.015208} {\bibfield
  {journal} {\bibinfo  {journal} {Phys. Rev. E}\ }\textbf {\bibinfo {volume}
  {105}},\ \bibinfo {pages} {015208} (\bibinfo {year} {2022})}\BibitemShut
  {NoStop}%
\bibitem [{\citenamefont {Storer}(1968)}]{Storer:JMP:1968}%
  \BibitemOpen
  \bibfield  {author} {\bibinfo {author} {\bibfnamefont {R.~G.}\ \bibnamefont
  {Storer}},\ }\href {https://doi.org/10.1063/1.1664666} {\bibfield  {journal}
  {\bibinfo  {journal} {Journal of Mathematical Physics}\ }\textbf {\bibinfo
  {volume} {9}},\ \bibinfo {pages} {964} (\bibinfo {year} {1968})}\BibitemShut
  {NoStop}%
\bibitem [{\citenamefont {Militzer}(2016)}]{MILITZER201688}%
  \BibitemOpen
  \bibfield  {author} {\bibinfo {author} {\bibfnamefont {B.}~\bibnamefont
  {Militzer}},\ }\href
  {https://doi.org/https://doi.org/10.1016/j.cpc.2016.03.011} {\bibfield
  {journal} {\bibinfo  {journal} {Computer Physics Communications}\ }\textbf
  {\bibinfo {volume} {204}},\ \bibinfo {pages} {88} (\bibinfo {year}
  {2016})}\BibitemShut {NoStop}%
\bibitem [{\citenamefont {Fraser}\ \emph {et~al.}(1996)\citenamefont {Fraser},
  \citenamefont {Foulkes}, \citenamefont {Rajagopal}, \citenamefont {Needs},
  \citenamefont {Kenny},\ and\ \citenamefont {Williamson}}]{Fraser:PRB:1996}%
  \BibitemOpen
  \bibfield  {author} {\bibinfo {author} {\bibfnamefont {L.~M.}\ \bibnamefont
  {Fraser}}, \bibinfo {author} {\bibfnamefont {W.~M.~C.}\ \bibnamefont
  {Foulkes}}, \bibinfo {author} {\bibfnamefont {G.}~\bibnamefont {Rajagopal}},
  \bibinfo {author} {\bibfnamefont {R.~J.}\ \bibnamefont {Needs}}, \bibinfo
  {author} {\bibfnamefont {S.~D.}\ \bibnamefont {Kenny}},\ and\ \bibinfo
  {author} {\bibfnamefont {A.~J.}\ \bibnamefont {Williamson}},\ }\href
  {https://doi.org/10.1103/PhysRevB.53.1814} {\bibfield  {journal} {\bibinfo
  {journal} {Phys. Rev. B}\ }\textbf {\bibinfo {volume} {53}},\ \bibinfo
  {pages} {1814} (\bibinfo {year} {1996})}\BibitemShut {NoStop}%
\bibitem [{\citenamefont {Larkin}\ \emph {et~al.}(2022)\citenamefont {Larkin},
  \citenamefont {Filinov},\ and\ \citenamefont {Levashov}}]{math10132270}%
  \BibitemOpen
  \bibfield  {author} {\bibinfo {author} {\bibfnamefont {A.}~\bibnamefont
  {Larkin}}, \bibinfo {author} {\bibfnamefont {V.}~\bibnamefont {Filinov}},\
  and\ \bibinfo {author} {\bibfnamefont {P.}~\bibnamefont {Levashov}},\ }\href
  {https://doi.org/10.3390/math10132270} {\bibfield  {journal} {\bibinfo
  {journal} {Mathematics}\ }\textbf {\bibinfo {volume} {10}},\ \bibinfo {pages}
  {2270} (\bibinfo {year} {2022})}\BibitemShut {NoStop}%
\bibitem [{\citenamefont {B\"ohme}\ \emph {et~al.}(2023)\citenamefont
  {B\"ohme}, \citenamefont {Moldabekov}, \citenamefont {Vorberger},\ and\
  \citenamefont {Dornheim}}]{Bohme:PRE:2023}%
  \BibitemOpen
  \bibfield  {author} {\bibinfo {author} {\bibfnamefont {M.}~\bibnamefont
  {B\"ohme}}, \bibinfo {author} {\bibfnamefont {Z.~A.}\ \bibnamefont
  {Moldabekov}}, \bibinfo {author} {\bibfnamefont {J.}~\bibnamefont
  {Vorberger}},\ and\ \bibinfo {author} {\bibfnamefont {T.}~\bibnamefont
  {Dornheim}},\ }\href {https://doi.org/10.1103/PhysRevE.107.015206} {\bibfield
   {journal} {\bibinfo  {journal} {Phys. Rev. E}\ }\textbf {\bibinfo {volume}
  {107}},\ \bibinfo {pages} {015206} (\bibinfo {year} {2023})}\BibitemShut
  {NoStop}%
\bibitem [{\citenamefont {Killian}\ \emph {et~al.}(1999)\citenamefont
  {Killian}, \citenamefont {Kulin}, \citenamefont {Bergeson}, \citenamefont
  {Orozco}, \citenamefont {Orzel},\ and\ \citenamefont
  {Rolston}}]{Killian:PRL:1999}%
  \BibitemOpen
  \bibfield  {author} {\bibinfo {author} {\bibfnamefont {T.~C.}\ \bibnamefont
  {Killian}}, \bibinfo {author} {\bibfnamefont {S.}~\bibnamefont {Kulin}},
  \bibinfo {author} {\bibfnamefont {S.~D.}\ \bibnamefont {Bergeson}}, \bibinfo
  {author} {\bibfnamefont {L.~A.}\ \bibnamefont {Orozco}}, \bibinfo {author}
  {\bibfnamefont {C.}~\bibnamefont {Orzel}},\ and\ \bibinfo {author}
  {\bibfnamefont {S.~L.}\ \bibnamefont {Rolston}},\ }\href
  {https://doi.org/10.1103/PhysRevLett.83.4776} {\bibfield  {journal} {\bibinfo
   {journal} {Phys. Rev. Lett.}\ }\textbf {\bibinfo {volume} {83}},\ \bibinfo
  {pages} {4776} (\bibinfo {year} {1999})}\BibitemShut {NoStop}%
\bibitem [{\citenamefont {Kulin}\ \emph {et~al.}(2000)\citenamefont {Kulin},
  \citenamefont {Killian}, \citenamefont {Bergeson},\ and\ \citenamefont
  {Rolston}}]{Kulin:PRL:2000}%
  \BibitemOpen
  \bibfield  {author} {\bibinfo {author} {\bibfnamefont {S.}~\bibnamefont
  {Kulin}}, \bibinfo {author} {\bibfnamefont {T.~C.}\ \bibnamefont {Killian}},
  \bibinfo {author} {\bibfnamefont {S.~D.}\ \bibnamefont {Bergeson}},\ and\
  \bibinfo {author} {\bibfnamefont {S.~L.}\ \bibnamefont {Rolston}},\ }\href
  {https://doi.org/10.1103/PhysRevLett.85.318} {\bibfield  {journal} {\bibinfo
  {journal} {Phys. Rev. Lett.}\ }\textbf {\bibinfo {volume} {85}},\ \bibinfo
  {pages} {318} (\bibinfo {year} {2000})}\BibitemShut {NoStop}%
\bibitem [{\citenamefont {Killian}\ \emph {et~al.}(2007)\citenamefont
  {Killian}, \citenamefont {Pattard}, \citenamefont {Pohl},\ and\ \citenamefont
  {Rost}}]{KILLIAN200777}%
  \BibitemOpen
  \bibfield  {author} {\bibinfo {author} {\bibfnamefont {T.}~\bibnamefont
  {Killian}}, \bibinfo {author} {\bibfnamefont {T.}~\bibnamefont {Pattard}},
  \bibinfo {author} {\bibfnamefont {T.}~\bibnamefont {Pohl}},\ and\ \bibinfo
  {author} {\bibfnamefont {J.}~\bibnamefont {Rost}},\ }\href
  {https://doi.org/https://doi.org/10.1016/j.physrep.2007.04.007} {\bibfield
  {journal} {\bibinfo  {journal} {Physics Reports}\ }\textbf {\bibinfo {volume}
  {449}},\ \bibinfo {pages} {77} (\bibinfo {year} {2007})}\BibitemShut
  {NoStop}%
\bibitem [{\citenamefont {Guo}\ \emph {et~al.}(2010)\citenamefont {Guo},
  \citenamefont {Lu},\ and\ \citenamefont {Han}}]{Guo:PRE:2010}%
  \BibitemOpen
  \bibfield  {author} {\bibinfo {author} {\bibfnamefont {L.}~\bibnamefont
  {Guo}}, \bibinfo {author} {\bibfnamefont {R.~H.}\ \bibnamefont {Lu}},\ and\
  \bibinfo {author} {\bibfnamefont {S.~S.}\ \bibnamefont {Han}},\ }\href
  {https://doi.org/10.1103/PhysRevE.81.046406} {\bibfield  {journal} {\bibinfo
  {journal} {Phys. Rev. E}\ }\textbf {\bibinfo {volume} {81}},\ \bibinfo
  {pages} {046406} (\bibinfo {year} {2010})}\BibitemShut {NoStop}%
\bibitem [{\citenamefont {Dornheim}(2019)}]{Dornheim:PRE:2019}%
  \BibitemOpen
  \bibfield  {author} {\bibinfo {author} {\bibfnamefont {T.}~\bibnamefont
  {Dornheim}},\ }\href {https://doi.org/10.1103/PhysRevE.100.023307} {\bibfield
   {journal} {\bibinfo  {journal} {Phys. Rev. E}\ }\textbf {\bibinfo {volume}
  {100}},\ \bibinfo {pages} {023307} (\bibinfo {year} {2019})}\BibitemShut
  {NoStop}%
\bibitem [{\citenamefont {Demyanov}\ and\ \citenamefont
  {Levashov}(2022{\natexlab{a}})}]{Demyanov:ContrPlasPhys:2022}%
  \BibitemOpen
  \bibfield  {author} {\bibinfo {author} {\bibfnamefont {G.~S.}\ \bibnamefont
  {Demyanov}}\ and\ \bibinfo {author} {\bibfnamefont {P.~R.}\ \bibnamefont
  {Levashov}},\ }\href {https://doi.org/https://doi.org/10.1002/ctpp.202200100}
  {\bibfield  {journal} {\bibinfo  {journal} {Contributions to Plasma Physics}\
  }\textbf {\bibinfo {volume} {62}},\ \bibinfo {pages} {e202200100} (\bibinfo
  {year} {2022}{\natexlab{a}})}\BibitemShut {NoStop}%
\bibitem [{\citenamefont {Ewald}(1921)}]{Ewald:1921}%
  \BibitemOpen
  \bibfield  {author} {\bibinfo {author} {\bibfnamefont {P.~P.}\ \bibnamefont
  {Ewald}},\ }\href {https://doi.org/https://doi.org/10.1002/andp.19213690304}
  {\bibfield  {journal} {\bibinfo  {journal} {Annalen der Physik}\ }\textbf
  {\bibinfo {volume} {369}},\ \bibinfo {pages} {253} (\bibinfo {year}
  {1921})}\BibitemShut {NoStop}%
\bibitem [{\citenamefont {de~Leeuw}\ \emph {et~al.}(1980)\citenamefont
  {de~Leeuw}, \citenamefont {Perram},\ and\ \citenamefont
  {Smith}}]{Leeuw:1980}%
  \BibitemOpen
  \bibfield  {author} {\bibinfo {author} {\bibfnamefont {S.~W.}\ \bibnamefont
  {de~Leeuw}}, \bibinfo {author} {\bibfnamefont {J.~W.}\ \bibnamefont
  {Perram}},\ and\ \bibinfo {author} {\bibfnamefont {E.~R.}\ \bibnamefont
  {Smith}},\ }\href {https://doi.org/https://doi.org/10.1098/rspa.1980.0135}
  {\bibfield  {journal} {\bibinfo  {journal} {Proceedings of the Royal Society
  of London. A. Mathematical and Physical Sciences}\ }\textbf {\bibinfo
  {volume} {373}},\ \bibinfo {pages} {27} (\bibinfo {year} {1980})}\BibitemShut
  {NoStop}%
\bibitem [{\citenamefont {Demyanov}\ and\ \citenamefont
  {Levashov}(2022{\natexlab{b}})}]{Demyanov:JPhysA:2022}%
  \BibitemOpen
  \bibfield  {author} {\bibinfo {author} {\bibfnamefont {G.~S.}\ \bibnamefont
  {Demyanov}}\ and\ \bibinfo {author} {\bibfnamefont {P.~R.}\ \bibnamefont
  {Levashov}},\ }\href {https://doi.org/10.1088/1751-8121/ac870b} {\bibfield
  {journal} {\bibinfo  {journal} {Journal of Physics A: Mathematical and
  Theoretical}\ }\textbf {\bibinfo {volume} {55}},\ \bibinfo {pages} {385202}
  (\bibinfo {year} {2022}{\natexlab{b}})}\BibitemShut {NoStop}%
\bibitem [{\citenamefont {Demyanov}\ and\ \citenamefont
  {Levashov}(2022{\natexlab{c}})}]{DemyanovOCP:PRE:2022}%
  \BibitemOpen
  \bibfield  {author} {\bibinfo {author} {\bibfnamefont {G.~S.}\ \bibnamefont
  {Demyanov}}\ and\ \bibinfo {author} {\bibfnamefont {P.~R.}\ \bibnamefont
  {Levashov}},\ }\href {https://doi.org/10.1103/PhysRevE.106.015204} {\bibfield
   {journal} {\bibinfo  {journal} {Phys. Rev. E}\ }\textbf {\bibinfo {volume}
  {106}},\ \bibinfo {pages} {015204} (\bibinfo {year}
  {2022}{\natexlab{c}})}\BibitemShut {NoStop}%
\bibitem [{\citenamefont {Yakub}\ and\ \citenamefont
  {Ronchi}(2003)}]{Yakub:2003}%
  \BibitemOpen
  \bibfield  {author} {\bibinfo {author} {\bibfnamefont {E.}~\bibnamefont
  {Yakub}}\ and\ \bibinfo {author} {\bibfnamefont {C.}~\bibnamefont {Ronchi}},\
  }\href {https://doi.org/10.1063/1.1624364} {\bibfield  {journal} {\bibinfo
  {journal} {The Journal of Chemical Physics}\ }\textbf {\bibinfo {volume}
  {119}},\ \bibinfo {pages} {11556} (\bibinfo {year} {2003})}\BibitemShut
  {NoStop}%
\bibitem [{\citenamefont {Yakub}\ and\ \citenamefont
  {Ronchi}(2005)}]{Yakub:2005}%
  \BibitemOpen
  \bibfield  {author} {\bibinfo {author} {\bibfnamefont {E.}~\bibnamefont
  {Yakub}}\ and\ \bibinfo {author} {\bibfnamefont {C.}~\bibnamefont {Ronchi}},\
  }\href {https://doi.org/10.1007/s10909-005-5451-5} {\bibfield  {journal}
  {\bibinfo  {journal} {Journal of Low Temperature Physics}\ }\textbf {\bibinfo
  {volume} {139}},\ \bibinfo {pages} {633} (\bibinfo {year}
  {2005})}\BibitemShut {NoStop}%
\bibitem [{\citenamefont {Yakub}(2006)}]{Yakub:JPA:2006}%
  \BibitemOpen
  \bibfield  {author} {\bibinfo {author} {\bibfnamefont {E.}~\bibnamefont
  {Yakub}},\ }\href@noop {} {\bibfield  {journal} {\bibinfo  {journal} {Journal
  of Physics A: Mathematical and General}\ }\textbf {\bibinfo {volume} {39}},\
  \bibinfo {pages} {4643} (\bibinfo {year} {2006})}\BibitemShut {NoStop}%
\bibitem [{\citenamefont {Ichimaru}(1982)}]{Ichimaru:1982}%
  \BibitemOpen
  \bibfield  {author} {\bibinfo {author} {\bibfnamefont {S.}~\bibnamefont
  {Ichimaru}},\ }\href {https://doi.org/10.1103/RevModPhys.54.1017} {\bibfield
  {journal} {\bibinfo  {journal} {Rev. Mod. Phys.}\ }\textbf {\bibinfo {volume}
  {54}},\ \bibinfo {pages} {1017} (\bibinfo {year} {1982})}\BibitemShut
  {NoStop}%
\bibitem [{\citenamefont {Fukuda}\ and\ \citenamefont
  {Nakamura}(2022)}]{Fukuda2022}%
  \BibitemOpen
  \bibfield  {author} {\bibinfo {author} {\bibfnamefont {I.}~\bibnamefont
  {Fukuda}}\ and\ \bibinfo {author} {\bibfnamefont {H.}~\bibnamefont
  {Nakamura}},\ }\href {https://doi.org/10.1007/s12551-022-01029-2} {\bibfield
  {journal} {\bibinfo  {journal} {Biophysical Reviews}\ }\textbf {\bibinfo
  {volume} {14}},\ \bibinfo {pages} {1315} (\bibinfo {year}
  {2022})}\BibitemShut {NoStop}%
\bibitem [{\citenamefont {Jha}\ \emph {et~al.}(2010)\citenamefont {Jha},
  \citenamefont {Sknepnek}, \citenamefont {Guerrero-Garc{\'i}a},\ and\
  \citenamefont {Olvera de~la Cruz}}]{Jha:2010}%
  \BibitemOpen
  \bibfield  {author} {\bibinfo {author} {\bibfnamefont {P.~K.}\ \bibnamefont
  {Jha}}, \bibinfo {author} {\bibfnamefont {R.}~\bibnamefont {Sknepnek}},
  \bibinfo {author} {\bibfnamefont {G.~I.}\ \bibnamefont
  {Guerrero-Garc{\'i}a}},\ and\ \bibinfo {author} {\bibfnamefont
  {M.}~\bibnamefont {Olvera de~la Cruz}},\ }\href@noop {} {\bibfield  {journal}
  {\bibinfo  {journal} {Journal of Chemical Theory and Computation}\ }\textbf
  {\bibinfo {volume} {6}},\ \bibinfo {pages} {3058} (\bibinfo {year}
  {2010})}\BibitemShut {NoStop}%
\bibitem [{\citenamefont {Feynman}(1972)}]{Feynman:1972:SMS}%
  \BibitemOpen
  \bibfield  {author} {\bibinfo {author} {\bibfnamefont {R.~P.}\ \bibnamefont
  {Feynman}},\ }\href@noop {} {\emph {\bibinfo {title} {{Statistical mechanics:
  a set of lectures by R. P. Feynman}}}},\ Frontiers in physics\ (\bibinfo
  {year} {1972})\ pp.\ \bibinfo {pages} {xii + 354},\ \bibinfo {note} {notes
  taken by R. Kikuchi and H. A. Feiveson. Edited by Jacob Shaham}\BibitemShut
  {NoStop}%
\bibitem [{\citenamefont {Kelbg}(1963{\natexlab{b}})}]{Kelbg:78:1963}%
  \BibitemOpen
  \bibfield  {author} {\bibinfo {author} {\bibfnamefont {G.}~\bibnamefont
  {Kelbg}},\ }\href {https://doi.org/https://doi.org/10.1002/andp.19634670703}
  {\bibfield  {journal} {\bibinfo  {journal} {Annalen der Physik}\ }\textbf
  {\bibinfo {volume} {467}},\ \bibinfo {pages} {354} (\bibinfo {year}
  {1963}{\natexlab{b}})}\BibitemShut {NoStop}%
\bibitem [{\citenamefont {Demyanov}\ and\ \citenamefont
  {Levashov}(2022{\natexlab{d}})}]{Demyanov:Kelbg:2022}%
  \BibitemOpen
  \bibfield  {author} {\bibinfo {author} {\bibfnamefont {G.~S.}\ \bibnamefont
  {Demyanov}}\ and\ \bibinfo {author} {\bibfnamefont {P.~R.}\ \bibnamefont
  {Levashov}},\ }\href {https://doi.org/10.48550/ARXIV.2205.09885} {\bibinfo
  {title} {{Derivation of the Kelbg potential/functional}}},\ \bibinfo
  {howpublished} {\url{https://arxiv.org/abs/2205.09885}} (\bibinfo {year}
  {2022}{\natexlab{d}})\BibitemShut {NoStop}%
\bibitem [{\citenamefont {Ebeling}\ \emph {et~al.}(1967)\citenamefont
  {Ebeling}, \citenamefont {Hoffmann},\ and\ \citenamefont
  {Kelbg}}]{Ebeling:CPP:1967}%
  \BibitemOpen
  \bibfield  {author} {\bibinfo {author} {\bibfnamefont {W.}~\bibnamefont
  {Ebeling}}, \bibinfo {author} {\bibfnamefont {H.~J.}\ \bibnamefont
  {Hoffmann}},\ and\ \bibinfo {author} {\bibfnamefont {G.}~\bibnamefont
  {Kelbg}},\ }\href {https://doi.org/https://doi.org/10.1002/ctpp.19670070307}
  {\bibfield  {journal} {\bibinfo  {journal} {Beiträge aus der Plasmaphysik}\
  }\textbf {\bibinfo {volume} {7}},\ \bibinfo {pages} {233} (\bibinfo {year}
  {1967})}\BibitemShut {NoStop}%
\bibitem [{\citenamefont {Lieb}(2004)}]{Lieb:2004}%
  \BibitemOpen
  \bibfield  {author} {\bibinfo {author} {\bibfnamefont {E.~H.}\ \bibnamefont
  {Lieb}},\ }\bibinfo {title} {{The Stability of Matter and Quantum
  Electrodynamics}},\ in\ \href {https://doi.org/10.1007/978-3-642-18623-3_7}
  {\emph {\bibinfo {booktitle} {Fundamental Physics --- Heisenberg and Beyond:
  Werner Heisenberg Centennial Symposium ``Developments in Modern
  Physics''}}},\ \bibinfo {editor} {edited by\ \bibinfo {editor} {\bibfnamefont
  {G.~W.}\ \bibnamefont {Buschhorn}}\ and\ \bibinfo {editor} {\bibfnamefont
  {J.}~\bibnamefont {Wess}}}\ (\bibinfo  {publisher} {Springer Berlin
  Heidelberg},\ \bibinfo {address} {Berlin, Heidelberg},\ \bibinfo {year}
  {2004})\ pp.\ \bibinfo {pages} {53--68}\BibitemShut {NoStop}%
\bibitem [{\citenamefont {Filinov}\ \emph {et~al.}(2004)\citenamefont
  {Filinov}, \citenamefont {Golubnychiy}, \citenamefont {Bonitz}, \citenamefont
  {Ebeling},\ and\ \citenamefont {Dufty}}]{FilinovA:PRE:2004}%
  \BibitemOpen
  \bibfield  {author} {\bibinfo {author} {\bibfnamefont {A.~V.}\ \bibnamefont
  {Filinov}}, \bibinfo {author} {\bibfnamefont {V.~O.}\ \bibnamefont
  {Golubnychiy}}, \bibinfo {author} {\bibfnamefont {M.}~\bibnamefont {Bonitz}},
  \bibinfo {author} {\bibfnamefont {W.}~\bibnamefont {Ebeling}},\ and\ \bibinfo
  {author} {\bibfnamefont {J.~W.}\ \bibnamefont {Dufty}},\ }\href
  {https://doi.org/10.1103/PhysRevE.70.046411} {\bibfield  {journal} {\bibinfo
  {journal} {Phys. Rev. E}\ }\textbf {\bibinfo {volume} {70}},\ \bibinfo
  {pages} {046411} (\bibinfo {year} {2004})}\BibitemShut {NoStop}%
\bibitem [{\citenamefont {Debye}\ and\ \citenamefont {E.}(1923)}]{Debye:1923}%
  \BibitemOpen
  \bibfield  {author} {\bibinfo {author} {\bibfnamefont {P.}~\bibnamefont
  {Debye}}\ and\ \bibinfo {author} {\bibfnamefont {H.}~\bibnamefont {E.}},\
  }\href {https://knowledge.electrochem.org/estir/hist/hist-12-Debye-1.pdf}
  {\bibfield  {journal} {\bibinfo  {journal} {Physikalische Zeitschrift}\
  }\textbf {\bibinfo {volume} {9}},\ \bibinfo {pages} {185} (\bibinfo {year}
  {1923})}\BibitemShut {NoStop}%
\bibitem [{\citenamefont {Wang}\ and\ \citenamefont
  {Landau}(2001)}]{Wang:PRL:2001}%
  \BibitemOpen
  \bibfield  {author} {\bibinfo {author} {\bibfnamefont {F.}~\bibnamefont
  {Wang}}\ and\ \bibinfo {author} {\bibfnamefont {D.~P.}\ \bibnamefont
  {Landau}},\ }\href@noop {} {\bibfield  {journal} {\bibinfo  {journal}
  {Physical review letters}\ }\textbf {\bibinfo {volume} {86}},\ \bibinfo
  {pages} {2050} (\bibinfo {year} {2001})}\BibitemShut {NoStop}%
\bibitem [{\citenamefont {Galassi}\ \emph {et~al.}(2021)\citenamefont {Galassi}
  \emph {et~al.}}]{galassi2021scientific}%
  \BibitemOpen
  \bibfield  {author} {\bibinfo {author} {\bibfnamefont {M.}~\bibnamefont
  {Galassi}} \emph {et~al.},\ }\href {https://www.gnu.org/software/gsl/}
  {\bibinfo {title} {{GNU Scientific Library Reference Manual (3rd Ed.)}}}
  (\bibinfo {year} {2021})\BibitemShut {NoStop}%
\end{thebibliography}


\end{document}